\documentclass[a4paper,fleqn,usenatbib]{mnras}

\usepackage{mathptmx}

\usepackage[T1]{fontenc}
\usepackage{ae,aecompl}

\usepackage{graphicx}	
\usepackage{amsmath}	
\usepackage{amssymb}	
\usepackage{textcomp}
\usepackage{color}
\usepackage{multicol}
\usepackage{longtable}
\usepackage{rotating}
\usepackage{pdflscape}
\usepackage{natbib}
\usepackage{enumitem}
\usepackage{appendix}
\usepackage{url}
\usepackage{float}
\usepackage{tabularx}
\usepackage{booktabs}
\usepackage[caption = false]{subfig}
\usepackage{scalerel}

\newcommand\HI{H\protect\scaleto{$I$}{1.2ex}}
\newcommand\HII{H\protect\scaleto{$II$}{1.2ex}}

\title[Fornax A SFHs]{The star formation histories of galaxies in different stages of pre-processing in the Fornax A group\thanks{based on observations made with the South African Large Telescope (SALT).}}

\author[Loubser et al.]{S. I. Loubser$^{1,2}$\thanks{E-mail:Ilani.Loubser@nwu.ac.za (SIL)}, K. Mosia$^{1}$, P. Serra$^{3}$, D. Kleiner$^{4,3}$, R.F. Peletier$^{5}$, R.C. Kraan-Korteweg$^{6}$, 
\newauthor{E. Iodice$^{7}$, A. Loni$^{8}$, P. Kamphuis$^{9}$ and N. Zabel$^{6}$} \\
$^{1}$Centre for Space Research, North-West University, Potchefstroom 2520, South Africa\\
$^{2}$National Institute for Theoretical and Computational Sciences (NITheCS), Potchefstroom 2520, South Africa \\
$^{3}$INAF - Osservatorio Astronomico di Cagliari, Via della Scienza 5, I-09047 Selargius (CA), Italy \\
$^{4}$Netherlands Institute for Radio Astronomy (ASTRON), Oude Hoogeveensedijk 4, 7991 PD Dwingeloo, the Netherlands \\
$^{5}$Kapteyn Astronomical Institute, University of Groningen, P. O. Box 800, 9700 AV Groningen, The Netherlands \\
$^{6}$Department of Astronomy, University of Cape Town, Private Bag X3, Rondebosch 7701, South Africa \\
$^{7}$INAF - Astronomical Observatory of Capodimonte, Salita Moiariello 16, 80131, Naples, Italy \\
$^{8}$Armagh Observatory and Planetarium, College Hill, Armagh BT61 9DG, UK \\
$^{9}$Ruhr University Bochum, Faculty of Physics and Astronomy, Astronomical Institute (AIRUB) 44780 Bochum, Germany}

\date{Accepted 2023 November 23. Received 2023 November 23; in original form 2023 October 31}

\pubyear{2023}

\begin{document}
\label{firstpage}
\pagerange{\pageref{firstpage}--\pageref{lastpage}}
\maketitle

\begin{abstract}
We study the recent star formation histories of ten galaxies in the Fornax A galaxy group, on the outskirts of the Fornax cluster. The group galaxies are gas-rich, and their neutral atomic hydrogen (\HI{}) was studied in detail with observations from the MeerKAT telescope. This allowed them to be classified into different stages of pre-processing (early, ongoing, advanced). We use long-slit spectra obtained with the South African Large Telescope (SALT) to analyse stellar population indicators to constrain quenching timescales and to compare these to the \HI{} gas content of the galaxies. The H$\alpha$ equivalent width, EW(H$\alpha$), suggest that the pre-processing stage is closely related to the recent (< 10 Myr) specific Star Formation Rate (sSFR). The early-stage galaxy (NGC 1326B) is not yet quenched in its outer parts, while the ongoing-stage galaxies mostly have a distributed population of very young stars, though less so in their outer parts. The galaxies in the advanced stage of pre-processing show very low recent sSFR in the outer parts. Our results suggest that NGC 1326B, FCC 35 and FCC 46 underwent significantly different histories from secular evolution during the last Gyr. The fact that most galaxies are on the secular evolution sequence implies that pre-processing has a negligible effect on these galaxies compared to secular evolution. We find EW(H$\alpha$) to be a useful tool for classifying the stage of pre-processing in group galaxies. The recent sSFR and \HI{} morphology show that galaxies in the Fornax A vicinity are pre-processing from the outside in.   
\end{abstract}


\begin{keywords}
galaxies: evolution, galaxies: groups: individual: Fornax A, galaxies: ISM, galaxies: star formation
\end{keywords}


\section{Introduction}
\label{Section:introduction}


The physical process(es) that describe the transition of star-forming, blue galaxies to red galaxies are of fundamental importance to our understanding of galaxy evolution. Galaxies evolve along the star formation main sequence until star formation ceases, and the galaxy joins the passive red population. This halt in star formation may be due to a number of different physical processes \citep{Peng2010, Peng2012}. If there is a sharp break in the star formation history of a galaxy, implying a rapid transition to quiescence (< 1 Gyr), it is often referred to as `quenching'. This is in contrast to `ageing' which describes the normal star formation sequence in which a blue galaxy will eventually end up as a red galaxy with old stellar populations, without the need for a particular event that impedes star formation. Physical processes that lead to quenching are strongly determined by the environment. Although the influence of the environment has been well demonstrated in galaxy clusters \citep{Balogh1999, Boselli2016}, environmentally driven star formation quenching also occurs in groups (e.g. \citealt{Barsanti2018, Davies2019, Wang2022, Bidaran2022}). Different mechanisms are expected to quench star formation on different timescales (e.g. \citealt{Schawinski2014, Cortese2021}). Studying these timescales can constrain the physical processes that govern galaxy evolution. 


The Fornax galaxy cluster is still actively assembling \citep{Drinkwater2001, Iodice2017}. This, together with its proximity to us (20 Mpc), means that galaxy interactions and processes can be studied in great detail, and hence Fornax continues to be the focus of several major observing programmes across different wavelength regimes with Southern-hemisphere facilities. Campaigns include deep optical imaging from the Fornax Deep Survey (FDS; \citealt{Iodice2016, Venhola2018, Peletier2020}), deep \HI{} observations from the MeerKAT Fornax Survey (MFS; \citealt{Serra2016, Serra2023}), observations with ALMA \citep{Zabel2019, Morokuma2022}, integral field spectroscopy from the Fornax3D Survey (F3D; \citealt{Sarzi2018}), as well as the SAMI-Fornax Dwarf Survey \citep{Romero-Gomez2023a}. 

\HI{} can reveal the effects of physical processes, e.g. ram pressure, gas stripping, thermal heating, and tidal interactions (e.g., \citealt{Cowie1977, Nulsen1982, Koribalski2004, Rasmussen2008, Chung2009, Yoon2017, deBlok2018, Kleiner2019, Ramatsoku2020, Kleiner2021}), before the effects of these mechanisms manifest themselves in the stellar properties. The MFS survey is revealing a wealth of interesting findings regarding the effects on galaxies in the Fornax cluster environment. \citet{Serra2023} presents a sample of six galaxies with long, one-sided, starless \HI{} tails radially orientated (in projection) within the cluster. The properties of the \HI{} tails represent the first unambiguous evidence of the ram pressure that shapes the distribution of \HI{} in the Fornax cluster. Low-mass galaxies are especially susceptible to environmental effects, and the study of \HI{} in Fornax dwarf galaxies presented in \citet{Kleiner2023} suggests rapid removal of \HI{} from Fornax dwarfs, which produces a population of quiescent early-type dwarfs in the cluster.

Ram-pressure stripping is not widespread in poor clusters and galaxy groups, since it requires a dense intracluster medium (ICM) and large velocities of galaxies relative to it \citep{Gunn1972, Boselli2006}. The higher velocity dispersion in clusters leads to shorter interaction times, which reduces the effect of tidal interactions on galaxies \citep{Boselli2006}. In group environments, with lower velocity dispersions, mergers and strangulation are more prevalent (e.g. \citealt{Barnes1985, Moore1996, Zabludoff1998, McGee2009, Loubser2022, Jung2022}), and results show slower timescales for the quenching of star formation \citep{DeLucia2012, Wetzel2013}. Furthermore, the properties of group galaxies appear to correlate with the group halo mass and virial radius, also suggesting that quenching in groups is different from quenching in clusters \citep{Weinmann2006, Woo2013, Haines2015}. 

It is well known that galaxies in cluster environments are more likely to have suppressed star formation rates and less cold gas than galaxies of similar stellar mass in less dense environments. However, the suppression of star formation in the outer regions of clusters cannot be reproduced by models in which star formation is quenched in infalling galaxies only once they enter the cluster, but is consistent with some of them being first (gently) quenched within galaxy groups \citep{Haines2015}. This is also reproduced by simulations \citep{Bahe2013}. In particular, spiral galaxies with low star-formation rates in the outskirts of clusters requires the presence of external (environmental) mechanisms that can transform and quench galaxies before they fall into the cluster \citep{Zabludoff1996, Fujita2004, Porter2008, Haines2013, Fossati2019}. This non-secular evolution of galaxies that occurs in the group environment prior to entering a cluster is widely referred to as `pre-processing'.

Two virial radii south-west of the Fornax cluster centre lie the galaxy group, Fornax A, centred around NGC 1316\footnote{NGC 1316 is often also referred to as Fornax A. For clarity we refer to the group as Fornax A, and the central galaxy as NGC 1316 throughout the rest of the paper.} \citep{Schweizer1980, Mackie1998, Goudfrooij2001, Iodice2017, Raj2020, Kleiner2021}. The Fornax A group appears to be in an early stage of assembly with respect to the cluster core \citep{Raj2020}. The central velocity of the Fornax A group is 1778 km s$^{-1}$ with a velocity dispersion of 204 km s$^{-1}$ \citep{Maddox2019}. The environment is not as dense as that of the cluster core (with a velocity dispersion of 318 km s$^{-1}$; \citealt{Maddox2019}), and unlike Fornax, the photometric properties of galaxies do not exhibit any clear trend with group-centric distances \citep{Iodice2019, Raj2020}. It is suggested that NGC 1316 itself formed about 1 to 2 Gyr ago through a merger between a lenticular and a Milky Way-like galaxy \citep{Lanz2010, Serra2019}. Together with subsequent intragroup interactions \citep{Schweizer1980, Iodice2017}, this event supplied the intragroup medium (IGrM) with neutral and ionised gas \citep{Kleiner2021}. Six of the nine late-type galaxies in Fornax A show an up-bending break in their light profiles (i.e., steeper towards the centre), suggestive either of strangulation slowly stopping star formation in their outskirts, or enhanced star formation in the outer discs (see discussion in \citealt{Raj2020}). Many of the details of the physical processes at work on galaxies in Fornax A are not yet clear \citep{Raj2020, Kleiner2021}.

The Fornax A group is an ideal system to study pre-processing in group environments. It is located at the cluster-centric distance (two virial radii) where pre-processing is believed to occur \citep{Lewis2002, Fujita2004, Mahajan2012, Bahe2013, Haines2015}. Using MeerKAT commissioning data, \citet{Kleiner2021} classified the Fornax A galaxies detected in \HI{} into different stages of pre-processing (early, ongoing, advanced) according to their neutral hydrogen (\HI{}) morphology, content, and position relative to gas scaling relations (atomic and molecular). 


Constraining time scales of quenching can be very informative. For instance, starvation implies longer time scales for a galaxy to cease its star formation (on the order of a few Gyr; \citealt{Wetzel2013, Boselli2014}) compared to shorter time scales due to active gas removal, e.g., ram-pressure stripping (on the order of a few hundreds of Myr; see \citealt{Cortese2021} for a discussion). To constrain time scales, we need to probe the stellar populations of galaxies. This is particularly useful where we can combine or compare stellar populations with detailed studies of the cold gas distribution and kinematics probing relatively recent gas removal (see \citet{Loni2023} for an example of a Fornax member, NGC 1436). Since observations only provide a single snapshot during the evolution of a galaxy, we need to devise parameters that can describe the specific Star Formation Rate (sSFR) of a given galaxy over different time scales. For example, the H$\alpha$ emission line from \HII{} regions traces the recent SFR on the order of the last 10 Myr, while the H$\delta$ or D4000 \AA{} absorption features and the $g-r$ colour roughly trace the SFR averaged over the last 800 Myr \citep{Balogh1999, Kauffmann2003}. Together, these optical features allow for a view into the change in the star formation rate (e.g., \citealt{Wang2020, Corcho2021, Weibel2023, Corcho2023, Corcho2023b}).

In this paper, we study the stellar populations of ten galaxies in the Fornax A galaxy group. Nine of these galaxies were classified into different stages of pre-processing based on highly resolved MeerKAT observations \citep{Kleiner2021}. Here, we analyse stellar population indicators to compare the stellar populations of the galaxies to their gas content, and constrain quenching time scales. We measure profiles of the equivalent width of H$\alpha$ (EW(H$\alpha$)), construct an ageing diagram of EW(H$\alpha$) against $g-r$ colours for the galaxies, and fit stellar population models to describe the star formation histories of the galaxies, both in their centres and their outskirts.

In Section \ref{data}, we describe our South African Large Telescope (SALT) observations, as well as the existing optical photometric data and \HI{} data that we draw upon. In Section \ref{measurements}, we present our emission line measurements, in particular the equivalent width of H$\alpha$. We convert our measurements into parameters that probe stellar populations of different ages in Section \ref{parameters}, and do full-spectrum fitting of stellar population models in Section \ref{stelpopmodels}. We combine our findings with previous results from multiwavelength data, and use it to interpret the process(es) at work in Fornax A in Section \ref{interpret}. We summarise our conclusions in Section \ref{summary}. 


\section{Data}
\label{data}

\subsection{SALT observations and data reduction} 

\begin{figure*}
\captionsetup[subfloat]{farskip=-0pt,captionskip=-0pt}
\centering
\subfloat{\includegraphics[scale=0.23]{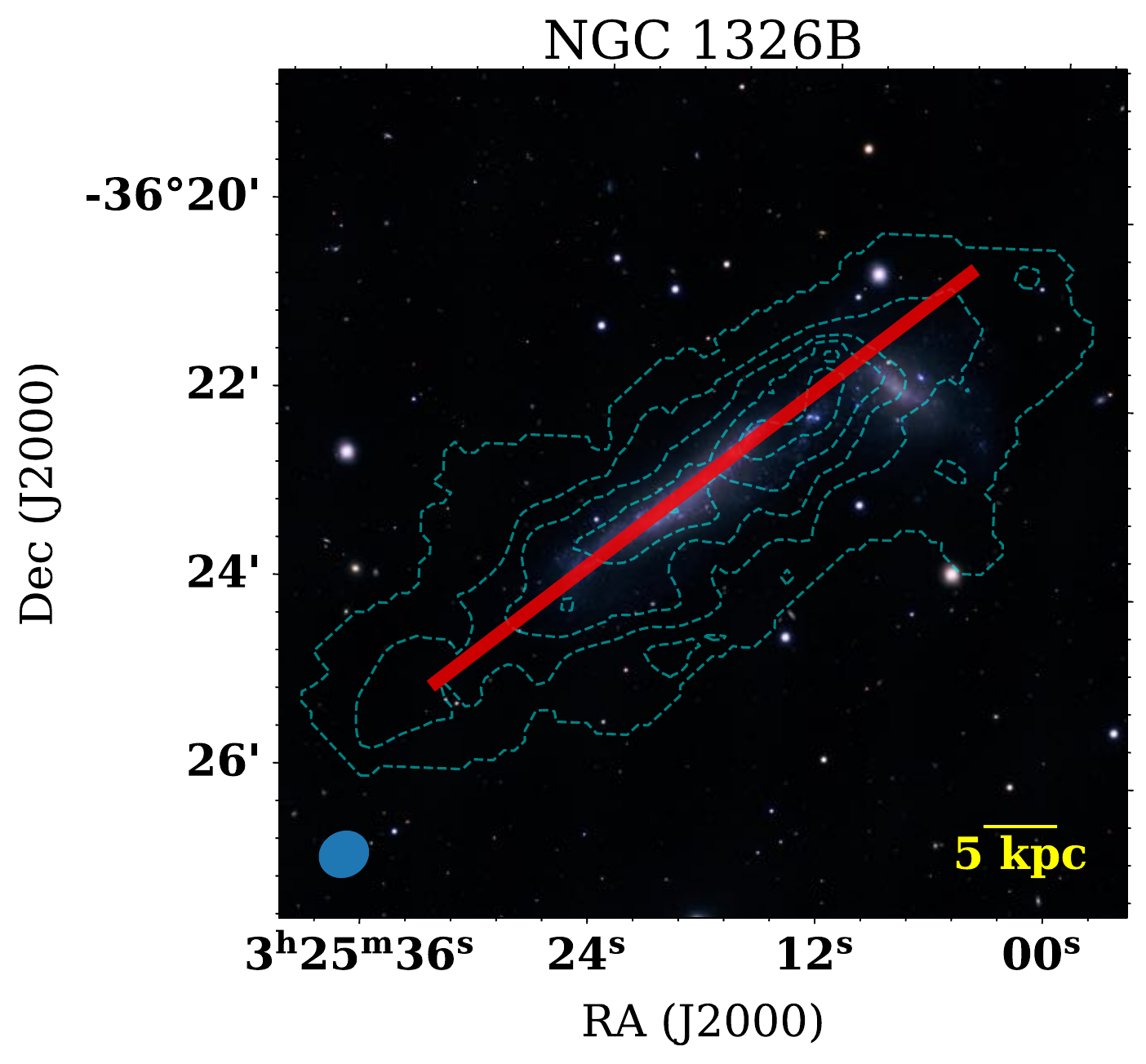}} 
\ \ \ \ \subfloat{\includegraphics[scale=0.23]{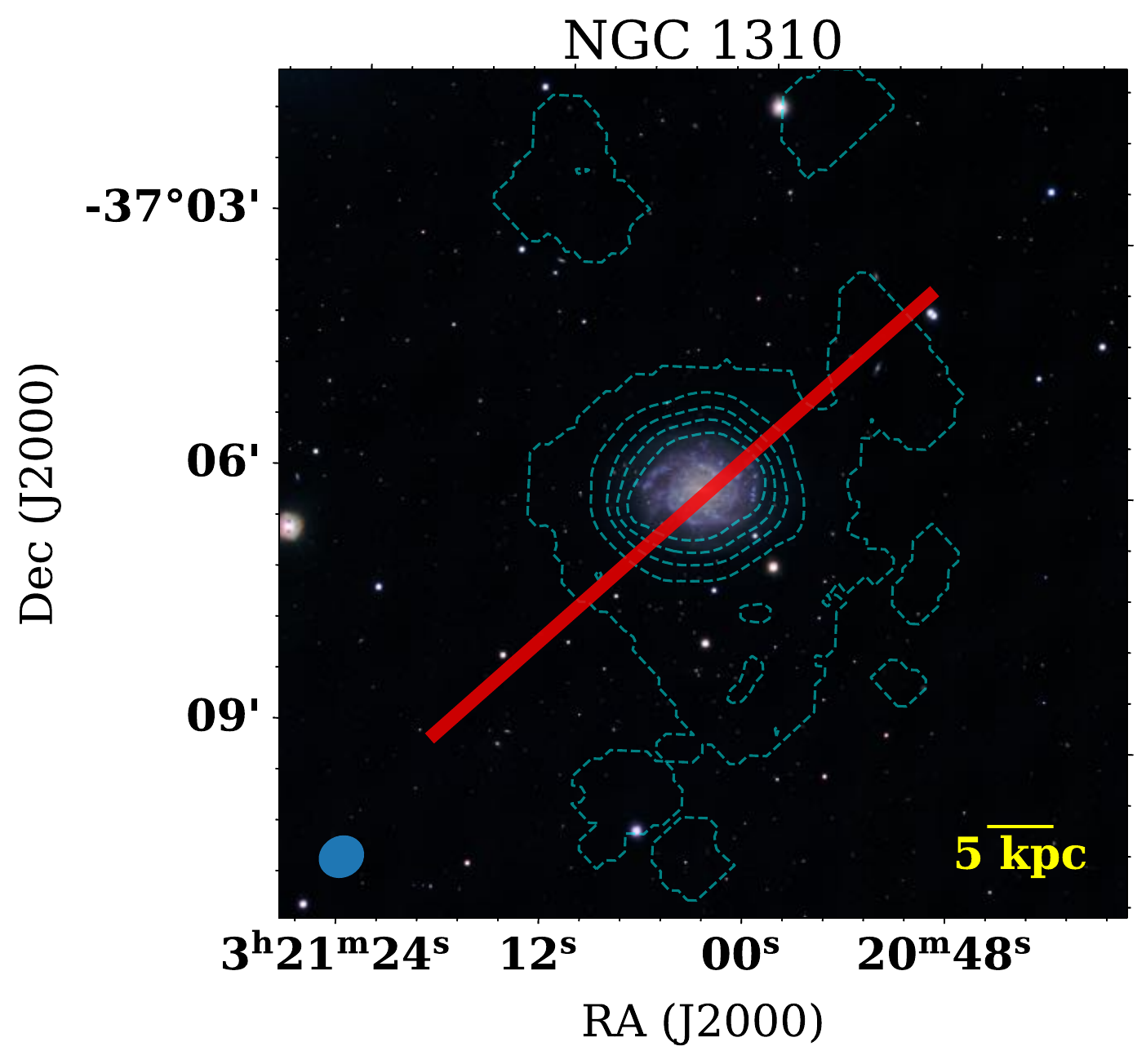}} 
\ \ \ \ \subfloat{\includegraphics[scale=0.23]{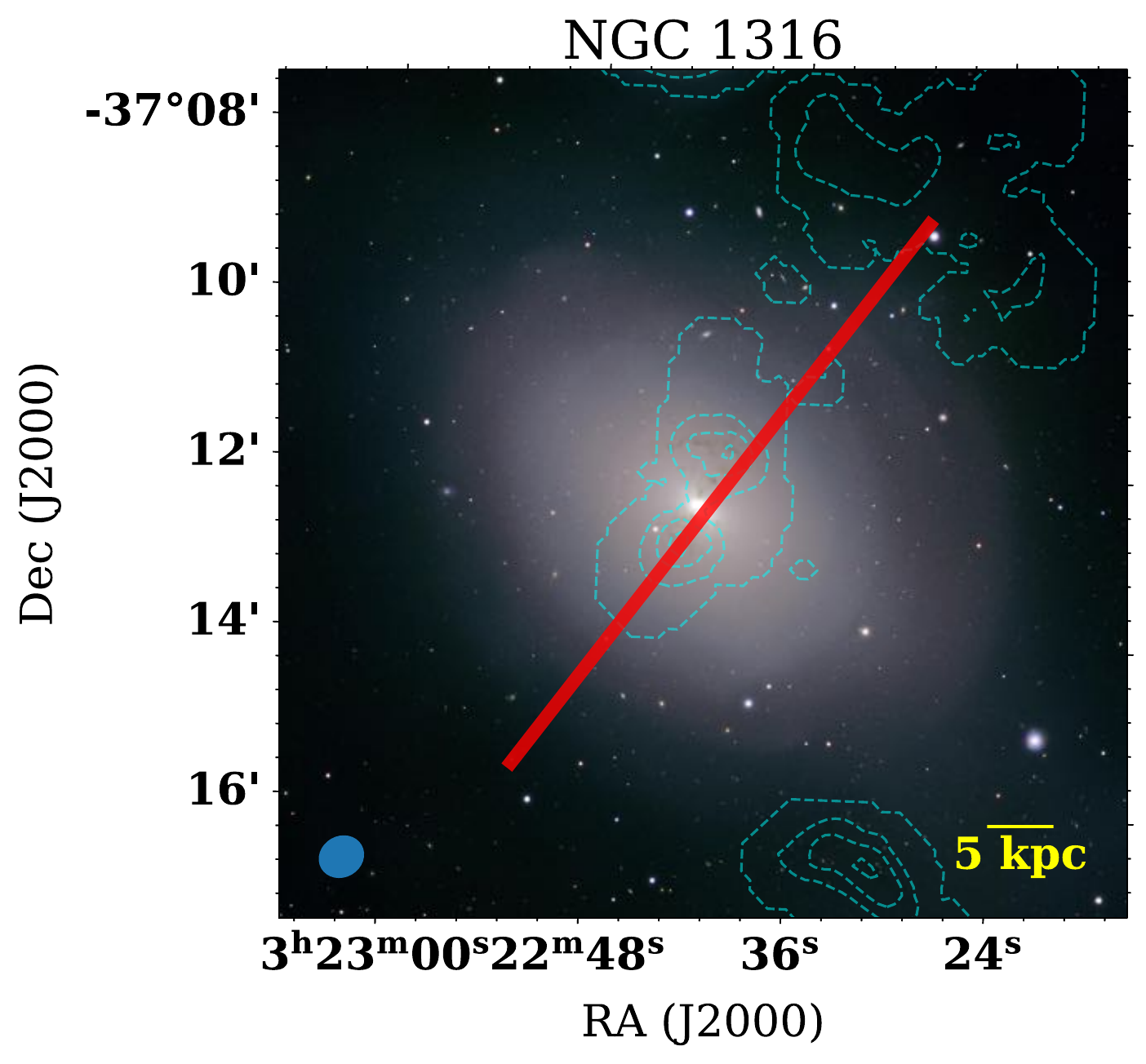}} \\
\subfloat{\includegraphics[scale=0.23]{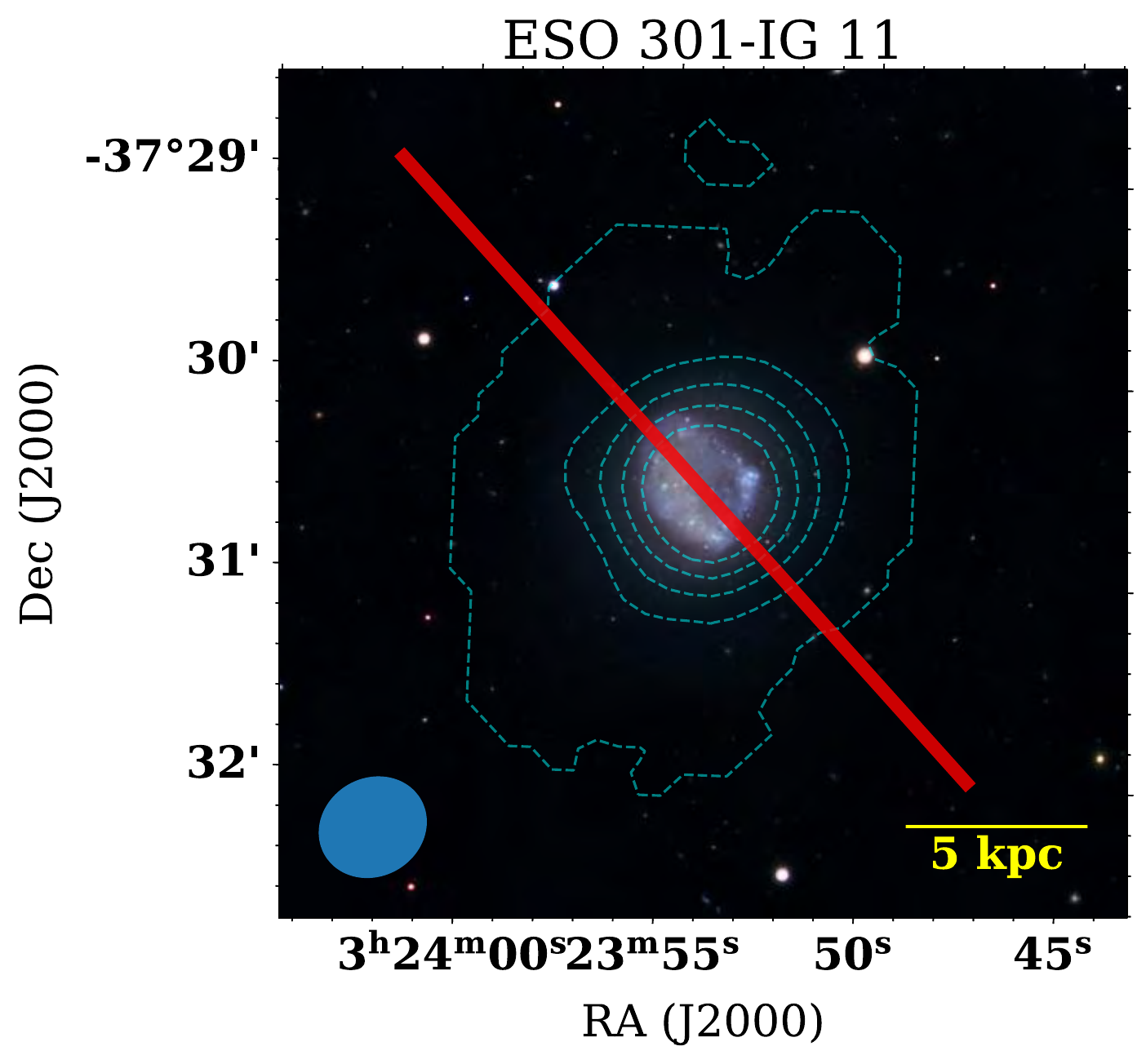}}
\ \ \ \ \subfloat{\includegraphics[scale=0.23]{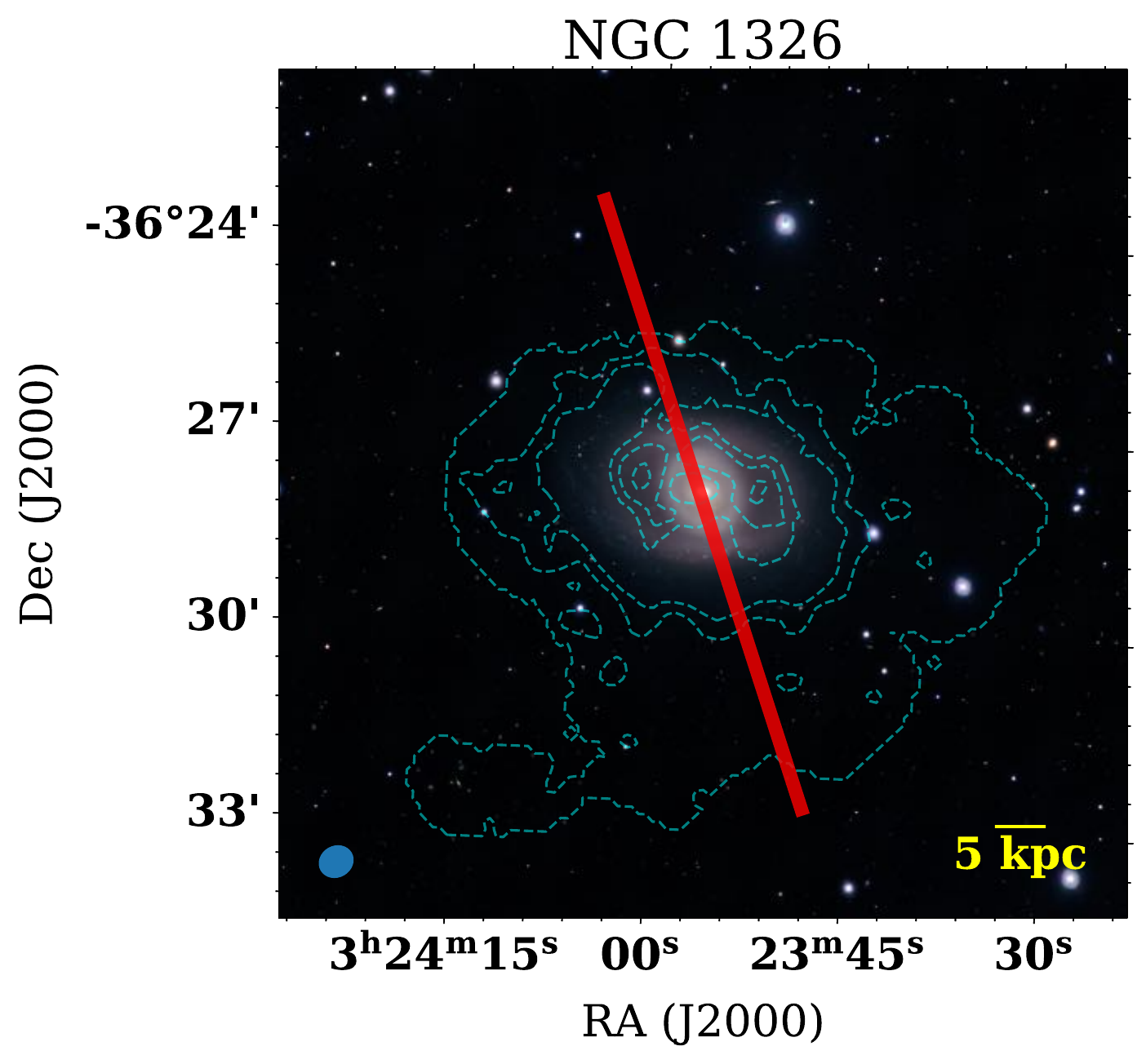}} 
\ \ \ \ \subfloat{\includegraphics[scale=0.23]{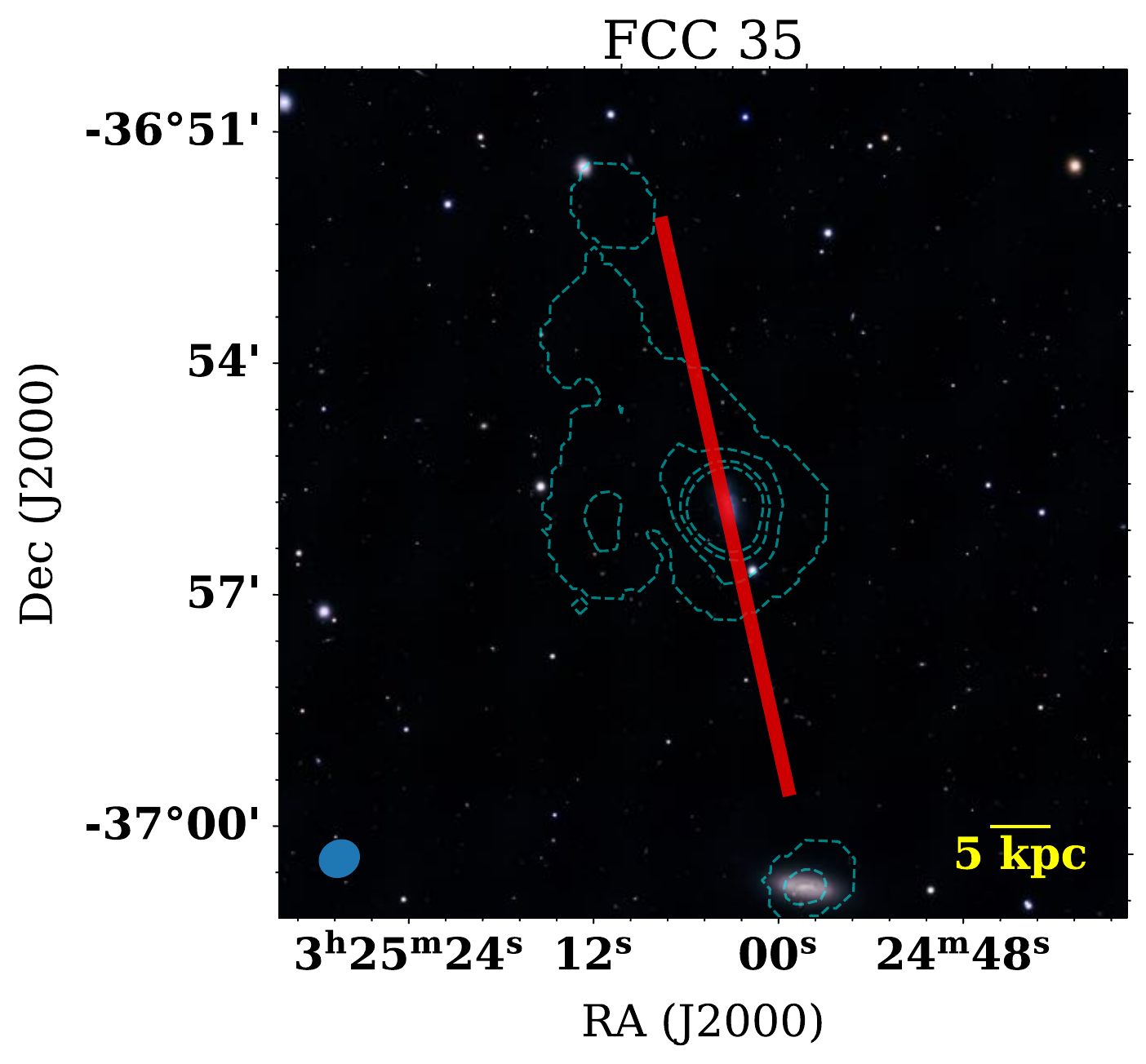}} \\
\subfloat{\includegraphics[scale=0.23]{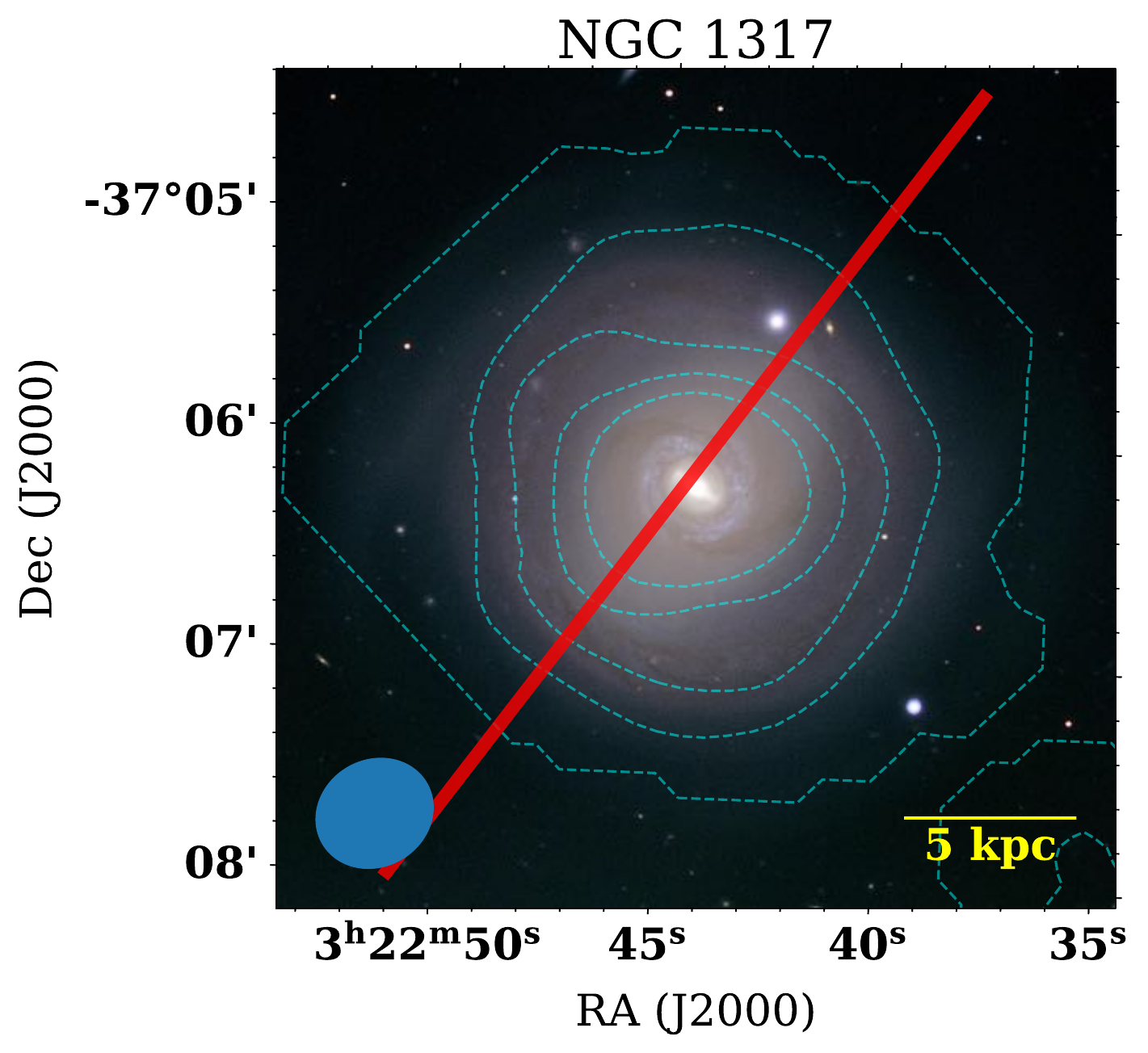}}
\subfloat{\includegraphics[scale=0.23]{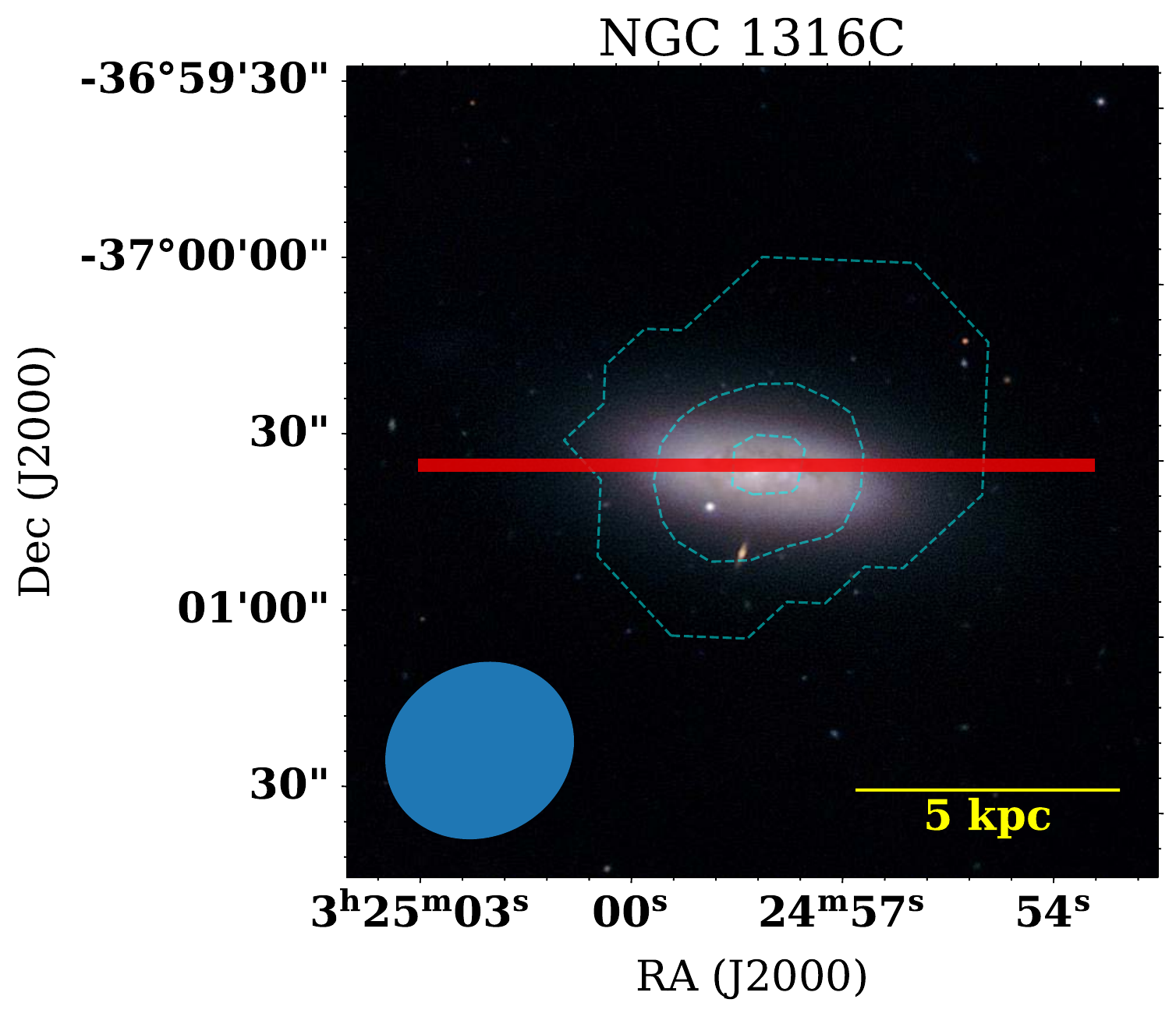}}
\subfloat{\includegraphics[scale=0.23]{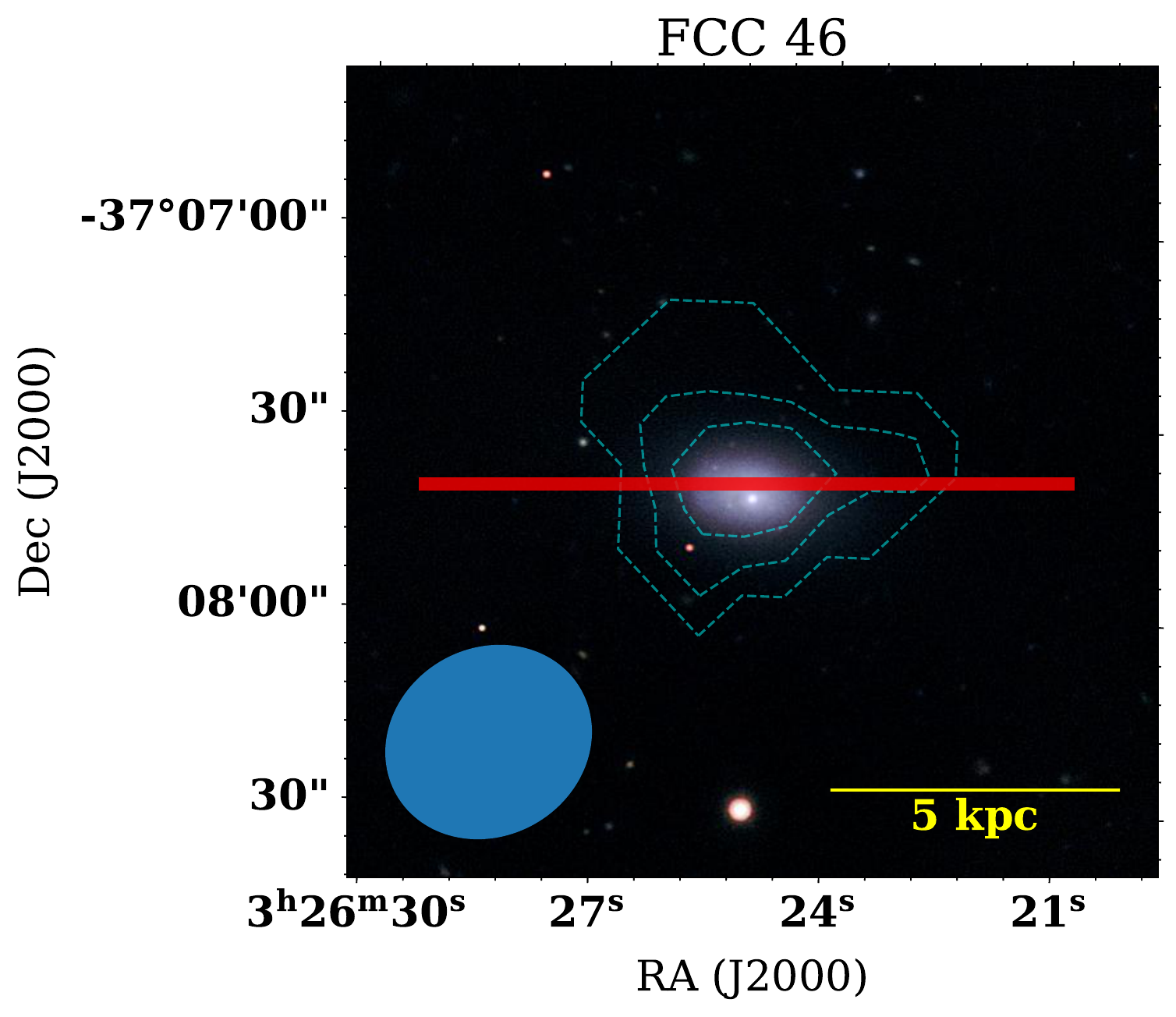}} \\
   \caption{Optical, composite ($g$-, $r$-, and $i$-band) FDS image of the galaxy with \HI{} from \citet{Kleiner2021} overlaid as contours. The position and orientation of the SALT slits are indicated in red, and the \HI{} beam size is shown in the bottom left corner.}
\label{fig:NGC1326BSFH}
\end{figure*}

\begin{figure}
\centering
\subfloat{\includegraphics[scale=0.26, angle=90]{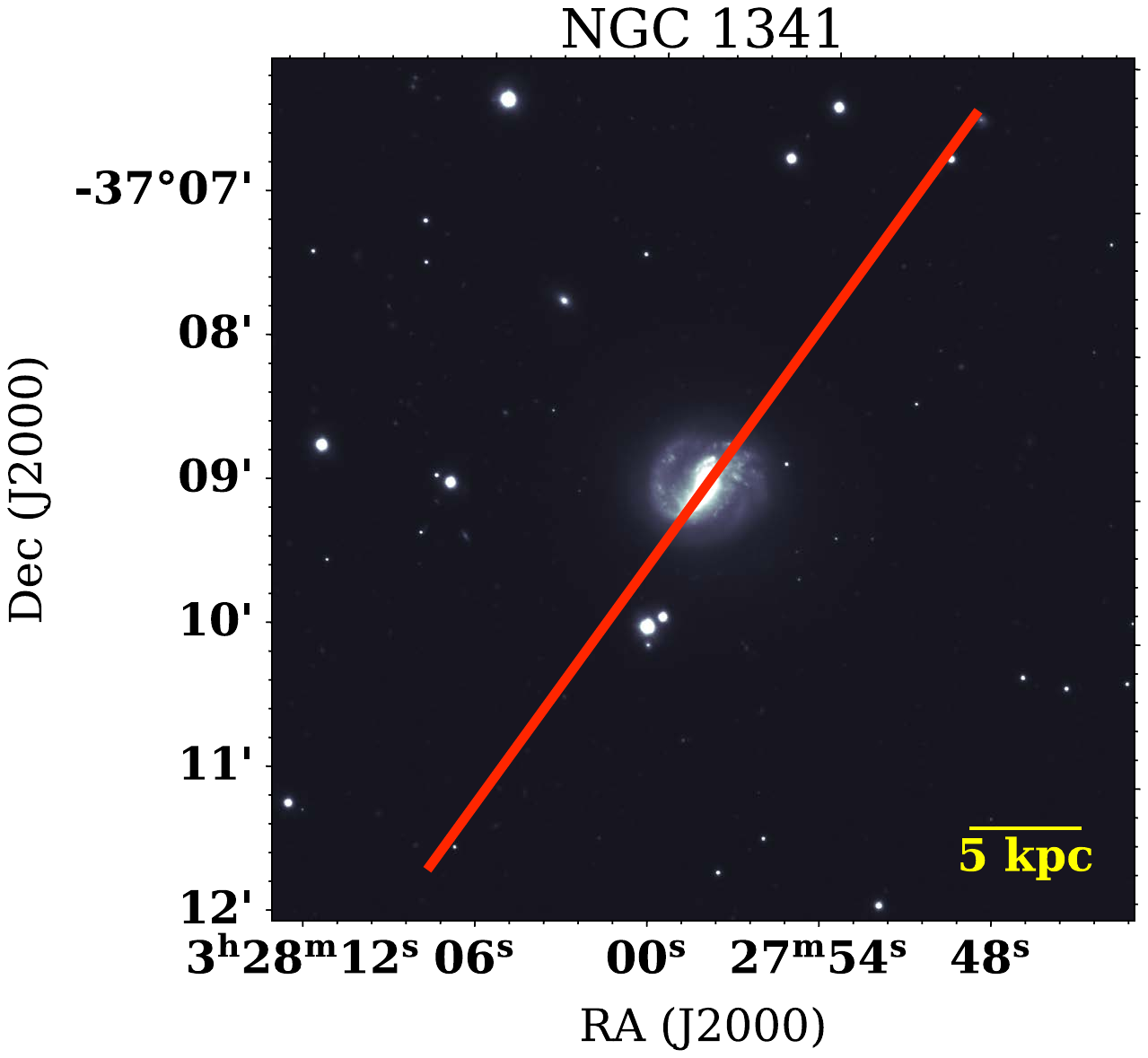}}\\
   \caption{Optical FDS image for NGC 1341 that was outside the field of view of \citet{Kleiner2021}. The position and orientation of the SALT slit are indicated in red.} 
\label{fig:NGC1341}
\end{figure}

We observed the ten Fornax A galaxies listed in Table \ref{table1} using the Robert Stobie Spectrograph (RSS) on SALT \citep{Burgh2003, Kobulnicky2003}, programme numbers 2019-2-MLT-002 and 2021-1-SCI-019 (PI: Loubser). We use the RSS long-slit mode, with the 8\arcmin\ slit orientated as shown in Fig. \ref{fig:NGC1326BSFH}, and in Fig. \ref{fig:NGC1341}. The slits were aligned in the direction of the major axis, except for NGC 1316 where the slit was aligned with the \HI{} gas morphology to probe the jet/interstellar medium interaction (for a separate study). We set the PG1300 grating at an angle of 44.5\degr, which corresponds to a spatial scale of 0.127\arcsec and a spectral scale of 0.33 \AA{} per unbinned pixel. The rest wavelength range covered by the spectra is 4780 to 6790 \AA{}. Using 2 $\times$ 2 binning, we obtain science exposures of 3 $\times$ 820 seconds per target (for all ten galaxies, all three exposures per target were taken consecutively), in intermediate seeing conditions (up to a maximum 2.5\arcsec), and in grey time. We also observe a Th--Ar arc for wavelength calibration directly after each set of science exposures, and a spectrophotometric standard star for relative flux calibration during the observing semester.  

Basic corrections and calibrations such as the overscan, gain, cross-talk corrections and mosaicking are performed by the SALT science pipeline, PySALT\footnote{https://pysalt.salt.ac.za} \citep{Crawford2010}, developed in the Python/PyRAF environment. We also perform subsequent data reduction steps in PySALT by following the standard long-slit data reduction techniques. Cosmic rays were removed from the two-dimensional spectrum using the SALT \texttt{crclean} algorithm. The two CCD gaps were filled with interpolated pixel values; however, we avoided these two wavelength ranges during measurements and spectral fitting. We perform wavelength calibration, sky subtraction, and flux calibration on the two-dimensional spectra. The three exposures for each target were then combined by taking the median and applying a 3$\sigma$ clipping algorithm.

\begin{table*}
\caption{The Fornax A galaxies observed with SALT. We also list their FCC and FDS catalogue names. The pre-processing classification is from \citet{Kleiner2021}, except for NGC 1341 which was outside their field of view. \HI{} masses are from \citet{Kleiner2021}, except NGC 1341 which was estimated as described in Section \ref{HImass} (and marked with a $\star$).}       
\label{table1}      
\centering                         
\begin{tabular}{l  l  l  l  l  c l} 
\hline
Name & FCC & FDS & RA & Dec &  Pre-processing & \HI{} mass \\
        	 &  &  & (deg) & (deg) &  & (M$_{\sun}$)  \\
\hline 
NGC 1326B & FCC 39 &   FDS25\_0001    &   51.3331    &    --36.3849    & early & 4.6E+09  \\
NGC 1310 & FCC 13 & FDS28\_0420  &   50.2643    &     --37.1017    & ongoing & 4.8E+08  \\
NGC 1316 & FCC 21 &  FDS26\_0001 & 50.6741  &	--37.2082  & ongoing & 6.8E+07  	\\
ESO 301-IG11 & FCC 28 & FDS26\_0000  &  50.9767     & --37.5100      & ongoing	 & 1.4E+08 	\\
NGC 1326 & FCC 29 & FDS25\_0000  &    50.9842   &    --36.4647     & ongoing	& 2.3E+09   \\
FCC 35 & FCC 35 & FDS25\_0008  &  51.2673     & --36.9277    & ongoing  & 3.3E+08	\\
NGC 1317 & FCC 22 & FDS26\_0254   &   50.6845    &   --37.1038     & advanced  & 2.8E+08   \\
NGC 1316C & FCC 33 & FDS26\_0003  &   51.2432    &    --37.0096     & advanced   & 1.7E+07 	\\
FCC 46 & FCC 46 &  FDS22\_0244  &   51.6043    &  --37.1278   & advanced	&  1.2E+07	\\
NGC 1341 & 	FCC 62 &  FDS22\_0000  &   51.9932    &   --37.1493	     & not classified	 & 2.7E+08$^{\star}$  	\\
\hline                                   
\end{tabular}
\end{table*}				

\begin{table*}
\caption{Optical properties of the Fornax A galaxies observed on SALT. We measure the redshift ($V$) from the SALT spectra (with a spectral resolution of 35 km s$^{-1}$) and compare it to previous measurements from the literature (compiled by \citealt{Maddox2019}). The references for the redshift data in the literature are M19 \citep{Maddox2019}, F1989 \citep{Ferguson1989}, and the Two Micron All Sky \citep{Skrutskie2006, Jarrett2000} Redshift Survey (2MRS; \citealt{Huchra2012}). We also list the SALT integrated equivalent width measurements of H$\alpha$ and the radius of the aperture. The stellar masses are from the FDS survey \citep{Raj2020, Iodice2017, Venhola2018}. Galactic extinction $E(B-V)$ is taken from \citet{Schlafly2011}.  The photometric properties ($r$ and $g-r$ magnitude, effective half-light radius ($R_{e}$) in arcsec) are from the FDS survey \citep{Raj2020, Iodice2017, Venhola2018}.}  
\label{table2}      
\centering                         
\begin{tabular}{l l l c c c c c c} 
\hline
Name  & $V$ SALT & $V$ Lit & Log EW(H$\alpha$)  & $M_{*}$ & $E(B-V)$ &  $r$  & $(g-r)$ & $R_{e}$  \\
           & (km s$^{-1}$) & (km s$^{-1}$) &  & (M$_{\sun}$) & (mag) &  (mag) & observed  &  (arcsec) \\
\hline 
NGC 1326B & 957 & 1005 (F1989) & 1.0250 (1.0$R_{e}$) &  1.8E+09 & 0.018 &  --18.92  &  0.34   & 53.0 $\pm$ 2.0  \\
NGC 1310 & 1817 & 1807 (2MRS) & 1.0373 (1.0$R_{e}$)  &  4.7E+09  & 0.021 &  --19.4  & 0.55  &  27.6 $\pm$ 0.2  \\	  
NGC 1316 & 1810 & 1760 (2MRS) & 0.2706 (0.6$R_{e}$)  &  6.7E+11   & 0.018 &  --23.6	& 0.74	& 87.0 $\pm$ 2.0  \\
ESO 301-IG11 & 1320 & 1396 (2MRS) & 1.2037 (1.0$R_{e}$) &  2.9E+09  & 0.021 & --18.4   &  0.56 &  22.1 $\pm$ 0.5  \\
NGC 1326 & 1320 & 1363 (F1989) & 1.0390 (0.3$R_{e}$)  &  2.9E+10  & 0.017  &  --21.51  & 0.78   &	48.2 $\pm$ 1.0 \\
FCC 35 & 1863 & 1827 (M19) & 1.7890 (1.0$R_{e}$) &  1.7E+08  & 0.018 &  --16.55 &  0.18 &  17.1 $\pm$ 0.4 \\
NGC 1317 & 1998 & 1941 (2MRS)  & 0.1576 (0.9$R_{e}$)   &  1.7E+10  & 0.018 &  --21.34  &  0.79 &  35.4 $\pm$ 0.1  \\
NGC 1316C & 2043 & 1974 (M19) & 0.9815 (1.0$R_{e}$) & 1.4E+09   & 0.019  & --18.08  &  0.68	&  22.6 $\pm$ 0.5 \\
FCC 46 & 2313 & 2268 (M19) & 0.1649 (1.0$R_{e}$)  & 2.0E+08 & 0.017 &  --17.27 &  0.53	&  \ 8.5 $\pm$ 0.8 \\
NGC 1341 & 1908 & 1881 (M19) &	1.3707 (1.0$R_{e}$) 	&   5.5E+09  & 0.014  &  --19.43  & 0.53 &	20.6 $\pm$ 0.4	 \\
\hline                                   
\end{tabular}
\end{table*}		

\subsection{Photometric data}
\label{Photdata}

For the Fornax A galaxies we use FDS data by \citet{Raj2020} for all galaxies, except for NGC 1316 and FCC 46 which were not part of their sample. For NGC 1316 we use the FDS measurements from \citet{Iodice2017}, and for the dwarf galaxy FCC 46 we use the FDS measurements from \citet{Venhola2018}. In particular, we use $g-r$ colours measured from FDS by \citet{Raj2020, Iodice2017, Venhola2018}. They are listed in Table \ref{table2}\footnote{\citet{Su2022} consolidated previous FDS photometric catalogues for consistent measurements. Their sample contains eight of our ten galaxies, and we compared the photometric measurements used here from different sources with their consistent measurements. The mean difference between the $g-r$ colours used here and the ones from \citet{Su2022} is only 0.02 magnitudes, the largest difference being 0.07 mag for NGC 1316. The difference has a negligible effect on our results.}. We obtain the estimated stellar mass $M_{*}$ from the FDS photometric data and the empirical relation from \citet{Taylor2011} which assumes a Chabrier \citep{Chabrier2003} initial mass function (IMF) 
\begin{equation}
\log_{10} (\frac{M_{*}}{M_{\odot}}) =1.15+0.70(g-i)-0.4M_{i}
\end{equation}
where $M_{i}$ is the absolute magnitude in the $i$-band. The stellar masses are indicated in Table \ref{table2}. The stellar mass of the early-type central group galaxy, NGC 1316, is given as between 5.2 and 8.3 $\times 10^{11}\ M_{\odot}$ in \citet{Iodice2017}. For our purposes, it suffices to use the mean of 6.7 $\times 10^{11}\ M_{\odot}$ in Table \ref{table2}.

The effective half-light radii, $R_{e}$ (in arcsec), of the Fornax A galaxies are also taken from \citet{Raj2020, Iodice2017, Venhola2018}, and derived from the $r$-band FDS data. We extract EW(H$\alpha$) along the slit out to at least $R_{e}$ for most galaxies, except NGC 1326 (0.3$R_{e}$), NGC 1316 (0.6$R_{e}$) and NGC 1317 (0.9$R_{e}$). In these three cases, the equivalent width of H$\alpha$ became too noisy to accurately measure beyond these radii. 

\subsection{MeerKAT \HI{}}
\label{HImass}

The \HI{} observations of Fornax A were taken during the commissioning of the MeerKAT telescope\footnote{Operated by the South African Radio Astronomy Observatory (SARAO).}, in preparation for the MeerKAT Fornax Survey (MFS), and are described in \citet{Kleiner2021}. All \HI{} mass detections reported in Table \ref{table1} are from \citet{Kleiner2021}, except NGC 1341, which was outside the field of view of the MeerKAT commissioning observations, but was previously detected in \HI{} by \citet{Courtois2015}. We estimate the \HI{} mass of NGC 1341 using the observations reported in \citet{Courtois2015} (their table 3), and similar to \citet{Kleiner2021}, we use equation 50 in \citet{Meyer2017} and a distance of 20 Mpc to Fornax A. We emphasise that the observations do not have the same sensitivity as the MeerKAT commissioning data and we use this \HI{} mass only as an estimate.

\citet{Kleiner2021}, in their MeerKAT study of Fornax A, define the pre-processing stages as: i) early -- galaxies that have not yet experienced pre-processing, have extended \HI\ discs and a high \HI\ content with a H$_{2}$-to-\HI\ ratio that is an order of magnitude lower than the median for their stellar mass; ii) ongoing -- galaxies that are currently being pre-processed, display \HI\ tails and truncated \HI\ discs with typical gas fractions and H$_{2}$-to-\HI\ ratios; iii) advanced -- galaxies are \HI\ deficient, no \HI\ in the outer disc, and H$_{2}$-to-\HI\ ratios that are an order of magnitude higher than the median for their stellar mass.

The MFS is currently being executed and is mapping the distribution and kinematics of \HI{} in the Fornax cluster using the MeerKAT telescope. The survey footprint covers the central region of the cluster out to the virial radius, and extends out to two virial radii towards the south west to include the Fornax A group. The MFS observations improve the sensitivity and resolution of previous \HI{} observations by at least an order of magnitude \citep{Serra2023}. We include the analysis of the SALT spectra for NGC 1341 here, although its \HI{} mass (with the same sensitivity as for the other Fornax A galaxies) will only be measured upon completion of the MFS survey. The MFS design, observations, and \HI{} data reduction are described in detail in \citet{Serra2023}. 

For the H$_{2}$ and CO measurements that allowed \citet{Kleiner2021} to classify the different stages of pre-processing, we refer to \citet{Morokuma2019, Zabel2019, Morokuma2022, Kleiner2021}.

\section{Measurements from optical data}
\label{measurements}

\subsection{Emission line measurements}
\label{emission}

We fit the stellar continuum and measure any emission lines present in our SALT spectra using the Penalised Pixel-Fitting\footnote{https://www-astro.physics.ox.ac.uk/$\sim$mxc/software/} (pPXF; \citealt{Cappellari2004, Cappellari2017}) and the Gas and Absorption Line Fitting\footnote{https://star.herts.ac.uk/$\sim$sarzi/} (GandALF; \citealt{Sarzi2006}) codes, respectively, using single-age stellar-population templates from the MILES library of \citet{Vazdekis2015}\footnote{http://miles.iac.es/}. During the pPXF stellar template fitting, we first mask regions potentially affected by emission, as well as the two SALT CCD chip gaps. Upon determining the optimal combination of stellar templates, we then fit the emission lines with Gaussian functions. We fit for the Balmer H$\alpha$ and H$\beta$ recombination lines, and the [O III]$\lambda\lambda$4959, 5007, [O II]$\lambda\lambda$6300, 6363, [N II]$\lambda\lambda$6548, 6583, and [S II]$\lambda\lambda$6713, 6730, forbidden lines. We use a multiplicative polynomial of the 6$^{\rm th}$ degree to adjust the shape of the continuum to account for flux calibration differences. Therefore, we do not derive the stellar reddening using the shape of the continuum. SALT, given its design, has a varying pupil size during observations, which precludes accurate absolute flux calibration. We i) only use emission line ratios of lines adjacent to each other (Fig. \ref{fig:BPT}), and ii) note whether or not the lines were detected above an amplitude-to-noise ratio (A/N) of 2 (Table \ref{table3}):

\begin{figure}
\centering
\subfloat{\includegraphics[scale=0.32]{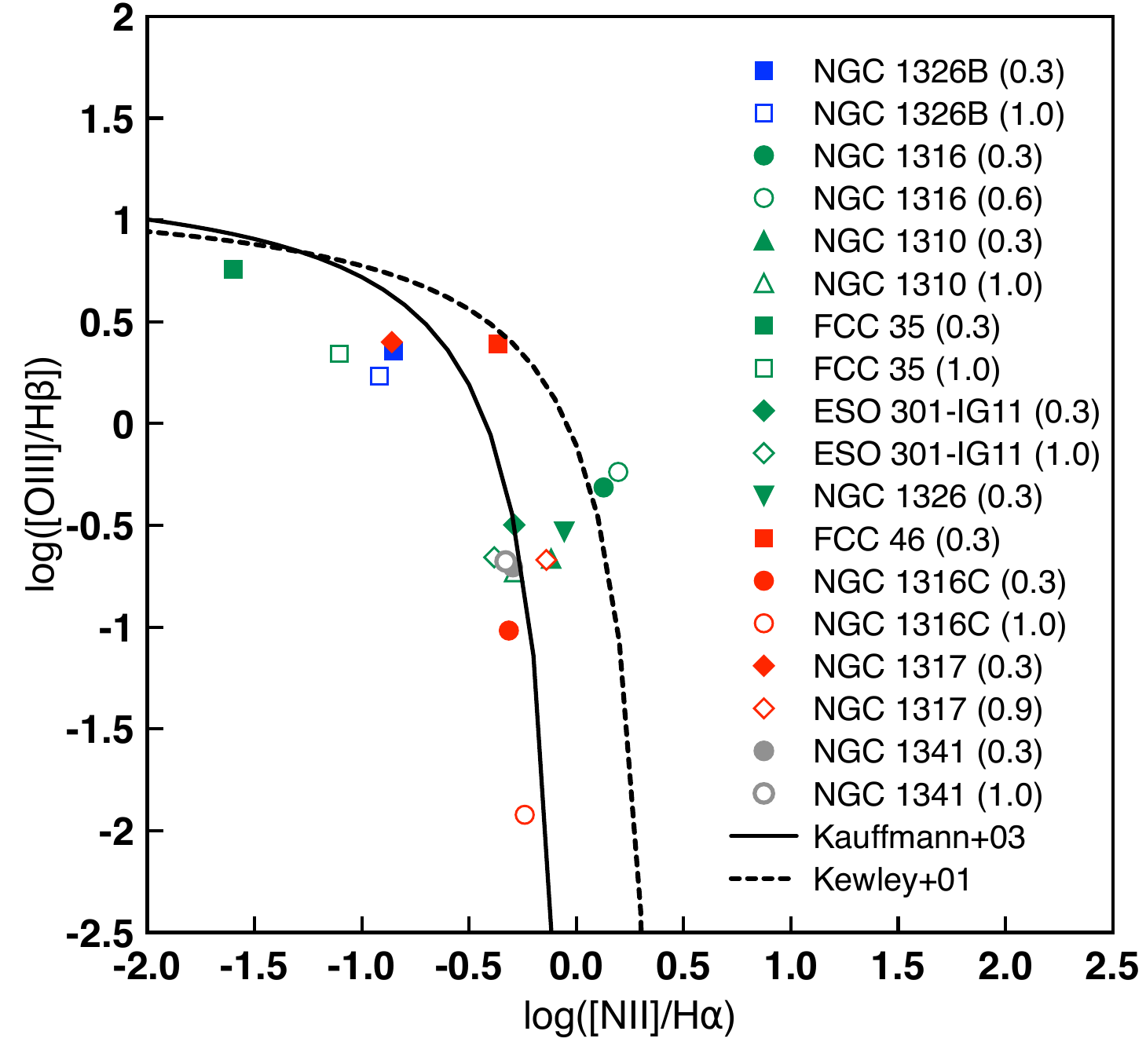}}\\
   \caption{The Fornax A galaxies on the BPT diagram \citep{Baldwin1981}. The demarcation lines of \citet{Kauffmann2003} and \citet{Kewley2001} are indicated. We plot all galaxies with filled symbols for line ratios within 0.3$R_{e}$ and empty symbols for the apertures from 0.3$R_{e}$ to the maximum $R_{e}$ used. The three different stages of pre-processing are given in blue, green, and red, for early, ongoing, and advanced stages, while NGC 1341 (grey) was not classified. For NGC 1326 we only measure the central aperture, and for FCC 46 we do not detect any emission lines in the outer aperture.} 
\label{fig:BPT}
\end{figure}

i) We are interested in the H$\alpha$ equivalent width as an indicator of star formation over the last 10 Myr, and assume that H$\alpha$ emission is directly associated with star formation. Therefore, we checked the line ratios indicating star formation on the Baldwin-Phillips-Terlevich (BPT) diagram \citep{Baldwin1981} presented in Fig. \ref{fig:BPT}. We also indicate the demarcation lines by \citet{Kauffmann2003} and \citet{Kewley2001}. Line ratios from AGN lie above the \citet{Kewley2001} line, and for star formation below the \citet{Kauffmann2003} line, with the composite part of the diagram in between the two lines. We plot all the galaxies on the BPT-diagram (with filled symbols for 0.3$R_{e}$ apertures and empty symbols for the apertures that range from 0.3$R_{e}$ to 1.0$R_{e}$). However, not all detected lines (H$\alpha$, H$\beta$, [N II], and [O III]) were detected above an A/N of 2 (Table \ref{table3}). We also indicate the three different stages of pre-processing by using different colours (blue for early, green for ongoing, red for advanced, and grey for NGC 1341 which was not classified). There are some caveats to the BPT diagram, such as that not all lines were detected above an A/N of 2, and that other ionisation mechanisms, e.g., shocks, can also lead to AGN-like emission \citep{Kewley2006}. Nevertheless, only the data points of NGC 1316 fall within the AGN part of the BPT diagram (as expected, see, e.g. \citealt{Maccagni2021}), while the rest lie in the star-forming or composite sections of the diagram. We therefore assume that, apart from NGC 1316, the ionising photons in the galaxies originate from the underlying stellar population. In particular, the contribution to the H$\alpha$ emission is assumed to originate entirely from young massive O and B stars (e.g., \citealt{Kennicutt1998}). There might be some particular cases, e.g. post-AGB stars, for which H$\alpha$ emission can be produced by other processes, but in statistical terms and on integrated scales the emission from young stars dominates \citep{Corcho2023}.

ii) Line detections can also be used as a qualitative time scale indicator for quenching, and we indicate the lines present in the spectra (above an A/N of 2) in Table \ref{table3}. If star formation is ongoing, then O stars keep [O III] present. If star formation is quenched, O stars are the first to disappear, whereas B stars keep hydrogen ionised for longer. The ratio [O III]/H$\alpha$ can be considered an indicator of recent quenching \citep{Citro2017, Quai2018}, but we refrain from using this ratio because of the lack of absolute flux calibration and we use the presence of the emission lines only as a qualitative indication of quenching.

\begin{table}
\caption{Emission line detections (with A/N > 2) in the SALT spectra of Fornax A galaxies. Central (``c") refers to lines detected within 0.3$R_{e}$, while outer (``o") to lines detected in 0.3$R_{e}$ to the maximum $R_{e}$ used (see Table \ref{table2}). For NGC 1326 we only measure the central aperture. For FCC 46 we do not detect any emission lines in the outer aperture, and for NGC 1316 we do not detect any emission lines with A/N > 2 in the outer aperture.}    
\label{table3}      
\centering                         
\begin{tabular}{l c c} 
\hline
Name & c/o  & Lines detected  \\
\hline 
NGC 1326B & c & H$\alpha$, H$\beta$, [O III]  \\
                    & o & H$\alpha$, H$\beta$, [O III], [S II] \\
NGC 1310  & c & H$\alpha$, [S II], [N II]  \\	
                   & o & H$\alpha$, H$\beta$, [S II], [N II]  \\     
NGC 1316 & c & H$\alpha$, [N II] \\
                  & o &  ---  \\
ESO 301-IG11 & c & H$\alpha$, H$\beta$, [S II], [N II]  \\
                        & o & H$\alpha$, H$\beta$, [S II], [N II]  \\ 
NGC 1326 & c & H$\alpha$, H$\beta$, [S II], [N II]  \\
FCC 35 & c & H$\alpha$, H$\beta$, [S II], [O III], [N II]	\\
                                       &  o &  H$\alpha$, [S II], [O III]         \\
NGC 1317 & c & H$\alpha$, H$\beta$, [O III] \\
                  & o & H$\alpha$, [S II], [N II] \\
NGC 1316C & c & H$\alpha$, H$\beta$, [S II], [N II]  \\
                    & o & H$\alpha$, [S II], [N II]  \\
FCC 46 & c & H$\alpha$, [O III]	\\
                                       & o & ---   \\
NGC 1341 & c & H$\alpha$, H$\beta$, [S II], [N II] 	\\
                 & o & H$\alpha$, H$\beta$, [S II], [N II]  \\
\hline                                   
\end{tabular}
\end{table}		

\subsection{EW(H$\alpha$) measurements}

We use H$\alpha$ emission line equivalent width (EW) measurements as an indicator of star formation to probe quenching along the long-slit. The EW of the nebular lines is measured in the rest frame by dividing the line flux by a measure of the surrounding continuum. To allow direct comparison with previous work, we measure the EW(H$\alpha$) following the definition in \citet{Corcho2023, Corcho2023b}: 

\begin{equation}
EW(H\alpha) \equiv \int^{6575}_{6550} \left( \frac{F{_\lambda}(\lambda)}{\frac{F_{B} \lambda_{R} - F_{R} \lambda_{B}}{\lambda_{R} - \lambda_{B}} + \lambda \frac{F_{R} - F_{B}}{\lambda_{R} - \lambda_{B}}} - 1 \right) d\lambda
\end{equation}

where $F_{B}$ and $F_{R}$ correspond to the mean flux per unit wavelength computed in the 6470 -- 6530 \AA{} and 6600 -- 6660 \AA{} bands, with central wavelengths $\lambda_{B}$ = 6500 \AA{} and $\lambda_{R}$ = 6630 \AA{}, respectively. With this definition, positive and negative values of EW denote emission and absorption, respectively. The limiting detectable EW(H$\alpha$) measurement depends on the A/N of the line and the surrounding continuum, as well as the velocity dispersion of the line \citep{Sarzi2006}. For a barely detected H$\alpha$ line (A/N = 2) with dispersion 80 km s$^{-1}$, the limiting EW is 1.5 \AA{} for a continuum with S/N = 6 \citep{Belfiore2018}. The higher EW(H$\alpha$) corresponds to younger stellar populations. We list the integrated equivalent width measurements of H$\alpha$ and the radius of the aperture in Table \ref{table2}, and use it in Figs. \ref{fig:Hasummary} to \ref{fig:Hadegree}.

\subsection{$g-r$ colours from the FDS survey}
\label{colour}

We use $g-r$ colours from the FDS survey \citep{Raj2020, Iodice2017, Venhola2018}, as included in Table \ref{table2}. Optical colours such as $g-r$ may be substantially affected by dust extinction. The flux calibration of SALT is not accurate enough to use the shape of the continuum to correct for dust extinction (see Section \ref{emission}). For the same reason, and because H$\beta$ line emission is often weak, we do not use the H$\alpha$ and H$\beta$ emission line measurements to correct for dust extinction in $g-r$ colours. Typically, colour excesses for internal extinction on resolved regions of CALIFA galaxies (of all types) range between $E(g-r)$ = 0.1 and 0.3 magnitudes \citep{Corcho2021}. Therefore, we rather use a $(g-r)_{\rm intrinsic} = (g-r)_{\rm observed} - E(g-r)$ correction where $E(g-r)$ is 0.2 $\pm$ 0.1. This is sufficient for our purposes to interpret the ageing diagram (AD) in Section \ref{ADdiagram}, where we indicate both the observed colours and the estimated correction on the diagram. The estimated correction will not change our main conclusions from Section \ref{ADdiagram}. Galactic extinction $E(B-V)$, taken from \citet{Schlafly2011}, is small and is given in Table \ref{table2} for reference. 

The FDS data allow for the colour profiles to be extracted out to several effective radii. We use the azimuthally averaged $g-r$ colour, but we also examined the surface photometry profiles ($g$ and $r$), particularly within the central $R_{e}$, of \citet{Raj2020}. Any relative changes between the $g$ and $r$ profiles within $R_{e}$ (see also the $g-r$ colour maps presented in \citealt{Raj2020}), or differences compared to the average $g-r$ colour, are significantly smaller than the uncertainty on dust correction or the spread of the ageing sequence or the expected variations in H$\alpha$. 


\section{Quenching time scale indicators}
\label{parameters}

\subsection{The presence of [O III]}

As mentioned in Section \ref{emission}, if star formation is ongoing and short-lived supermassive O and early B stars are present, then O stars keep the emission line [O III] present. If star formation is quenching, O stars are the first to disappear, while B stars keep hydrogen ionised. As Table \ref{table3} shows, we detect [O III] in NGC 1326B, FCC 35 (central and outer regions), and in NGC 1317 and FCC 46 (central region; see also \citealt{Romero-Gomez2023a}), implying -- at least qualitatively -- that star formation is ongoing or recent in these galaxies. These are the galaxies that are removed from the secular evolution sequence (defined as the continuous evolution of a galaxy driven by the consumption of gas through uninterrupted star formation until quiescence is reached) as discussed in Section \ref{ADdiagram}. 

\subsection{The ageing diagram}
\label{ADdiagram}

\begin{figure*}
\centering
\subfloat{\includegraphics[scale=0.44]{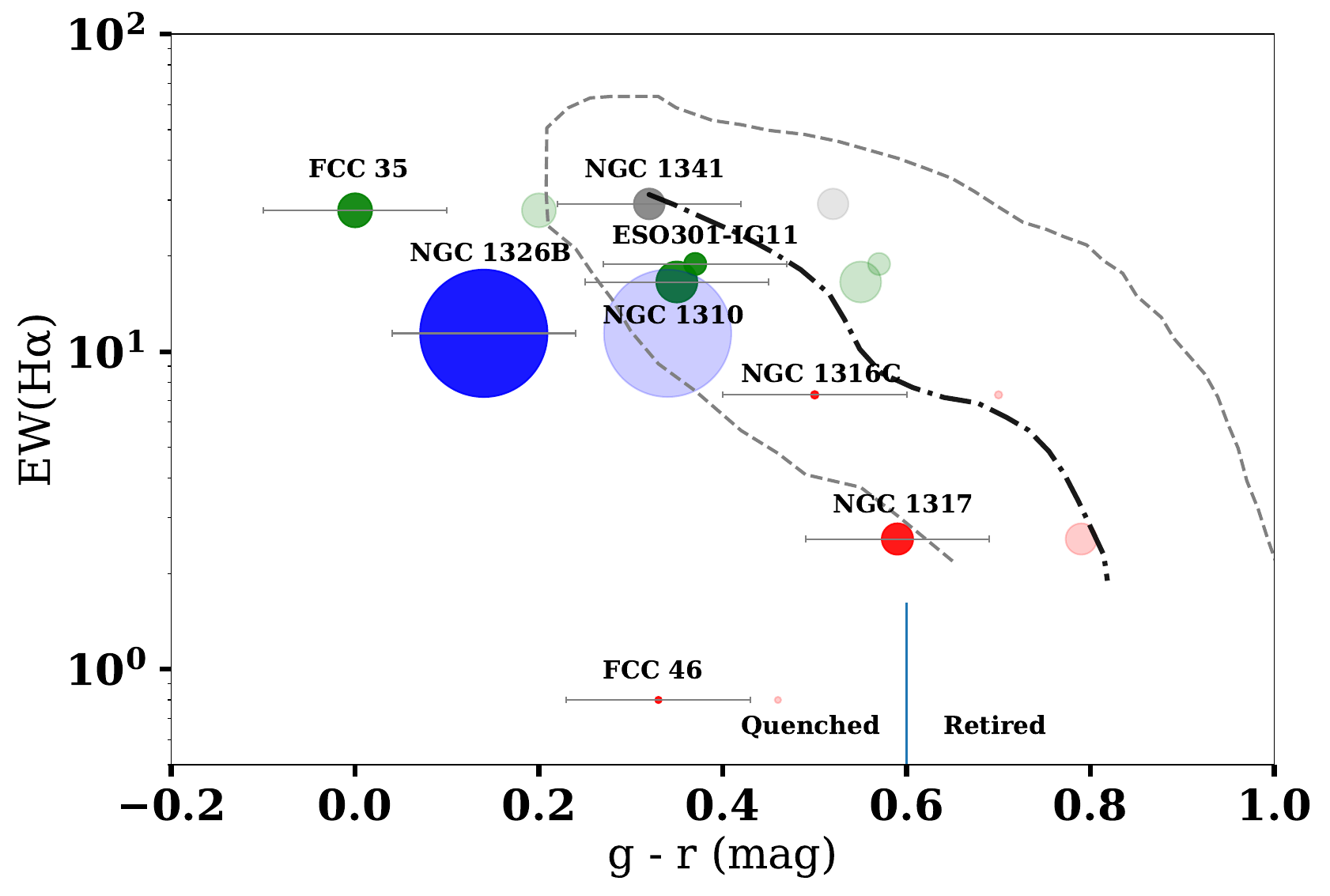}}\\
   \caption{The AD for Fornax A galaxies based on apertures of 0.3$R_{e}$ to 1.0$R_{e}$. The black dot-dashed curve indicates the secular ageing sequence (and its spread in grey) of \citet{Corcho2023}. The colour of the data points corresponds to their pre-processing category (blue for early, green for ongoing, red for advanced, and grey for uncategorised), and the size of the symbol to the amount of \HI{}, from the smallest red circle (1.4E+08 M$_{\odot}$, FCC 46) to the large blue circle (6.2E+09 M$_{\odot}$, NGC 1326B). The solid symbols include an estimated correction of $E(g-r)$ = 0.2 $\pm$ 0.1, and transparent symbols indicate the observed $g-r$ colour, without any correction for the possible effect of dust extinction, as discussed in Section \ref{colour}. The vertical blue line separates the `Quenched' and `Retired' parts of the diagram as described in Section \ref{ADdiagram}.}
\label{fig:Hasummary}
\end{figure*}

We place our galaxies on an AD (see Fig. \ref{fig:Hasummary}) that describes the correspondence between the fraction of stars formed during the last $\sim$10 Myr, as traced by the equivalent width EW(H$\alpha$), and the fraction of stellar mass formed on scales of $\sim$1 Gyr, using the optical colour $g-r$ \citep{Corcho2021, Corcho2023, Corcho2023b}. The expectation is that quenching proceeds outside-in, even though this process is more subtle in groups than in clusters. Hence we plot the EW(H$\alpha$) for the apertures ranging from 0.3$R_{e}$ to 1.0$R_{e}$ on the AD and colour the data points by their pre-processing category (blue for early, green for ongoing, red for advanced, and grey for uncategorised). The size of the symbol corresponds to the amount of \HI{} observed with MeerKAT, from the smallest red circle (1.4E+08 M$_{\odot}$, FCC 46) to the large blue circle (6.2E+09 M$_{\odot}$, NGC 1326B), except for NGC 1341 which was beyond the footprint of \citet{Kleiner2021}, and we estimate the \HI{} mass from the observations reported in \citet{Courtois2015}. The H$\alpha$ emission of NGC 1316 is more difficult to interpret due to its significant AGN activity, and we only measure the EW(H$\alpha$) out to 0.6$R_{e}$. Therefore, NGC 1316 is not shown in the plot. NGC 1326 is also not plotted since we only probed the central part of the galaxy.   

This diagram allows insight into the recent changes in the sSFRs of galaxies and allows us to separate galaxies governed by secular evolution (ageing) from systems whose star formation was interrupted during the last $\sim$Gyr (quenching). The vertical axis of the AD represents an estimate of the average sSFR over the last few Myr and traces the mass fraction of short-lived O and B stars that are able to ionise the interstellar medium (ISM). The horizontal axis of the AD represents an estimate of the average sSFR over $\sim$0.1 -- 1 Gyr and traces the fraction of intermediate-age stellar populations, dominated by A-type stars. Ageing can be understood as the sequence of secular evolution (from blue emission to red absorption), indicated in Fig. \ref{fig:Hasummary}, which takes place over several Gyr. Sudden quenching of star formation implies a faster transition through the blue absorption domain on a timescale of the order of $\sim$300 Myr \citep{Corcho2023}. We compare our results in the AD with those in \citet{Corcho2023}, and indicate their secular ageing sequence in black (and its spread in grey). They used large, statistical samples of more than 9000 galaxies from the Calar Alto Legacy Integral Field Area (CALIFA) and Mapping Nearby Galaxies at Apache Point Observatory (MaNGA) surveys, in combination with predictions from IllustrisTNG-100, for emission within 1.5$R_{e}$ in stellar mass. The spread in grey covers around 90 per cent of these samples. 

Most environmental quenching processes can not only diminish star formation, but can also enhance star formation efficiency, which will lead to the depletion of the gas reservoir in very short timescales \citep{Zinn2013, Cresci2015, Corcho2023b}. The ageing sequence describes objects whose star formation varies smoothly over time (the last $\sim3$ Gyr, \citealt{Abramson2016}), and implies the normal slow transition from a blue, star forming galaxy to a red, old galaxy. These objects end in the `Retired' part of the diagram. Galaxies that experienced quenching episodes during the last $\sim$ Gyr, will feature significantly lower values of EW(H$\alpha$), due to the lack of O and B stars, while still displaying a stellar continuum dominated by intermediate stellar populations (see, e.g., \citealt{Corcho2023, Corcho2023b}). These galaxies will end in the `Quenched' part of the AD.  

Figure \ref{fig:Hasummary} suggests that the sSFR (as measured in the 0.3$R_{e}$ -- 1.0$R_{e}$ apertures) on scales of the last few Myrs (on the y-axis) is related to the pre-processing stage. The sSFR of NGC 1326B (early stage) is particularly high beyond 1.0$R_{e}$ (as seen in Fig. \ref{fig:Hasummary2}) but not as pronounced within 1.0$R_{e}$. The AD further suggests that FCC 35, NGC 1326B, and FCC 46 have histories different from secular ageing during the last Gyr (i.e., they are more `blue' than expected and have diminished recent sSFR as measured from EW(H$\alpha$)). The transparent symbols in Fig. \ref{fig:Hasummary} indicate the observed $g-r$ colour ($(g-r)_{\rm observed}$), without any correction for the possible effect of dust extinction, and the solid symbols indicate the estimated $(g-r)_{\rm intrinsic}$, as discussed in Section \ref{colour}. The star-forming galaxies NGC 1326B and FCC 35 are likely to have more significant internal extinction and located leftward from the secular evolution sequence, whereas FCC 46 will have less internal extinction and the estimated $(g-r)_{\rm intrinsic}$ is likely too blue. 

These three galaxies (FCC 35, NGC 1326B, and FCC 46) are all from different pre-processing classes, and the amount of recent star formation (y-axis) corresponds to the stage of pre-processing, with FCC 46 at an advanced stage of pre-processing according to its \HI{} content and morphology. We also note that the recent sSFR of FCC 35 is one of the highest in our sample, but very asymmetric (i.e., with a higher SFR on one side of the galaxy). 

Statistical studies using the AD \citep{Corcho2023, Corcho2023b} show that although quenched galaxies are fairly rare (3 to 10\% at $z<0.1$), they are more likely to be lower mass systems in dense environments. We investigate the relationship with stellar mass in Section \ref{stellarmass}.

\subsection{Equivalent width H$\alpha$ radial profiles}

\begin{figure}
\centering
\subfloat{\includegraphics[scale=0.35]{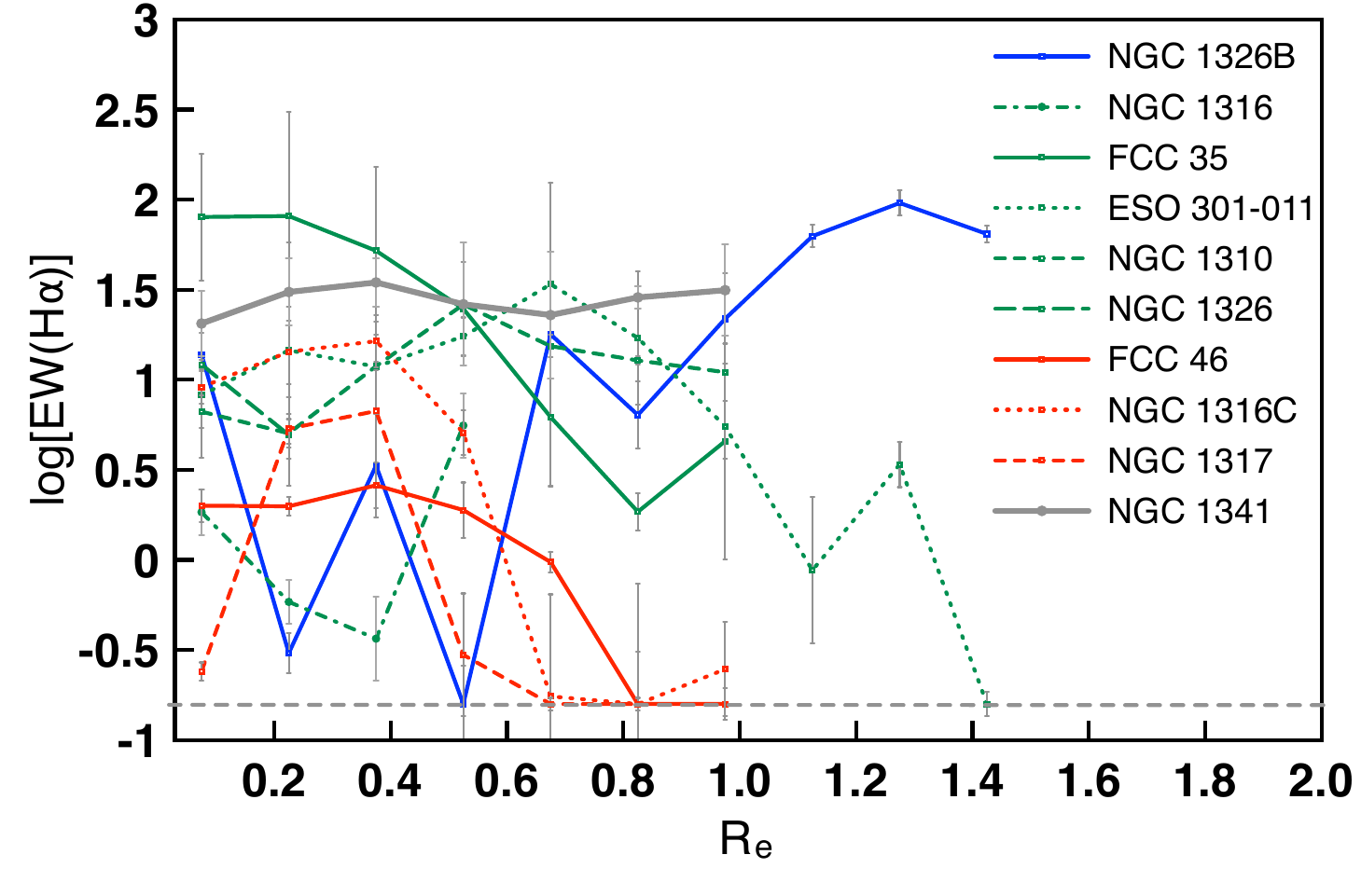}}\\
   \caption{EW(H$\alpha$) profiles in steps of 0.15$R_{e}$. The plot illustrates the spatial change in the recent sSFR along the slit position. The grey dashed line indicates where H$\alpha$ is an absorption line (below the grey dashed line).}
\label{fig:Hasummary2}
\end{figure}

Integral field spectroscopic observations reveal that not all regions within a single galaxy are necessarily concentrated in one location on an ageing diagram, such as in Fig. \ref{fig:Hasummary}, but are often broadly extended along the ageing sequence \citep{Corcho2021}. We show the EW(H$\alpha$) profiles in steps of 0.15$R_{e}$ in Fig. \ref{fig:Hasummary2}. Errors were obtained by measuring and comparing both sides of the galaxy from the centre and taking the variance of the two measurements of the same aperture on both sides of the galaxy.  

The only galaxy that shows a clear EW(H$\alpha$) profile that decreases towards the centre is the early-stage pre-processing galaxy NGC 1326B, suggesting it is not yet quenched in the outer parts. From a cosmological context, we expect that the central regions of star-forming galaxies are formed at earlier times (``inside-out" growth), and are more evolved with a lower sSFR than outer regions, which can still be gas rich \citep{deJong1996, Bell2000, Munoz-Mateos2007, Gonzalez-Delgado2015, Belfiore2018}. However, if star-forming galaxies are experiencing strangulation, their outer disks would not have cold gas to sustain star formation.

NGC 1341 (unknown stage of pre-processing), NGC 1310, and ESO301-IG11 (both ongoing stage of pre-processing) have flat EW(H$\alpha$) profiles within 1.0$R_{e}$, inferring that their population of very young stars (<10 Myr) is distributed throughout the galaxy.  ESO301-IG11, which we could measure further than 1.0$R_{e}$ shows a much smaller fraction of very young stars outside of 1.0$R_{e}$ than the early-stage galaxy NGC 1326B. 

FCC 35 (ongoing stage of pre-processing), FCC 46, NGC 1316C, and NGC 1317 (all three advanced stage of pre-processing) still show some fraction of very young stars, but only in the centres of the galaxies. A large fraction of very young stars is found only in the case of FCC 35. 

The observed outside-in pre-processing strongly favours environmental processes, as opposed to internally triggered quenching mechanisms such as AGN or supernovae \citep{Croton2006, Fitts2017}. 

\subsection{The effect of stellar mass}
\label{stellarmass}

\begin{figure}
\centering
\subfloat{\includegraphics[scale=0.28]{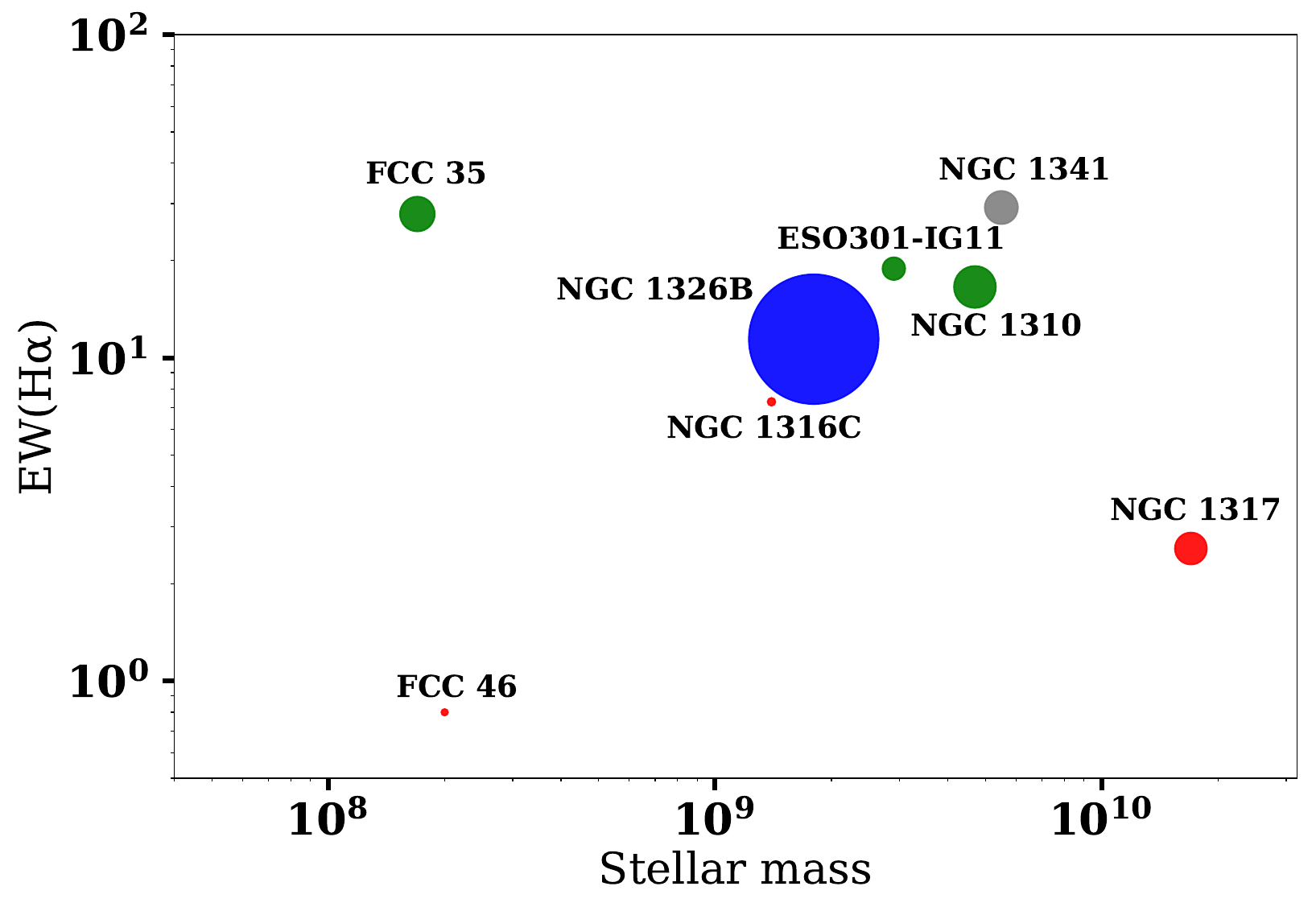}}\\
   \caption{The EW(H$\alpha$) against stellar mass (in M$_{\odot}$) for Fornax A galaxies is plotted using the 0.3$R_{e}$ to 1.0$R_{e}$ apertures. Symbol colours and sizes are the same as in Fig. \ref{fig:Hasummary}.}
\label{fig:Hamass}
\end{figure}

Dwarf galaxies will react differently to the IGrM than more massive galaxies. Figure \ref{fig:Hamass} displays the EW(H$\alpha$) against stellar mass for Fornax A galaxies, using apertures of 0.3$R_{e}$ to 1.0$R_{e}$. Both the ongoing and advanced pre-processing galaxies have a wide spread over stellar mass. Therefore, the signatures we see in the outer parts of galaxies cannot be attributed to stellar mass alone. However, the galaxies removed from the ageing sequence, and in particular the dwarf galaxies (FCC 35 and FCC 46), are the lowest stellar mass galaxies in our sample. 

\subsection{The effect of spatial distribution}

\begin{figure}
\centering
\subfloat{\includegraphics[scale=0.28]{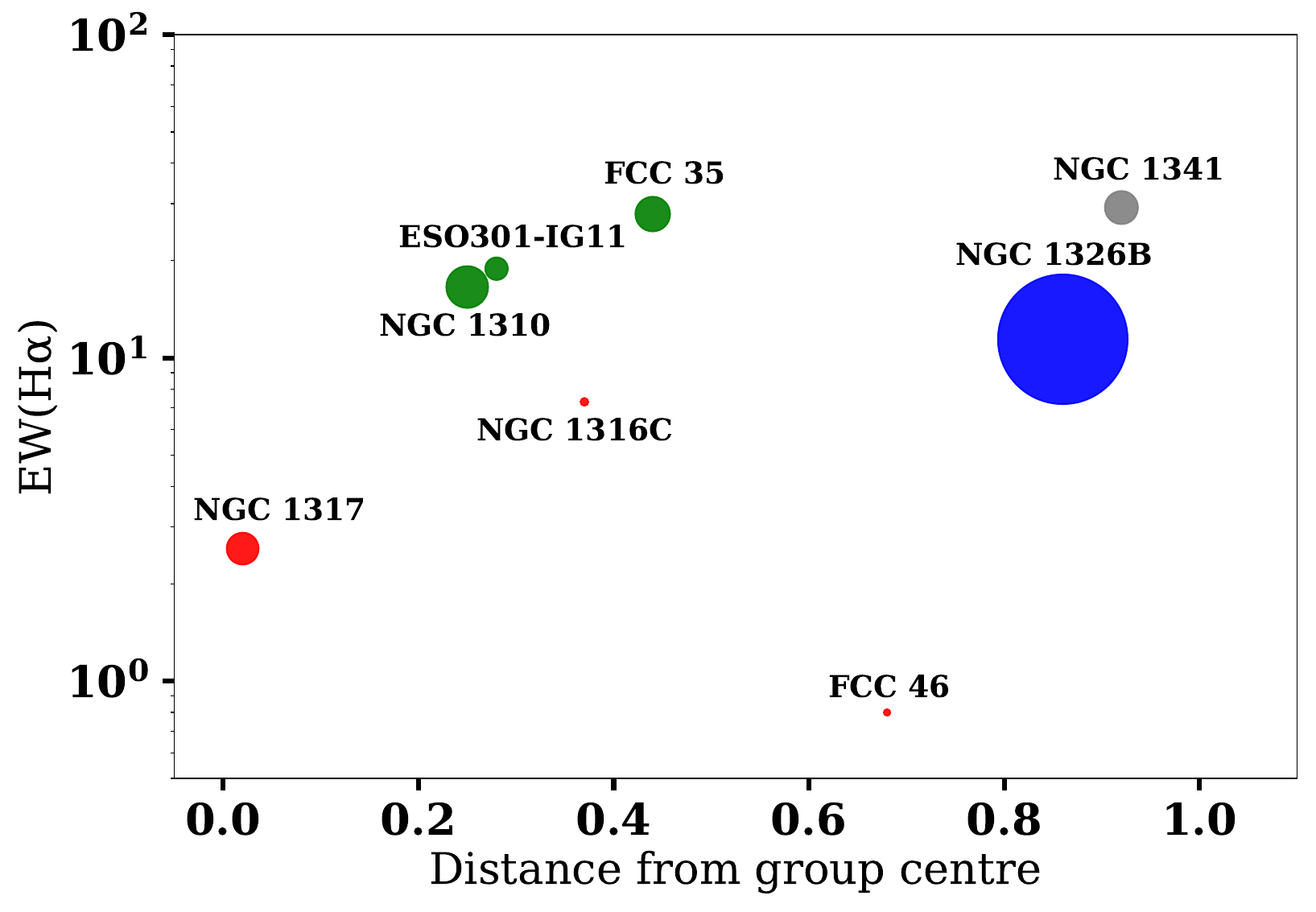}}\\
   \caption{The EW(H$\alpha$) against projected distance (in degrees) from the group centre for Fornax A galaxies is plotted using the 0.3$R_{e}$ to 1.0$R_{e}$ apertures. Symbol colours and sizes are the same as in Fig. \ref{fig:Hasummary}.}
\label{fig:Hadegree}
\end{figure}

We consider the spatial projected distribution of the galaxies with respect to the centre of the group, as that can also influence the likelihood of pre-processing. Figure \ref{fig:Hadegree} illustrates the projected distances (in degrees) from the centre (NGC 1316) as given in \citet{Raj2020}. We find no obvious correlation between the projected distances and the pre-processing stage or recent sSFR. \citet{Raj2020} also find no clear trend in the properties of Fornax A galaxies with group-centric distances in their photometric analysis. This is also the case when we consider galaxies with the closest projected distances to the cluster (see Fig. 2 in \citealt{Kleiner2021}). However, the galaxies removed from the ageing sequence (FCC 35, NGC 1326B, and FCC 46) are some of the galaxies farthest away from the group centre, but closest to the main Fornax cluster. 


\section{Fitting stellar population models}
\label{stelpopmodels}

We now turn our attention to the stellar absorption features and fit evolutionary population synthesis models (e.g., \citealt{Heavens2000, Worthey1994, CidFernandes2005, Ocvirk2006, Koleva2009, Vazdekis2010, Maraston2011}). While this can describe star formation episodes over longer time scales, it is based on assumptions rather than a model-independent approach such as the AD, and both methods should be regarded as complementary. 

\begin{figure}
\captionsetup[subfloat]{farskip=-0pt,captionskip=-0pt}
\centering
\subfloat{\includegraphics[scale=0.45, trim=65 225 85 286.5, clip]{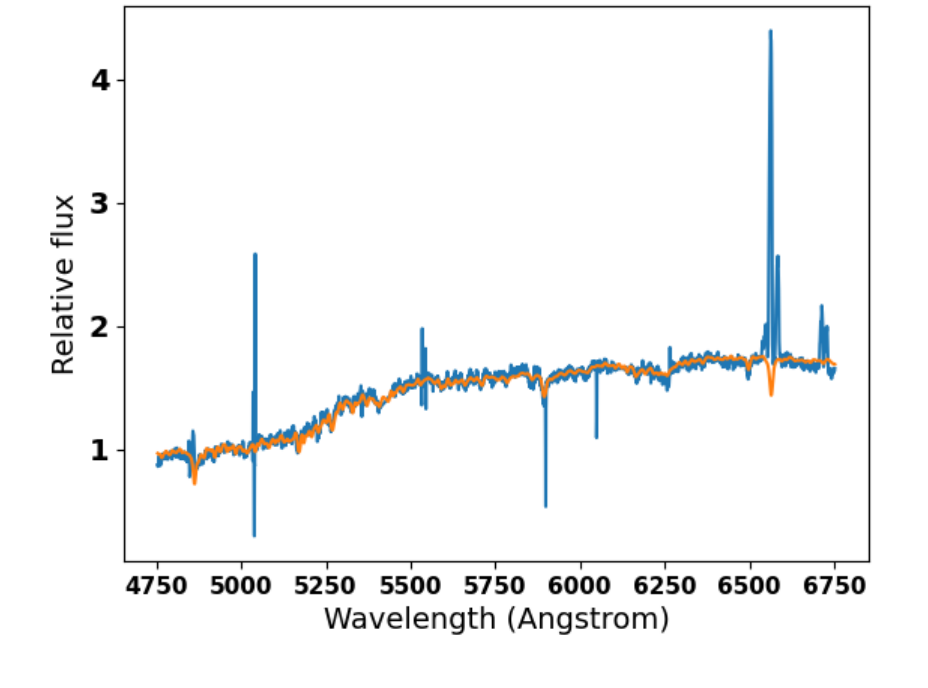}}\\
\subfloat{\includegraphics[scale=0.45, trim=60 240 80 280, clip]{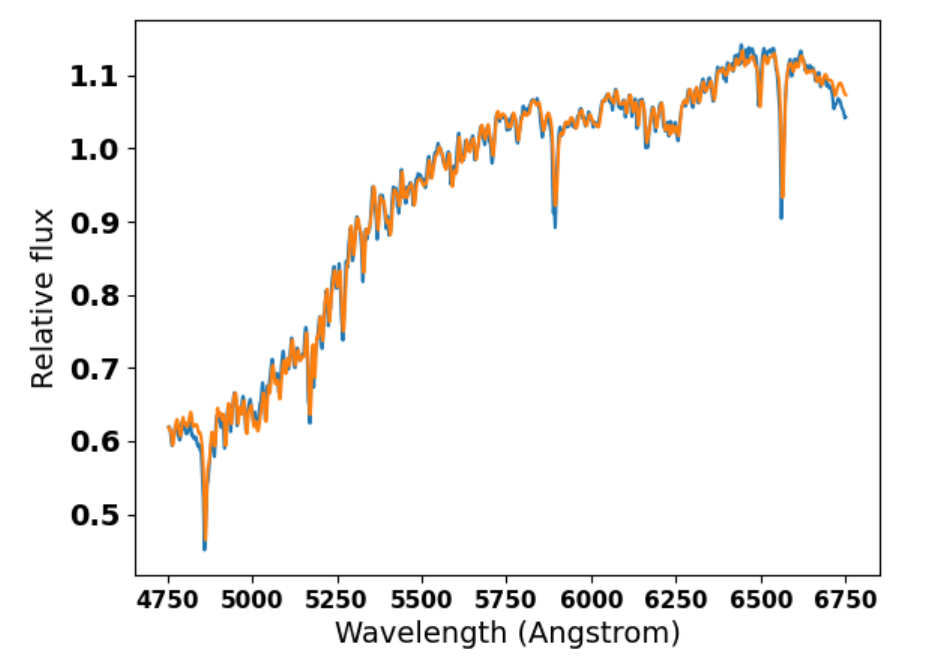}}\\ 
   \caption{NGC 1316C inner spectrum (0.3$R_{e}$) before (top) and after (bottom) emission line and sky line residual removal. The observed spectrum is shown in blue, and the best-fitting stellar absorption template combination is shown in orange (using the \citet{Maraston2011} models, the ELODIE stellar library \citep{Prugniel2007}, and a Kroupa IMF \citep{Kroupa2001}). The fluxes are relative as no absolute flux calibration was performed, and in particular the flux in the bottom plot is normalised.}
\label{fig:NGC1316C emission}
\end{figure}

We use the Fitting IteRativEly For Likelihood analYsis (FIREFLY, \citealt{Wilkinson2017}) code\footnote{http://www.icg.port.ac.uk/firefly/} to fit stellar population models. FIREFLY is a chi-squared minimisation fitting code that fits combinations of single-burst stellar population models to spectra, following an iterative best-fitting process controlled by the Bayesian information criterion. Stellar population synthesis assumes that the stellar populations in galaxies consist of the sum of Single Stellar Populations (SSPs), populations that consist of stars all born at the same time and with the same metallicity. All solutions within a statistical cut are retained with their weight, as fully described in \citet{Wilkinson2017}. No additive or multiplicative polynomials are employed to adjust the spectral shape. All spectra were corrected for foreground Galactic extinction using the reddening maps of \citet{Schlegel1998}, even though the Galactic extinction is small (see Table \ref{table3}). Before using FIREFLY, we subtract the emission lines using the GandALF Gaussian fits described in Section \ref{emission}. In Fig. \ref{fig:NGC1316C emission}, we show the inner spectrum (the aperture within 0.3$R_{e}$) of NGC 1316C as an example both before and after subtraction of the emission lines and some residual cosmic rays or sky lines. 

The SSP analysis assumes that the observed spectrum is conformed of the sum of discrete populations (single-burst stellar populations), in contrast to continuous star formation. This time-averaged approximation is an idealised representation that is only valid for certain galaxy populations, such as early-type galaxies that are believed to have experienced relatively short and intense bursts of star formation in the past. In reality, many galaxies undergo multiple episodes of star formation and may have more complex star formation histories. Nevertheless, SSPs can still provide useful insights into the average properties of the stellar population, particularly when studying integrated light. However, it should be interpreted with additional constraints, such as multiwavelength observations or spatially resolved studies. Here, we focus on the differences between the inner and outer apertures (i.e., gradients rather than absolute properties), which eliminates, to some extent, the systematic uncertainty inherent in different stellar population models and ingredients \citep{Loubser2012, Ratsimbazafy2017}. 

We use models by \citet{Maraston2011} that are calculated keeping the energetics fixed but vary the input stellar spectra, as well as models by \citet{Vazdekis2015}. We experiment with three different empirical libraries, namely MILES \citep{Sanchez2006}, STELIB \citep{LeBorgne2004}, and ELODIE \citep{Prugniel2007}. Testing different combinations of models and libraries allows us to understand the systematic uncertainty involved in the fit of the model \citep{Groenewald2014, Loubser2016, Loubser2021}. Here, we present the combination of \citet{Maraston2011} and ELODIE \citep{Prugniel2007}. However, using \citet{Vazdekis2015} and MILES \citep{Sanchez2006} does not alter the conclusions. Because we do not expect to be able to capture variations in the initial mass function (IMF) through spectral fitting, we use the Kroupa IMF \citep{Kroupa2001} models. Again, the results hardly change if we use a Salpeter IMF \citep{Salpeter1955}. 

\citet{Wilkinson2017} extensively tested the effect of the adopted wavelength range when fitting stellar population models by using mock galaxy spectra and a star cluster for which the age and metallicity are known. They concluded that ELODIE-based models lead to consistent ages when a large wavelength range is used, but fail to do so when bluer wavelength ranges are not taken into account (approximately below 4300 \AA{}, the region particularly sensitive to features from younger stars). Similar to the MUSE analysis reported in \citet{Loni2023} for the Fornax galaxy NGC 1436, we exclude stellar templates below 1 Gyr; this not only improves the model fits, but also improves the consistency between the results we obtain using different model and library combinations. 

\begin{figure}
\centering
\subfloat{\includegraphics[scale=0.34]{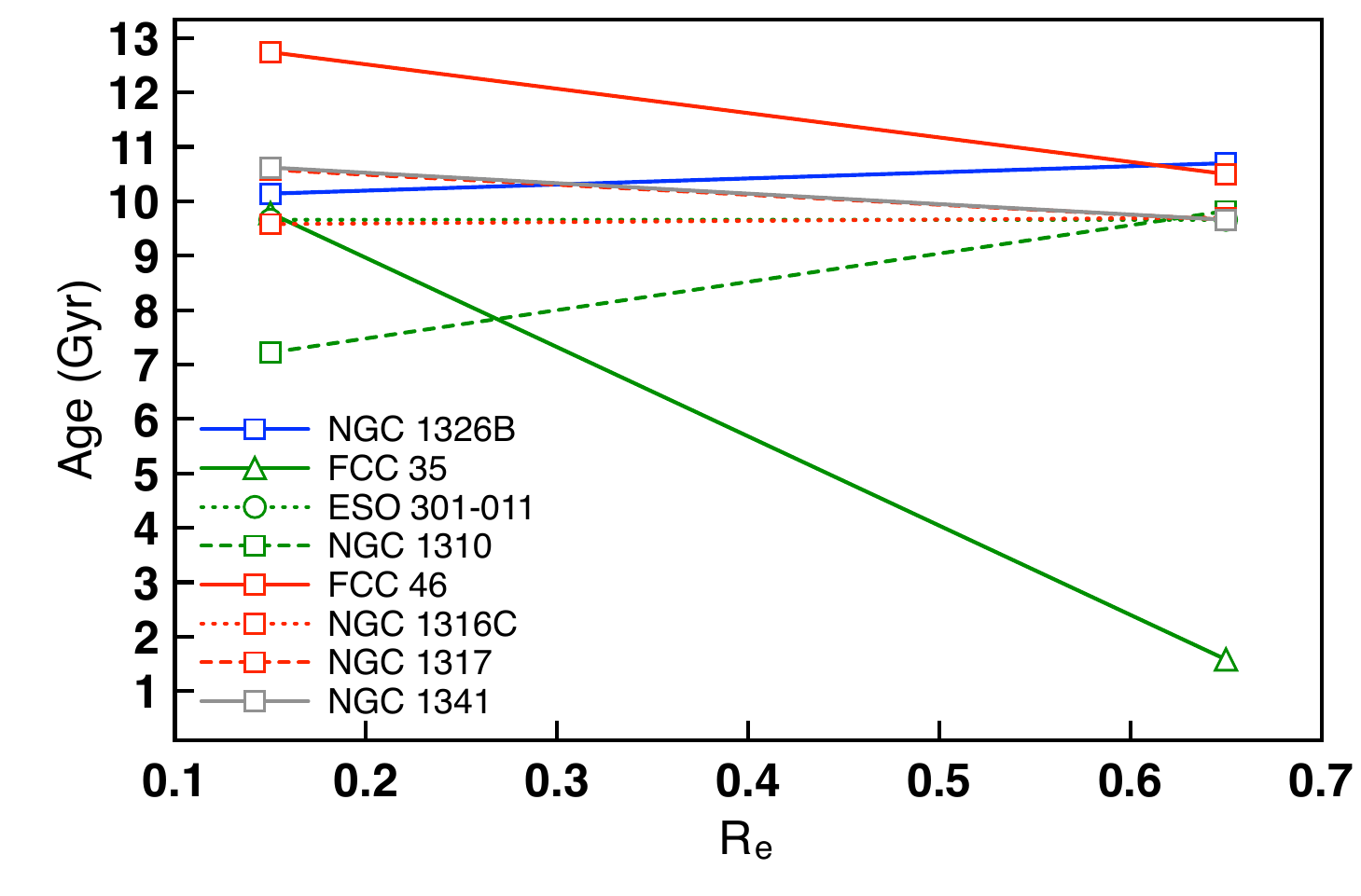}}\\
   \caption{Mass-weighted, SSP-equivalent age gradients between 0.15 (within 0.3$R_{e}$ apertures) and 0.65$R_{e}$ (0.3$R_{e}$ to 1.0$R_{e}$ apertures), coloured by pre-processing stage.}
\label{fig:Agesplots}
\end{figure}

We fit the central apertures (within 0.3$R_{e}$) and the outer apertures (0.3$R_{e}$ to 1.0$R_{e}$, both sides of the galaxy combined) of all galaxies to determine their mass-weighted, SSP-equivalent ages. We interpret the difference as an indicator of age gradients, rather than trying to derive a full star formation history with limited information. Stellar templates below 1 Gyr were excluded, although for NGC 1316 we found that templates for very young stars (< 1 Gyr) had to be considered to obtain good fits. Because it is a complex AGN host galaxy, this does not necessarily imply the presence of younger stellar populations, and we do not draw conclusions from the stellar population fit of NGC 1316. We only extract a central aperture for NGC 1326. Figure \ref{fig:Agesplots} displays the gradients, coloured by pre-processing stage. For completeness, we also show the best-fitting FIREFLY combinations of single-burst stellar population models to our spectra (inner and outer apertures) in Appendix \ref{SFHs}. 

We only see a significant age gradient for FCC 35, and notable age gradients for NGC 1310 and FCC 46 (as the uncertainties inherent in age determination increase for older SSP-equivalent ages, see \citet{Loubser2016}). NGC 1310 and FCC 35 are both in an ongoing stage of pre-processing, and FCC 46 is in an advanced stage of pre-processing. The mass-weighted, SSP-equivalent age of NGC 1310 is younger in the inner region, whereas for FCC 46 it is older in the inner region, although for FCC 46 both the inner and outer regions have ages older than 10 Gyr. For FCC 35, the mass-weighted SSP-equivalent age is much younger in the outer region. This gradient is opposite to the recent sSFR gradient measured using the EW(H$\alpha$) profiles, which is enhanced in the centre. The recent sSFR is also asymmetric, and clearly FCC 35 had a very dramatic star formation history and evolution (see discussion in Section \ref{individual}).  

\section{Pre-processing and quenching in Fornax A}
\label{interpret}

In this section, we discuss the galaxies individually and combine our results obtained from the stellar population indicators with the formation histories derived from all the multi-wavelength observations available.

In the discussion below, we also use the projected phase-space analysis for the Fornax A group by \citet{Raj2020}. They showed that most of the group members are located in the region of recent and intermediate infallers. The ancient infallers are the central galaxy NGC 1316 and the barred spiral galaxy NGC 1317. The bright spiral galaxies in the Fornax A group (NGC 1310, NGC 1326A, FCC 035, and NGC 1341) are intermediate infallers. NGC 1316C, ESO 301-IG11, NGC 1326 and NGC 1326B are recent infallers. 	

We also note that the crossing time of the group is $\sim R_{vir}/\sigma_{gr} = 0.38\ \rm Mpc / 204\ \rm km\ s^{-1} \sim 1.8\ \rm Gyr$, using $R_{vir}$ from \citet{Drinkwater2001} and $\sigma_{gr}$ from \citet{Maddox2019}. In contrast, the timescale for pre-processing is only 240 Myr for a $10^{8}$ M$_{\sun}$ \HI{} dwarf galaxy \citep{Kleiner2021}.

\subsection{Individual galaxies}
\label{individual}

\subsubsection{NGC 1326B} The galaxy is a recent infaller in the group \citep{Raj2020}, in an early stage of pre-processing \citep{Kleiner2021}, and has the highest \HI{} mass and \HI{} fraction of our sample plotted on the AD. The SALT slit was placed along the major axis. Therefore, we could extract EW(H$\alpha$) up to 1.5$R_{e}$. The AD implies a recent history (< 1 Gyr) significantly different from secular ageing, with very recent star formation still present at a moderately high rate. The EW(H$\alpha$) profile varies along the slit within 1.0$R_{e}$, but is consistently at its highest between 1.0$R_{e}$ and 1.5$R_{e}$. It is the only galaxy to show a clear EW(H$\alpha$) profile that decreases towards the centre, as the outermost disc (beyond 1.0$R_{e}$) has a high sSFR and is gas rich. 

The gradient from the mass-weighted, SSP-equivalent ages is relatively flat and only slightly younger in the centre. This agrees with the results from FDS imaging, that is, that the outer disc of the galaxy is redder than its inner disc with $(g - i)_{h_{out}-h_{in}}$ = 0.36 mag \citep{Raj2020}, although this is on a scale of more than 2.0$R_{e}$. It has an extended \HI{} symmetric disc, but low molecular gas content (with no CO detected by ALMA, \citealt{Morokuma2022}). It does not interact with NGC 1326A, and lies close to the virial radius of the group \citep{Drinkwater2001}, in the direction of the Fornax cluster.

\subsubsection{NGC 1310} NGC 1310 is an intermediate infaller in the group \citep{Raj2020}, and is in an ongoing pre-processing stage \citep{Kleiner2021}. The EW(H$\alpha$) profile has slight variations but hovers around log EW(H$\alpha$) $\sim$ 1 out to 1.0$R_{e}$, indicating a spatially-extended high recent sSFR, which is still consistent with the secular ageing sequence of the AD. The galaxy has extended \HI{} features that are disturbed and asymmetric, and the outer disc appears redder than the inner disc with average $(g - i)_{h_{out}-h_{in}}$ = 0.44 mag \citep{Raj2020} on a scale of more than 3.0$R_{e}$. Our gradient from the mass-weighted SSP-equivalent ages also shows a younger population at the centre.

\subsubsection{NGC 1316} We placed the slit in the direction of the gas detected and not on the major axis of the stellar halo and therefore derived an EW(H$\alpha$) measurement only to 0.6$R_{e}$. This giant radio galaxy, the brightest group galaxy (BGG), has a unique history in that it displays both ongoing and advanced stages of pre-processing \citep{Kleiner2021}. It had a merger 1 to 2 Gyr ago \citep{Schweizer1980, Serra2019}. Large amounts of \HI{} \citep{Mackie1998, Serra2019, Kleiner2021}, molecular gas \citep{Morokuma2019}, and dust \citep{Schweizer1980, Lanz2010} are detected. 

The H$\alpha$ emission is likely to be, at least in part, a result of the AGN as also shown in Fig. \ref{fig:BPT}. The mass-weighted SSP-equivalent age is also difficult to interpret. We had to include very young stellar templates to obtain a good fit to the observed spectra. This does, however, not necessarily imply massive fractions of very young stellar populations.

\subsubsection{ESO 301-IG11} This galaxy is in an ongoing stage of pre-processing \citep{Kleiner2021}. The slit was placed along the major axis, and we could extract EW(H$\alpha$) up to 1.5$R_{e}$. The integrated value is consistent with the secular ageing sequence of the AD. We measure a flat EW(H$\alpha$) profile within 1.0$R_{e}$, showing that the population of very young stars (<10 Myr) is distributed throughout the galaxy, with a much smaller fraction of very young stars outside of 1.0$R_{e}$. The galaxy is blue in colour, although the outer stellar disc is redder than the inner disc \citep{Raj2020}. The gradient from the mass-weighted SSP-equivalent ages is relatively flat.

\subsubsection{NGC 1326} This galaxy is a recent infaller into the group \citep{Raj2020}, and is in an ongoing stage of pre-processing \citep{Kleiner2021}. We could only measure the flux from within the central core of the galaxy (out to 0.3$R_{e}$) and not from the extended diffuse stellar halo. We therefore did not include this galaxy in the AD along with the integrated values from the other galaxies. NGC 1326 is the most massive late-type galaxy in the sample, and shows plenty of star-forming regions \citep{Raj2020}.

\subsubsection{FCC 35} This galaxy is an intermediate infaller in the group \citep{Raj2020}, and is in an ongoing stage of pre-processing \citep{Kleiner2021}. It has the highest EW(H$\alpha$) of the group sample (that is, very recent high sSFR). Its EW(H$\alpha$) profile decreases from the centre outward. The spectra show extremely strong and narrow emission lines indicative of a blue compact dwarf or an active starburst \HII{} galaxy. We also note that the recent sSFR of FCC 35 is very enhanced but asymmetric (i.e. more so on the side of the galaxy in the direction of the long \HI{} tail). Neither tidal nor hydrodynamical forces can be ruled out from the \HI{} analysis, but it is clearly currently / recently undergoing some interaction(s) \citep{Kleiner2021}.

This is one of the bluest galaxies \citep{Venhola2018, Raj2020}, with an outer disc bluer compared to its inner disc with $(g - i)_{h_{out}-h_{in}}$ = --0.36 mag. This is also evident in the mass-weighted SSP-equivalent age gradient, which is younger in the outer region. This gradient is opposite to the recent sSFR gradient suggested by the EW(H$\alpha$) profile, which is enhanced in the centre. It has a long \HI{} tail pointing away from the group centre, and is not \HI{} deficient, suggesting a recent displacement of gas \citep{Kleiner2021}.

\subsubsection{NGC 1317} NGC 1317 is an ancient infaller into the group \citep{Raj2020} in an advanced stage of pre-processing \citep{Kleiner2021}. The AD implies recent histories that are not too different from secular ageing, with some measurable very recent star formation, although among the lowest in the sample (like all three the advanced-stage galaxies). The \HI{} disc has settled, which suggests that the galaxy has not been affected by any recent (< 1 Gyr) environmental interactions. The galaxy has a red colour \citep{Raj2020}, and is \HI{} deficient (with \HI{} clouds nearby), as discussed in \citet{Kleiner2021}. 

The gradient from the mass-weighted SSP-equivalent ages is relatively flat. \citet{Raj2020} found no star formation beyond the inner 0.5$\arcmin$ disc. This agrees with the results from phase-space analysis that suggest that the galaxy has passed through the pericentre and lost its gas through interactions with the IGrM (see also \citealt{Serra2019}). 

This galaxy can be transitioning from a spiral to a lenticular galaxy, similar to NGC 1436 \citep{Loni2023}. Figure \ref{fig:SFH_NGC1317} shows an older outer disc and a younger inner region still forming stars, and Fig. \ref{fig:Hasummary2} also shows quite well how the EW(H$\alpha$) profile decreases at larger radii. The pre-processing is advanced, as the \HI{} disc is well settled, comparable to NGC 1436.

\subsubsection{NGC 1316C} This recent infaller in the group \citep{Raj2020} is in an advanced stage of pre-processing. It has little \HI{} (unresolved), in a settled disc, suggesting that the galaxy has not been affected by any recent environmental interactions. The EW(H$\alpha$) profile decreases outwards (between 0.5 and 1.0$R_{e}$), and the galaxy lies on the secular ageing sequence in the AD. The gradient from the mass-weighted SSP-equivalent ages is relatively flat.

\subsubsection{FCC 46} This early-type dwarf is in an advanced stage of pre-processing \citep{Kleiner2021}. It is a very small galaxy, and has very little \HI{} (unresolved) that appears kinematically decoupled from the stellar body. The AD implies a recent history significantly different from secular ageing and little to no recent star formation in the outer regions. The galaxy is quenched according to the AD. The EW(H$\alpha$) profile decreases sharply between 0.6 and 1.0$R_{e}$. The gradient from the mass-weighted, SSP-equivalent ages shows older stars in the central region, although both the central and outer regions have ages older than 10 Gyr. \HI{} analysis infer that it has recently experienced a star-forming event, such as a minor merger with a late-type dwarf where it accreted \HI{} that is now in a polar ring (kinematically decoupled from the stellar body) rotating around the optical minor axis \citep{Kleiner2021}.

\subsubsection{NGC 1341} This galaxy, an intermediate infaller in the group \citep{Raj2020}, has relatively high EW(H$\alpha$) over all radii, indicating that the very young stars are spatially distributed throughout the galaxy. It has an integrated value that is consistent with the secular ageing sequence on the AD. The galaxy is outside the field of view of the MeerKAT commissioning observations, but \HI{} has previously been detected by \citet{Courtois2015}. The outer disc of this galaxy is somewhat redder than its inner disc with $(g - i)_{h_{out}-h_{in}}$ = 0.08 mag \citep{Raj2020}, and the gradient from the mass-weighted SSP-equivalent ages is relatively flat. Based on the AD alone, we can suggest that the galaxy is in the early or ongoing stage of pre-processing. Further results based on the EW(H$\alpha$) profiles suggest that it is in an ongoing stage.

\subsection{The Fornax A group}

The most common physical mechanisms in group environments that can transform galaxies from star-forming to quiescent objects are merging and strangulation \citep{Barnes1985, Zabludoff1998, McGee2009}. The number of galaxies in the Fornax A group is nearly ten times lower than that of the Fornax cluster core \citep{Venhola2019, Raj2020}, where the transformation of star-forming galaxies into quiescent galaxies is more efficient. Fornax A has only one early-type galaxy, NGC 1316 (an S0). The absence of early-type galaxies in the group confirms that the group is in a different stage of evolution. The fact that most galaxies (five out of eight) are on the secular evolution sequence implies that pre-processing does not have an immediate effect on the stellar populations and overall a negligible effect compared to secular evolution. The EW(H$\alpha$) profiles show that the environmental transformation in Fornax A is quenching galaxies from the outside in. 

However, it should be noted that Fornax A, in its early mass assembly phase, is not representative of most nearby galaxy groups. One example of a similar group is IC 1459 \citep{Iodice2020}, where the central and peculiar S0 galaxy is quite similar to NGC 1316, and all other members are late-type galaxies with an abundance of \HI{}, as in Fornax A. In general, other nearby groups will be more advanced in their evolution and galaxy transformation.

\section{Summary}
\label{summary}

We study the stellar populations of ten galaxies in the Fornax A galaxy group. Nine of these galaxies were studied in detail with observations from the MeerKAT telescope and classified into different stages of pre-processing \citep{Kleiner2021}. We analysed their stellar population indicators to compare them to the gas content of the galaxies. To achieve this, we measured profiles of the equivalent width of H$\alpha$, constructed an ageing diagram of EW(H$\alpha$) against $g-r$ colours of the galaxies using the outer region EW(H$\alpha$) measurements, and fitted single-burst stellar population models to describe the relative star formation histories of the galaxies in their centres and in their outskirts.

We find that the very recent star formation corresponds closely to the stage of pre-processing, with the early and ongoing stage galaxies in the upper part of the AD, and the advanced stage galaxies in the lower half of the AD, which probe very recent sSFR on the y-axis (EW(H$\alpha$)). Using the colour $g-r$ as an indicator for sSFR over 0.1 to 1 Gyr, the AD shows that NGC 1326B (early), FCC 35 (ongoing), and FCC 46 (advanced) have histories significantly different from secular ageing within the last Gyr. It is possible that star formation was first enhanced, which led to the depletion of gas reservoirs. For the two dwarf galaxies (FCC 35 and FCC 46), this agrees with the conclusions of the \HI{} analysis \citep{Kleiner2021}. These two dwarf galaxies have the lowest stellar mass in our sample. 

The fact that these three galaxies (NGC 1326B, FCC 35 and FCC 46) are all from different pre-processing classes suggests that the stage of pre-processing does not relate to the position of a galaxy relative to the secular ageing sequence on the x-axis. The process(es) that moved these galaxies from the secular ageing sequence must act on timescales less than 1 Gyr. This strongly suggests an environmental effect such as ram pressure stripping, as strangulation, for example, would take several Gyr, and the galaxies will not deviate strongly from the ageing sequence. The fact that most galaxies (five out of eight) are on the secular evolution sequence implies that pre-processing has a negligible effect, at least on the stellar properties, compared to secular evolution. 

We also analysed the spatially-resolved measurements of EW(H$\alpha$). The only galaxy that shows an EW(H$\alpha$) profile that decreases towards the centre is the only early-stage pre-processing galaxy in our sample, NGC 1326B, which suggests that it is not yet quenched in the outer parts. Our results suggest that the stage of pre-processing is related to the distribution and mass fraction of the very young (<10 Myr stars): the early stage galaxy (NGC 1326B) has an increasing profile with high sSFR in the outer regions; the ongoing stage galaxies reveal a distributed population (all over the galaxy) of very young stars (except for FCC 35 which still has a very high sSFR in the centre); and the advanced stage of pre-processing galaxies show a decreasing profile with very low recent sSFR in the outer regions. Based on our measurements, we suggest that NGC 1341 (previously unclassified) is in an ongoing stage of pre-processing. The outside-in pre-processing strongly suggests environmental processes, as opposed to internally triggered quenching mechanisms such as AGN or supernovae.  


There is no correspondence between the pre-processing stage and the mass-weighted SSP equivalent age gradients ($>$ 1 Gyr). We only see a significant age gradient for FCC 35, and
notable age gradients for NGC 1310 and FCC 46. Both FCC 35 and NGC 1310 are in an ongoing stage of pre-processing. The mass-weighted SSP-equivalent age of NGC 1310 is younger in the inner region, while for FCC 35 it is much younger in the outer region. For FCC 35, this gradient is opposite to the recent sSFR gradient measured by EW(H$\alpha$), which is enhanced in the centre. The recent sSFR is also asymmetric. Clearly, FCC 35 had a very dramatic SFH and evolution.   

In summary, we show that the AD and EW(H$\alpha$) profiles can be useful tools to classify the stage of pre-processing in group galaxies, and we conclude that the environmental transformation in Fornax A is pre-processing the galaxies from the outside in.   

\section*{Acknowledgements} 

This work is based on research supported in part by the National Research Foundation (NRF) of South Africa (NRF Grant Number: 146053). Any opinion, finding, and conclusion or recommendation expressed in this material is that of the author(s), and the NRF does not accept any liability in this regard. S.I.L. also acknowledges the support from the Italian Ministry of Foreign Affairs and International Cooperation (MAECI) and the NRF as part of the ISARP RADIOSKY2023 Joint Research Scheme. This project has received funding from the European Research Council (ERC) under the European Union’s Horizon 2020 research and innovation programme (grant agreement no. 679627, ``FORNAX"). P.K. is supported by the BMBF project 05A20PC4 for D-MeerKAT. N.Z. is supported through the South African Research Chairs Initiative of the Department of Science and Innovation and the NRF.

All long-slit spectra observations reported in this paper were obtained with the South African Large Telescope (SALT) under programme numbers 2019-2-MLT-002 and 2021-1-SCI-019 (PI: Loubser). The MeerKAT telescope is operated by the South African Radio Astronomy Observatory, which is a facility of the NRF, an agency of the Department of Science and Innovation. This research used Astropy,\footnote{http://www.astropy.org} a community-developed core Python package for Astronomy \citep{Astropy2013, Astropy2018}.

\section*{Data availability}

The data underlying this article will be shared on reasonable request with the corresponding author. Data from the MeerKAT Fornax Survey are available at https://sites.google.com/inaf.it/meerkatfornaxsurvey. 



\bibliographystyle{mnras}
\bibliography{References} 

\begin{thebibliography}{}
\makeatletter
\relax
\def\mn@urlcharsother{\let\do\@makeother \do\$\do\&\do\#\do\^\do\_\do\%\do\~}
\def\mn@doi{\begingroup\mn@urlcharsother \@ifnextchar [ {\mn@doi@}
  {\mn@doi@[]}}
\def\mn@doi@[#1]#2{\def\@tempa{#1}\ifx\@tempa\@empty \href
  {http://dx.doi.org/#2} {doi:#2}\else \href {http://dx.doi.org/#2} {#1}\fi
  \endgroup}
\def\mn@eprint#1#2{\mn@eprint@#1:#2::\@nil}
\def\mn@eprint@arXiv#1{\href {http://arxiv.org/abs/#1} {{\tt arXiv:#1}}}
\def\mn@eprint@dblp#1{\href {http://dblp.uni-trier.de/rec/bibtex/#1.xml}
  {dblp:#1}}
\def\mn@eprint@#1:#2:#3:#4\@nil{\def\@tempa {#1}\def\@tempb {#2}\def\@tempc
  {#3}\ifx \@tempc \@empty \let \@tempc \@tempb \let \@tempb \@tempa \fi \ifx
  \@tempb \@empty \def\@tempb {arXiv}\fi \@ifundefined
  {mn@eprint@\@tempb}{\@tempb:\@tempc}{\expandafter \expandafter \csname
  mn@eprint@\@tempb\endcsname \expandafter{\@tempc}}}

\bibitem[\protect\citeauthoryear{{Abramson}, {Gladders}, {Dressler}, {Oemler},
  {Poggianti}  \& {Vulcani}}{{Abramson} et~al.}{2016}]{Abramson2016}
{Abramson} L.~E.,  {Gladders} M.~D.,  {Dressler} A.,  {Oemler} Augustus J.,
  {Poggianti} B.,   {Vulcani} B.,  2016, \mn@doi [\apj]
  {10.3847/0004-637X/832/1/7}, \href
  {https://ui.adsabs.harvard.edu/abs/2016ApJ...832....7A} {832, 7}

\bibitem[\protect\citeauthoryear{{Astropy Collaboration} et~al.,}{{Astropy
  Collaboration} et~al.}{2013}]{Astropy2013}
{Astropy Collaboration} et~al., 2013, \mn@doi [\aap]
  {10.1051/0004-6361/201322068}, \href
  {https://ui.adsabs.harvard.edu/abs/2013A&A...558A..33A} {558, A33}

\bibitem[\protect\citeauthoryear{{Astropy Collaboration} et~al.,}{{Astropy
  Collaboration} et~al.}{2018}]{Astropy2018}
{Astropy Collaboration} et~al., 2018, \mn@doi [\aj] {10.3847/1538-3881/aabc4f},
  \href {https://ui.adsabs.harvard.edu/abs/2018AJ....156..123A} {156, 123}

\bibitem[\protect\citeauthoryear{{Bah{\'e}}, {McCarthy}, {Balogh}  \&
  {Font}}{{Bah{\'e}} et~al.}{2013}]{Bahe2013}
{Bah{\'e}} Y.~M.,  {McCarthy} I.~G.,  {Balogh} M.~L.,   {Font} A.~S.,  2013,
  \mn@doi [\mnras] {10.1093/mnras/stt109}, \href
  {https://ui.adsabs.harvard.edu/abs/2013MNRAS.430.3017B} {430, 3017}

\bibitem[\protect\citeauthoryear{{Baldwin}, {Phillips}  \&
  {Terlevich}}{{Baldwin} et~al.}{1981}]{Baldwin1981}
{Baldwin} J.~A.,  {Phillips} M.~M.,   {Terlevich} R.,  1981, \mn@doi [PASP]
  {10.1086/130766}, \href {http://saaoads.chpc.ac.za/abs/1981PASP...93....5B}
  {93, 5}

\bibitem[\protect\citeauthoryear{{Balogh}, {Morris}, {Yee}, {Carlberg}  \&
  {Ellingson}}{{Balogh} et~al.}{1999}]{Balogh1999}
{Balogh} M.~L.,  {Morris} S.~L.,  {Yee} H. K.~C.,  {Carlberg} R.~G.,
  {Ellingson} E.,  1999, \mn@doi [ApJ] {10.1086/308056}, \href
  {http://saaoads.chpc.ac.za/abs/1999ApJ...527...54B} {527, 54}

\bibitem[\protect\citeauthoryear{{Barnes}}{{Barnes}}{1985}]{Barnes1985}
{Barnes} J.,  1985, \mn@doi [\mnras] {10.1093/mnras/215.3.517}, \href
  {https://ui.adsabs.harvard.edu/abs/1985MNRAS.215..517B} {215, 517}

\bibitem[\protect\citeauthoryear{{Barsanti} et~al.,}{{Barsanti}
  et~al.}{2018}]{Barsanti2018}
{Barsanti} S.,  et~al., 2018, \mn@doi [\apj] {10.3847/1538-4357/aab61a}, \href
  {https://ui.adsabs.harvard.edu/abs/2018ApJ...857...71B} {857, 71}

\bibitem[\protect\citeauthoryear{{Belfiore} et~al.,}{{Belfiore}
  et~al.}{2018}]{Belfiore2018}
{Belfiore} F.,  et~al., 2018, \mn@doi [\mnras] {10.1093/mnras/sty768}, \href
  {https://ui.adsabs.harvard.edu/abs/2018MNRAS.477.3014B} {477, 3014}

\bibitem[\protect\citeauthoryear{{Bell} \& {de Jong}}{{Bell} \& {de
  Jong}}{2000}]{Bell2000}
{Bell} E.~F.,  {de Jong} R.~S.,  2000, \mn@doi [\mnras]
  {10.1046/j.1365-8711.2000.03138.x}, \href
  {https://ui.adsabs.harvard.edu/abs/2000MNRAS.312..497B} {312, 497}

\bibitem[\protect\citeauthoryear{{Bidaran} et~al.,}{{Bidaran}
  et~al.}{2022}]{Bidaran2022}
{Bidaran} B.,  et~al., 2022, \mn@doi [\mnras] {10.1093/mnras/stac2005}, \href
  {https://ui.adsabs.harvard.edu/abs/2022MNRAS.515.4622B} {515, 4622}

\bibitem[\protect\citeauthoryear{{Boselli} \& {Gavazzi}}{{Boselli} \&
  {Gavazzi}}{2006}]{Boselli2006}
{Boselli} A.,  {Gavazzi} G.,  2006, \mn@doi [\pasp] {10.1086/500691}, \href
  {https://ui.adsabs.harvard.edu/abs/2006PASP..118..517B} {118, 517}

\bibitem[\protect\citeauthoryear{{Boselli}, {Cortese}, {Boquien}, {Boissier},
  {Catinella}, {Lagos}  \& {Saintonge}}{{Boselli} et~al.}{2014}]{Boselli2014}
{Boselli} A.,  {Cortese} L.,  {Boquien} M.,  {Boissier} S.,  {Catinella} B.,
  {Lagos} C.,   {Saintonge} A.,  2014, \mn@doi [\aap]
  {10.1051/0004-6361/201322312}, \href
  {https://ui.adsabs.harvard.edu/abs/2014A&A...564A..66B} {564, A66}

\bibitem[\protect\citeauthoryear{{Boselli} et~al.,}{{Boselli}
  et~al.}{2016}]{Boselli2016}
{Boselli} A.,  et~al., 2016, \mn@doi [\aap] {10.1051/0004-6361/201629221},
  \href {https://ui.adsabs.harvard.edu/abs/2016A&A...596A..11B} {596, A11}

\bibitem[\protect\citeauthoryear{{Burgh}, {Nordsieck}, {Kobulnicky},
  {Williams}, {O'Donoghue}, {Smith}  \& {Percival}}{{Burgh}
  et~al.}{2003}]{Burgh2003}
{Burgh} E.~B.,  {Nordsieck} K.~H.,  {Kobulnicky} H.~A.,  {Williams} T.~B.,
  {O'Donoghue} D.,  {Smith} M.~P.,   {Percival} J.~W.,  2003, in {Iye} M.,
  {Moorwood} A. F.~M.,  eds, ~SPIE Vol. 4841, Instrument Design and Performance
  for Optical/Infrared Ground-based Telescopes. pp 1463--1471,
  \mn@doi{10.1117/12.460312}

\bibitem[\protect\citeauthoryear{{Cappellari}}{{Cappellari}}{2017}]{Cappellari2017}
{Cappellari} M.,  2017, \mn@doi [\mnras] {10.1093/mnras/stw3020}, \href
  {https://ui.adsabs.harvard.edu/abs/2017MNRAS.466..798C} {466, 798}

\bibitem[\protect\citeauthoryear{{Cappellari} \& {Emsellem}}{{Cappellari} \&
  {Emsellem}}{2004}]{Cappellari2004}
{Cappellari} M.,  {Emsellem} E.,  2004, \mn@doi [PASP] {10.1086/381875}, \href
  {http://saaoads.chpc.ac.za/abs/2004PASP..116..138C} {116, 138}

\bibitem[\protect\citeauthoryear{{Chabrier}}{{Chabrier}}{2003}]{Chabrier2003}
{Chabrier} G.,  2003, \mn@doi [\pasp] {10.1086/376392}, \href
  {https://ui.adsabs.harvard.edu/abs/2003PASP..115..763C} {115, 763}

\bibitem[\protect\citeauthoryear{{Chung}, {van Gorkom}, {Kenney}, {Crowl}  \&
  {Vollmer}}{{Chung} et~al.}{2009}]{Chung2009}
{Chung} A.,  {van Gorkom} J.~H.,  {Kenney} J. D.~P.,  {Crowl} H.,   {Vollmer}
  B.,  2009, \mn@doi [\aj] {10.1088/0004-6256/138/6/1741}, \href
  {https://ui.adsabs.harvard.edu/abs/2009AJ....138.1741C} {138, 1741}

\bibitem[\protect\citeauthoryear{{Cid Fernandes}, {Mateus}, {Sodr{\'e}},
  {Stasi{\'n}ska}  \& {Gomes}}{{Cid Fernandes} et~al.}{2005}]{CidFernandes2005}
{Cid Fernandes} R.,  {Mateus} A.,  {Sodr{\'e}} L.,  {Stasi{\'n}ska} G.,
  {Gomes} J.~M.,  2005, \mn@doi [\mnras] {10.1111/j.1365-2966.2005.08752.x},
  \href {https://ui.adsabs.harvard.edu/abs/2005MNRAS.358..363C} {358, 363}

\bibitem[\protect\citeauthoryear{{Citro}, {Pozzetti}, {Quai}, {Moresco},
  {Vallini}  \& {Cimatti}}{{Citro} et~al.}{2017}]{Citro2017}
{Citro} A.,  {Pozzetti} L.,  {Quai} S.,  {Moresco} M.,  {Vallini} L.,
  {Cimatti} A.,  2017, \mn@doi [\mnras] {10.1093/mnras/stx932}, \href
  {https://ui.adsabs.harvard.edu/abs/2017MNRAS.469.3108C} {469, 3108}

\bibitem[\protect\citeauthoryear{{Corcho-Caballero}, {Casado}, {Ascasibar}  \&
  {Garc{\'\i}a-Benito}}{{Corcho-Caballero} et~al.}{2021}]{Corcho2021}
{Corcho-Caballero} P.,  {Casado} J.,  {Ascasibar} Y.,   {Garc{\'\i}a-Benito}
  R.,  2021, \mn@doi [\mnras] {10.1093/mnras/stab2503}, \href
  {https://ui.adsabs.harvard.edu/abs/2021MNRAS.507.5477C} {507, 5477}

\bibitem[\protect\citeauthoryear{{Corcho-Caballero}, {Ascasibar}, {S{\'a}nchez}
   \& {L{\'o}pez-S{\'a}nchez}}{{Corcho-Caballero} et~al.}{2023a}]{Corcho2023}
{Corcho-Caballero} P.,  {Ascasibar} Y.,  {S{\'a}nchez} S.~F.,
  {L{\'o}pez-S{\'a}nchez} {\'A}.~R.,  2023a, \mn@doi [\mnras]
  {10.1093/mnras/stad147}, \href
  {https://ui.adsabs.harvard.edu/abs/2023MNRAS.520..193C} {520, 193}

\bibitem[\protect\citeauthoryear{{Corcho-Caballero}, {Ascasibar}, {Cortese},
  {S{\'a}nchez}, {L{\'o}pez-S{\'a}nchez}, {Fraser-McKelvie}  \&
  {Zafar}}{{Corcho-Caballero} et~al.}{2023b}]{Corcho2023b}
{Corcho-Caballero} P.,  {Ascasibar} Y.,  {Cortese} L.,  {S{\'a}nchez} S.~F.,
  {L{\'o}pez-S{\'a}nchez} {\'A}.~R.,  {Fraser-McKelvie} A.,   {Zafar} T.,
  2023b, \mn@doi [\mnras] {10.1093/mnras/stad2096}, \href
  {https://ui.adsabs.harvard.edu/abs/2023MNRAS.524.3692C} {524, 3692}

\bibitem[\protect\citeauthoryear{{Cortese}, {Catinella}  \& {Smith}}{{Cortese}
  et~al.}{2021}]{Cortese2021}
{Cortese} L.,  {Catinella} B.,   {Smith} R.,  2021, \mn@doi [\pasa]
  {10.1017/pasa.2021.18}, \href
  {https://ui.adsabs.harvard.edu/abs/2021PASA...38...35C} {38, e035}

\bibitem[\protect\citeauthoryear{{Courtois} \& {Tully}}{{Courtois} \&
  {Tully}}{2015}]{Courtois2015}
{Courtois} H.~M.,  {Tully} R.~B.,  2015, \mn@doi [\mnras]
  {10.1093/mnras/stu2405}, \href
  {https://ui.adsabs.harvard.edu/abs/2015MNRAS.447.1531C} {447, 1531}

\bibitem[\protect\citeauthoryear{{Cowie} \& {McKee}}{{Cowie} \&
  {McKee}}{1977}]{Cowie1977}
{Cowie} L.~L.,  {McKee} C.~F.,  1977, \mn@doi [\apj] {10.1086/154911}, \href
  {https://ui.adsabs.harvard.edu/abs/1977ApJ...211..135C} {211, 135}

\bibitem[\protect\citeauthoryear{{Crawford} et~al.,}{{Crawford}
  et~al.}{2010}]{Crawford2010}
{Crawford} S.~M.,  et~al., 2010, in Observatory Operations: Strategies,
  Processes, and Systems III. p. 773725, \mn@doi{10.1117/12.857000}

\bibitem[\protect\citeauthoryear{{Cresci} et~al.,}{{Cresci}
  et~al.}{2015}]{Cresci2015}
{Cresci} G.,  et~al., 2015, \mn@doi [\apj] {10.1088/0004-637X/799/1/82}, \href
  {https://ui.adsabs.harvard.edu/abs/2015ApJ...799...82C} {799, 82}

\bibitem[\protect\citeauthoryear{{Croton} et~al.,}{{Croton}
  et~al.}{2006}]{Croton2006}
{Croton} D.~J.,  et~al., 2006, \mn@doi [\mnras]
  {10.1111/j.1365-2966.2005.09675.x}, \href
  {https://ui.adsabs.harvard.edu/abs/2006MNRAS.365...11C} {365, 11}

\bibitem[\protect\citeauthoryear{{Davies} et~al.,}{{Davies}
  et~al.}{2019}]{Davies2019}
{Davies} L.~J.~M.,  et~al., 2019, \mn@doi [\mnras] {10.1093/mnras/sty3393},
  \href {https://ui.adsabs.harvard.edu/abs/2019MNRAS.483.5444D} {483, 5444}

\bibitem[\protect\citeauthoryear{{De Lucia}, {Weinmann}, {Poggianti},
  {Arag{\'o}n-Salamanca}  \& {Zaritsky}}{{De Lucia} et~al.}{2012}]{DeLucia2012}
{De Lucia} G.,  {Weinmann} S.,  {Poggianti} B.~M.,  {Arag{\'o}n-Salamanca} A.,
   {Zaritsky} D.,  2012, \mn@doi [\mnras] {10.1111/j.1365-2966.2012.20983.x},
  \href {https://ui.adsabs.harvard.edu/abs/2012MNRAS.423.1277D} {423, 1277}

\bibitem[\protect\citeauthoryear{{Drinkwater}, {Gregg}, {Holman}  \&
  {Brown}}{{Drinkwater} et~al.}{2001}]{Drinkwater2001}
{Drinkwater} M.~J.,  {Gregg} M.~D.,  {Holman} B.~A.,   {Brown} M.~J.~I.,  2001,
  \mn@doi [\mnras] {10.1046/j.1365-8711.2001.04646.x}, \href
  {https://ui.adsabs.harvard.edu/abs/2001MNRAS.326.1076D} {326, 1076}

\bibitem[\protect\citeauthoryear{{Ferguson}}{{Ferguson}}{1989}]{Ferguson1989}
{Ferguson} H.~C.,  1989, \mn@doi [\aj] {10.1086/115152}, \href
  {https://ui.adsabs.harvard.edu/abs/1989AJ.....98..367F} {98, 367}

\bibitem[\protect\citeauthoryear{{Fitts} et~al.,}{{Fitts}
  et~al.}{2017}]{Fitts2017}
{Fitts} A.,  et~al., 2017, \mn@doi [\mnras] {10.1093/mnras/stx1757}, \href
  {https://ui.adsabs.harvard.edu/abs/2017MNRAS.471.3547F} {471, 3547}

\bibitem[\protect\citeauthoryear{{Fossati}, {Fumagalli}, {Gavazzi},
  {Consolandi}, {Boselli}, {Yagi}, {Sun}  \& {Wilman}}{{Fossati}
  et~al.}{2019}]{Fossati2019}
{Fossati} M.,  {Fumagalli} M.,  {Gavazzi} G.,  {Consolandi} G.,  {Boselli} A.,
  {Yagi} M.,  {Sun} M.,   {Wilman} D.~J.,  2019, \mn@doi [\mnras]
  {10.1093/mnras/stz136}, \href
  {https://ui.adsabs.harvard.edu/abs/2019MNRAS.484.2212F} {484, 2212}

\bibitem[\protect\citeauthoryear{{Fujita}}{{Fujita}}{2004}]{Fujita2004}
{Fujita} Y.,  2004, \mn@doi [\pasj] {10.1093/pasj/56.1.29}, \href
  {https://ui.adsabs.harvard.edu/abs/2004PASJ...56...29F} {56, 29}

\bibitem[\protect\citeauthoryear{{Gonz{\'a}lez Delgado} et~al.,}{{Gonz{\'a}lez
  Delgado} et~al.}{2015}]{Gonzalez-Delgado2015}
{Gonz{\'a}lez Delgado} R.~M.,  et~al., 2015, \mn@doi [\aap]
  {10.1051/0004-6361/201525938}, \href
  {https://ui.adsabs.harvard.edu/abs/2015A&A...581A.103G} {581, A103}

\bibitem[\protect\citeauthoryear{{Goudfrooij}, {Mack}, {Kissler-Patig},
  {Meylan}  \& {Minniti}}{{Goudfrooij} et~al.}{2001}]{Goudfrooij2001}
{Goudfrooij} P.,  {Mack} J.,  {Kissler-Patig} M.,  {Meylan} G.,   {Minniti} D.,
   2001, \mn@doi [\mnras] {10.1046/j.1365-8711.2001.04154.x}, \href
  {https://ui.adsabs.harvard.edu/abs/2001MNRAS.322..643G} {322, 643}

\bibitem[\protect\citeauthoryear{{Groenewald} \& {Loubser}}{{Groenewald} \&
  {Loubser}}{2014}]{Groenewald2014}
{Groenewald} D.~N.,  {Loubser} S.~I.,  2014, \mn@doi [MNRAS]
  {10.1093/mnras/stu1319}, \href
  {http://saaoads.chpc.ac.za/abs/2014MNRAS.444..808G} {444, 808}

\bibitem[\protect\citeauthoryear{{Gunn} \& {Gott}}{{Gunn} \&
  {Gott}}{1972}]{Gunn1972}
{Gunn} J.~E.,  {Gott} J.~Richard I.,  1972, \mn@doi [\apj] {10.1086/151605},
  \href {https://ui.adsabs.harvard.edu/abs/1972ApJ...176....1G} {176, 1}

\bibitem[\protect\citeauthoryear{{Haines} et~al.,}{{Haines}
  et~al.}{2013}]{Haines2013}
{Haines} C.~P.,  et~al., 2013, \mn@doi [\apj] {10.1088/0004-637X/775/2/126},
  \href {https://ui.adsabs.harvard.edu/abs/2013ApJ...775..126H} {775, 126}

\bibitem[\protect\citeauthoryear{{Haines} et~al.,}{{Haines}
  et~al.}{2015}]{Haines2015}
{Haines} C.~P.,  et~al., 2015, \mn@doi [\apj] {10.1088/0004-637X/806/1/101},
  \href {https://ui.adsabs.harvard.edu/abs/2015ApJ...806..101H} {806, 101}

\bibitem[\protect\citeauthoryear{{Heavens}, {Jimenez}  \& {Lahav}}{{Heavens}
  et~al.}{2000}]{Heavens2000}
{Heavens} A.~F.,  {Jimenez} R.,   {Lahav} O.,  2000, \mn@doi [\mnras]
  {10.1046/j.1365-8711.2000.03692.x}, \href
  {https://ui.adsabs.harvard.edu/abs/2000MNRAS.317..965H} {317, 965}

\bibitem[\protect\citeauthoryear{{Huchra} et~al.,}{{Huchra}
  et~al.}{2012}]{Huchra2012}
{Huchra} J.~P.,  et~al., 2012, \mn@doi [\apjs] {10.1088/0067-0049/199/2/26},
  \href {https://ui.adsabs.harvard.edu/abs/2012ApJS..199...26H} {199, 26}

\bibitem[\protect\citeauthoryear{{Iodice} et~al.,}{{Iodice}
  et~al.}{2016}]{Iodice2016}
{Iodice} E.,  et~al., 2016, \mn@doi [\apj] {10.3847/0004-637X/820/1/42}, \href
  {https://ui.adsabs.harvard.edu/abs/2016ApJ...820...42I} {820, 42}

\bibitem[\protect\citeauthoryear{{Iodice} et~al.,}{{Iodice}
  et~al.}{2017}]{Iodice2017}
{Iodice} E.,  et~al., 2017, \mn@doi [\apj] {10.3847/1538-4357/aa6846}, \href
  {https://ui.adsabs.harvard.edu/abs/2017ApJ...839...21I} {839, 21}

\bibitem[\protect\citeauthoryear{{Iodice} et~al.,}{{Iodice}
  et~al.}{2019}]{Iodice2019}
{Iodice} E.,  et~al., 2019, \mn@doi [\aap] {10.1051/0004-6361/201935721}, \href
  {https://ui.adsabs.harvard.edu/abs/2019A&A...627A.136I} {627, A136}

\bibitem[\protect\citeauthoryear{{Iodice} et~al.,}{{Iodice}
  et~al.}{2020}]{Iodice2020}
{Iodice} E.,  et~al., 2020, \mn@doi [\aap] {10.1051/0004-6361/201936435}, \href
  {https://ui.adsabs.harvard.edu/abs/2020A&A...635A...3I} {635, A3}

\bibitem[\protect\citeauthoryear{{Jarrett}, {Chester}, {Cutri}, {Schneider},
  {Skrutskie}  \& {Huchra}}{{Jarrett} et~al.}{2000}]{Jarrett2000}
{Jarrett} T.~H.,  {Chester} T.,  {Cutri} R.,  {Schneider} S.,  {Skrutskie} M.,
   {Huchra} J.~P.,  2000, \mn@doi [\aj] {10.1086/301330}, \href
  {https://ui.adsabs.harvard.edu/abs/2000AJ....119.2498J} {119, 2498}

\bibitem[\protect\citeauthoryear{{Jung} et~al.,}{{Jung}
  et~al.}{2022}]{Jung2022}
{Jung} S.~L.,  et~al., 2022, \mn@doi [\mnras] {10.1093/mnras/stac1622}, \href
  {https://ui.adsabs.harvard.edu/abs/2022MNRAS.515...22J} {515, 22}

\bibitem[\protect\citeauthoryear{{Kauffmann} et~al.,}{{Kauffmann}
  et~al.}{2003}]{Kauffmann2003}
{Kauffmann} G.,  et~al., 2003, \mn@doi [MNRAS]
  {10.1046/j.1365-8711.2003.06291.x}, \href
  {http://saaoads.chpc.ac.za/abs/2003MNRAS.341...33K} {341, 33}

\bibitem[\protect\citeauthoryear{{Kennicutt}}{{Kennicutt}}{1998}]{Kennicutt1998}
{Kennicutt} Robert~C. J.,  1998, \mn@doi [\araa]
  {10.1146/annurev.astro.36.1.189}, \href
  {https://ui.adsabs.harvard.edu/abs/1998ARA&A..36..189K} {36, 189}

\bibitem[\protect\citeauthoryear{{Kewley}, {Dopita}, {Sutherland}, {Heisler}
  \& {Trevena}}{{Kewley} et~al.}{2001}]{Kewley2001}
{Kewley} L.~J.,  {Dopita} M.~A.,  {Sutherland} R.~S.,  {Heisler} C.~A.,
  {Trevena} J.,  2001, \mn@doi [ApJ] {10.1086/321545}, \href
  {http://saaoads.chpc.ac.za/abs/2001ApJ...556..121K} {556, 121}

\bibitem[\protect\citeauthoryear{{Kewley}, {Groves}, {Kauffmann}  \&
  {Heckman}}{{Kewley} et~al.}{2006}]{Kewley2006}
{Kewley} L.~J.,  {Groves} B.,  {Kauffmann} G.,   {Heckman} T.,  2006, \mn@doi
  [\mnras] {10.1111/j.1365-2966.2006.10859.x}, \href
  {https://ui.adsabs.harvard.edu/abs/2006MNRAS.372..961K} {372, 961}

\bibitem[\protect\citeauthoryear{{Kleiner} et~al.,}{{Kleiner}
  et~al.}{2019}]{Kleiner2019}
{Kleiner} D.,  et~al., 2019, \mn@doi [\mnras] {10.1093/mnras/stz2063}, \href
  {https://ui.adsabs.harvard.edu/abs/2019MNRAS.488.5352K} {488, 5352}

\bibitem[\protect\citeauthoryear{{Kleiner} et~al.,}{{Kleiner}
  et~al.}{2021}]{Kleiner2021}
{Kleiner} D.,  et~al., 2021, \mn@doi [\aap] {10.1051/0004-6361/202039898},
  \href {https://ui.adsabs.harvard.edu/abs/2021A&A...648A..32K} {648, A32}

\bibitem[\protect\citeauthoryear{{Kleiner} et~al.,}{{Kleiner}
  et~al.}{2023}]{Kleiner2023}
{Kleiner} D.,  et~al., 2023, \mn@doi [\aap] {10.1051/0004-6361/202346461},
  \href {https://ui.adsabs.harvard.edu/abs/2023A&A...675A.108K} {675, A108}

\bibitem[\protect\citeauthoryear{{Kobulnicky}, {Nordsieck}, {Burgh}, {Smith},
  {Percival}, {Williams}  \& {O'Donoghue}}{{Kobulnicky}
  et~al.}{2003}]{Kobulnicky2003}
{Kobulnicky} H.~A.,  {Nordsieck} K.~H.,  {Burgh} E.~B.,  {Smith} M.~P.,
  {Percival} J.~W.,  {Williams} T.~B.,   {O'Donoghue} D.,  2003, in {Iye} M.,
  {Moorwood} A. F.~M.,  eds,  Society of Photo-Optical Instrumentation
  Engineers (SPIE) Conference Series Vol. 4841, Instrument Design and
  Performance for Optical/Infrared Ground-based Telescopes. pp 1634--1644,
  \mn@doi{10.1117/12.460315}

\bibitem[\protect\citeauthoryear{{Koleva}, {Prugniel}, {Bouchard}  \&
  {Wu}}{{Koleva} et~al.}{2009}]{Koleva2009}
{Koleva} M.,  {Prugniel} P.,  {Bouchard} A.,   {Wu} Y.,  2009, \mn@doi [A\&A]
  {10.1051/0004-6361/200811467}, \href
  {http://saaoads.chpc.ac.za/abs/2009A\&A...501.1269K} {501, 1269}

\bibitem[\protect\citeauthoryear{{Koribalski} et~al.,}{{Koribalski}
  et~al.}{2004}]{Koribalski2004}
{Koribalski} B.~S.,  et~al., 2004, \mn@doi [\aj] {10.1086/421744}, \href
  {https://ui.adsabs.harvard.edu/abs/2004AJ....128...16K} {128, 16}

\bibitem[\protect\citeauthoryear{{Kroupa}}{{Kroupa}}{2001}]{Kroupa2001}
{Kroupa} P.,  2001, \mn@doi [\mnras] {10.1046/j.1365-8711.2001.04022.x}, \href
  {http://adsabs.harvard.edu/abs/2001MNRAS.322..231K} {322, 231}

\bibitem[\protect\citeauthoryear{{Lanz}, {Jones}, {Forman}, {Ashby}, {Kraft}
  \& {Hickox}}{{Lanz} et~al.}{2010}]{Lanz2010}
{Lanz} L.,  {Jones} C.,  {Forman} W.~R.,  {Ashby} M. L.~N.,  {Kraft} R.,
  {Hickox} R.~C.,  2010, \mn@doi [\apj] {10.1088/0004-637X/721/2/1702}, \href
  {https://ui.adsabs.harvard.edu/abs/2010ApJ...721.1702L} {721, 1702}

\bibitem[\protect\citeauthoryear{{Le Borgne}, {Rocca-Volmerange}, {Prugniel},
  {Lan{\c c}on}, {Fioc}  \& {Soubiran}}{{Le Borgne}
  et~al.}{2004}]{LeBorgne2004}
{Le Borgne} D.,  {Rocca-Volmerange} B.,  {Prugniel} P.,  {Lan{\c c}on} A.,
  {Fioc} M.,   {Soubiran} C.,  2004, \mn@doi [A\&A]
  {10.1051/0004-6361:200400044}, \href
  {http://saaoads.chpc.ac.za/abs/2004A%26A...425..881L} {425, 881}

\bibitem[\protect\citeauthoryear{{Lewis} et~al.,}{{Lewis}
  et~al.}{2002}]{Lewis2002}
{Lewis} I.,  et~al., 2002, \mn@doi [\mnras] {10.1046/j.1365-8711.2002.05558.x},
  \href {https://ui.adsabs.harvard.edu/abs/2002MNRAS.334..673L} {334, 673}

\bibitem[\protect\citeauthoryear{{Loni} et~al.,}{{Loni}
  et~al.}{2023}]{Loni2023}
{Loni} A.,  et~al., 2023, \mn@doi [\mnras] {10.1093/mnras/stad1422}, \href
  {https://ui.adsabs.harvard.edu/abs/2023MNRAS.523.1140L} {523, 1140}

\bibitem[\protect\citeauthoryear{{Loubser} \&
  {S{\'a}nchez-Bl{\'a}zquez}}{{Loubser} \&
  {S{\'a}nchez-Bl{\'a}zquez}}{2012}]{Loubser2012}
{Loubser} S.~I.,  {S{\'a}nchez-Bl{\'a}zquez} P.,  2012, \mn@doi [MNRAS]
  {10.1111/j.1365-2966.2012.21079.x}, \href
  {http://saaoads.chpc.ac.za/abs/2012MNRAS.425..841L} {425, 841}

\bibitem[\protect\citeauthoryear{{Loubser}, {Babul}, {Hoekstra}, {Mahdavi},
  {Donahue}, {Bildfell}  \& {Voit}}{{Loubser} et~al.}{2016}]{Loubser2016}
{Loubser} S.~I.,  {Babul} A.,  {Hoekstra} H.,  {Mahdavi} A.,  {Donahue} M.,
  {Bildfell} C.,   {Voit} G.~M.,  2016, \mn@doi [MNRAS]
  {10.1093/mnras/stv2784}, \href
  {http://adsabs.harvard.edu/abs/2016MNRAS.456.1565L} {456, 1565}

\bibitem[\protect\citeauthoryear{{Loubser}, {Hoekstra}, {Babul}, {Bah{\'e}}  \&
  {Donahue}}{{Loubser} et~al.}{2021}]{Loubser2021}
{Loubser} S.~I.,  {Hoekstra} H.,  {Babul} A.,  {Bah{\'e}} Y.~M.,   {Donahue}
  M.,  2021, \mn@doi [\mnras] {10.1093/mnras/staa3530}, \href
  {https://ui.adsabs.harvard.edu/abs/2021MNRAS.500.4153L} {500, 4153}

\bibitem[\protect\citeauthoryear{{Loubser}, {Lagos}, {Babul}, {O'Sullivan},
  {Jung}, {Olivares}  \& {Kolokythas}}{{Loubser} et~al.}{2022}]{Loubser2022}
{Loubser} S.~I.,  {Lagos} P.,  {Babul} A.,  {O'Sullivan} E.,  {Jung} S.~L.,
  {Olivares} V.,   {Kolokythas} K.,  2022, \mn@doi [\mnras]
  {10.1093/mnras/stac1781}, \href
  {https://ui.adsabs.harvard.edu/abs/2022MNRAS.515.1104L} {515, 1104}

\bibitem[\protect\citeauthoryear{{Maccagni} et~al.,}{{Maccagni}
  et~al.}{2021}]{Maccagni2021}
{Maccagni} F.~M.,  et~al., 2021, \mn@doi [\aap] {10.1051/0004-6361/202141143},
  \href {https://ui.adsabs.harvard.edu/abs/2021A&A...656A..45M} {656, A45}

\bibitem[\protect\citeauthoryear{{Mackie} \& {Fabbiano}}{{Mackie} \&
  {Fabbiano}}{1998}]{Mackie1998}
{Mackie} G.,  {Fabbiano} G.,  1998, \mn@doi [\aj] {10.1086/300203}, \href
  {https://ui.adsabs.harvard.edu/abs/1998AJ....115..514M} {115, 514}

\bibitem[\protect\citeauthoryear{{Maddox}, {Serra}, {Venhola}, {Peletier},
  {Loubser}  \& {Iodice}}{{Maddox} et~al.}{2019}]{Maddox2019}
{Maddox} N.,  {Serra} P.,  {Venhola} A.,  {Peletier} R.,  {Loubser} I.,
  {Iodice} E.,  2019, \mn@doi [\mnras] {10.1093/mnras/stz2530}, \href
  {https://ui.adsabs.harvard.edu/abs/2019MNRAS.490.1666M} {490, 1666}

\bibitem[\protect\citeauthoryear{{Mahajan}, {Raychaudhury}  \&
  {Pimbblet}}{{Mahajan} et~al.}{2012}]{Mahajan2012}
{Mahajan} S.,  {Raychaudhury} S.,   {Pimbblet} K.~A.,  2012, \mn@doi [\mnras]
  {10.1111/j.1365-2966.2012.22059.x}, \href
  {https://ui.adsabs.harvard.edu/abs/2012MNRAS.427.1252M} {427, 1252}

\bibitem[\protect\citeauthoryear{{Maraston} \& {Str{\"o}mb{\"a}ck}}{{Maraston}
  \& {Str{\"o}mb{\"a}ck}}{2011}]{Maraston2011}
{Maraston} C.,  {Str{\"o}mb{\"a}ck} G.,  2011, \mn@doi [MNRAS]
  {10.1111/j.1365-2966.2011.19738.x}, \href
  {http://saaoads.chpc.ac.za/abs/2011MNRAS.418.2785M} {418, 2785}

\bibitem[\protect\citeauthoryear{{McGee}, {Balogh}, {Bower}, {Font}  \&
  {McCarthy}}{{McGee} et~al.}{2009}]{McGee2009}
{McGee} S.~L.,  {Balogh} M.~L.,  {Bower} R.~G.,  {Font} A.~S.,   {McCarthy}
  I.~G.,  2009, \mn@doi [\mnras] {10.1111/j.1365-2966.2009.15507.x}, \href
  {https://ui.adsabs.harvard.edu/abs/2009MNRAS.400..937M} {400, 937}

\bibitem[\protect\citeauthoryear{{Meyer}, {Robotham}, {Obreschkow},
  {Westmeier}, {Duffy}  \& {Staveley-Smith}}{{Meyer} et~al.}{2017}]{Meyer2017}
{Meyer} M.,  {Robotham} A.,  {Obreschkow} D.,  {Westmeier} T.,  {Duffy} A.~R.,
   {Staveley-Smith} L.,  2017, \mn@doi [\pasa] {10.1017/pasa.2017.31}, \href
  {https://ui.adsabs.harvard.edu/abs/2017PASA...34...52M} {34, 52}

\bibitem[\protect\citeauthoryear{{Moore}, {Katz}, {Lake}, {Dressler}  \&
  {Oemler}}{{Moore} et~al.}{1996}]{Moore1996}
{Moore} B.,  {Katz} N.,  {Lake} G.,  {Dressler} A.,   {Oemler} A.,  1996,
  \mn@doi [\nat] {10.1038/379613a0}, \href
  {https://ui.adsabs.harvard.edu/abs/1996Natur.379..613M} {379, 613}

\bibitem[\protect\citeauthoryear{{Morokuma-Matsui} et~al.,}{{Morokuma-Matsui}
  et~al.}{2019}]{Morokuma2019}
{Morokuma-Matsui} K.,  et~al., 2019, \mn@doi [\pasj] {10.1093/pasj/psz067},
  \href {https://ui.adsabs.harvard.edu/abs/2019PASJ...71...85M} {71, 85}

\bibitem[\protect\citeauthoryear{{Morokuma-Matsui} et~al.,}{{Morokuma-Matsui}
  et~al.}{2022}]{Morokuma2022}
{Morokuma-Matsui} K.,  et~al., 2022, \mn@doi [\apjs]
  {10.3847/1538-4365/ac983b}, \href
  {https://ui.adsabs.harvard.edu/abs/2022ApJS..263...40M} {263, 40}

\bibitem[\protect\citeauthoryear{{Mu{\~n}oz-Mateos}, {Gil de Paz}, {Boissier},
  {Zamorano}, {Jarrett}, {Gallego}  \& {Madore}}{{Mu{\~n}oz-Mateos}
  et~al.}{2007}]{Munoz-Mateos2007}
{Mu{\~n}oz-Mateos} J.~C.,  {Gil de Paz} A.,  {Boissier} S.,  {Zamorano} J.,
  {Jarrett} T.,  {Gallego} J.,   {Madore} B.~F.,  2007, \mn@doi [\apj]
  {10.1086/511812}, \href
  {https://ui.adsabs.harvard.edu/abs/2007ApJ...658.1006M} {658, 1006}

\bibitem[\protect\citeauthoryear{{Nulsen}}{{Nulsen}}{1982}]{Nulsen1982}
{Nulsen} P.~E.~J.,  1982, \mn@doi [\mnras] {10.1093/mnras/198.4.1007}, \href
  {https://ui.adsabs.harvard.edu/abs/1982MNRAS.198.1007N} {198, 1007}

\bibitem[\protect\citeauthoryear{{Ocvirk}, {Pichon}, {Lan{\c{c}}on}  \&
  {Thi{\'e}baut}}{{Ocvirk} et~al.}{2006}]{Ocvirk2006}
{Ocvirk} P.,  {Pichon} C.,  {Lan{\c{c}}on} A.,   {Thi{\'e}baut} E.,  2006,
  \mn@doi [\mnras] {10.1111/j.1365-2966.2005.09182.x}, \href
  {https://ui.adsabs.harvard.edu/abs/2006MNRAS.365...46O} {365, 46}

\bibitem[\protect\citeauthoryear{{Peletier} et~al.,}{{Peletier}
  et~al.}{2020}]{Peletier2020}
{Peletier} R.,  et~al., 2020, \mn@doi [arXiv e-prints]
  {10.48550/arXiv.2008.12633}, \href
  {https://ui.adsabs.harvard.edu/abs/2020arXiv200812633P} {p. arXiv:2008.12633}

\bibitem[\protect\citeauthoryear{{Peng} et~al.,}{{Peng}
  et~al.}{2010}]{Peng2010}
{Peng} Y.-j.,  et~al., 2010, \mn@doi [\apj] {10.1088/0004-637X/721/1/193},
  \href {https://ui.adsabs.harvard.edu/abs/2010ApJ...721..193P} {721, 193}

\bibitem[\protect\citeauthoryear{{Peng}, {Lilly}, {Renzini}  \&
  {Carollo}}{{Peng} et~al.}{2012}]{Peng2012}
{Peng} Y.-j.,  {Lilly} S.~J.,  {Renzini} A.,   {Carollo} M.,  2012, \mn@doi
  [\apj] {10.1088/0004-637X/757/1/4}, \href
  {https://ui.adsabs.harvard.edu/abs/2012ApJ...757....4P} {757, 4}

\bibitem[\protect\citeauthoryear{{Porter}, {Raychaudhury}, {Pimbblet}  \&
  {Drinkwater}}{{Porter} et~al.}{2008}]{Porter2008}
{Porter} S.~C.,  {Raychaudhury} S.,  {Pimbblet} K.~A.,   {Drinkwater} M.~J.,
  2008, \mn@doi [\mnras] {10.1111/j.1365-2966.2008.13388.x}, \href
  {https://ui.adsabs.harvard.edu/abs/2008MNRAS.388.1152P} {388, 1152}

\bibitem[\protect\citeauthoryear{{Prugniel}, {Koleva}, {Ocvirk}, {Le Borgne}
  \& {Soubiran}}{{Prugniel} et~al.}{2007}]{Prugniel2007}
{Prugniel} P.,  {Koleva} M.,  {Ocvirk} P.,  {Le Borgne} D.,   {Soubiran} C.,
  2007, in {Vazdekis} A.,  {Peletier} R.,  eds,  IAU Symposium Vol. 241,
  Stellar Populations as Building Blocks of Galaxies. pp 68--72 (\mn@eprint
  {arXiv} {astro-ph/0703130}), \mn@doi{10.1017/S1743921307007454}

\bibitem[\protect\citeauthoryear{{Quai}, {Pozzetti}, {Citro}, {Moresco}  \&
  {Cimatti}}{{Quai} et~al.}{2018}]{Quai2018}
{Quai} S.,  {Pozzetti} L.,  {Citro} A.,  {Moresco} M.,   {Cimatti} A.,  2018,
  \mn@doi [\mnras] {10.1093/mnras/sty1045}, \href
  {https://ui.adsabs.harvard.edu/abs/2018MNRAS.478.3335Q} {478, 3335}

\bibitem[\protect\citeauthoryear{{Raj} et~al.,}{{Raj} et~al.}{2020}]{Raj2020}
{Raj} M.~A.,  et~al., 2020, \mn@doi [\aap] {10.1051/0004-6361/202038043}, \href
  {https://ui.adsabs.harvard.edu/abs/2020A&A...640A.137R} {640, A137}

\bibitem[\protect\citeauthoryear{{Ramatsoku} et~al.,}{{Ramatsoku}
  et~al.}{2020}]{Ramatsoku2020}
{Ramatsoku} M.,  et~al., 2020, \mn@doi [\aap] {10.1051/0004-6361/202037800},
  \href {https://ui.adsabs.harvard.edu/abs/2020A&A...636L...1R} {636, L1}

\bibitem[\protect\citeauthoryear{{Rasmussen}, {Ponman}, {Verdes-Montenegro},
  {Yun}  \& {Borthakur}}{{Rasmussen} et~al.}{2008}]{Rasmussen2008}
{Rasmussen} J.,  {Ponman} T.~J.,  {Verdes-Montenegro} L.,  {Yun} M.~S.,
  {Borthakur} S.,  2008, \mn@doi [\mnras] {10.1111/j.1365-2966.2008.13451.x},
  \href {https://ui.adsabs.harvard.edu/abs/2008MNRAS.388.1245R} {388, 1245}

\bibitem[\protect\citeauthoryear{{Ratsimbazafy}, {Loubser}, {Crawford},
  {Cress}, {Bassett}, {Nichol}  \& {V{\"a}is{\"a}nen}}{{Ratsimbazafy}
  et~al.}{2017}]{Ratsimbazafy2017}
{Ratsimbazafy} A.~L.,  {Loubser} S.~I.,  {Crawford} S.~M.,  {Cress} C.~M.,
  {Bassett} B.~A.,  {Nichol} R.~C.,   {V{\"a}is{\"a}nen} P.,  2017, \mn@doi
  [\mnras] {10.1093/mnras/stx301}, \href
  {https://ui.adsabs.harvard.edu/abs/2017MNRAS.467.3239R} {467, 3239}

\bibitem[\protect\citeauthoryear{{Romero-G{\'o}mez} et~al.,}{{Romero-G{\'o}mez}
  et~al.}{2023}]{Romero-Gomez2023a}
{Romero-G{\'o}mez} J.,  et~al., 2023, \mn@doi [\mnras] {10.1093/mnras/stad953},
  \href {https://ui.adsabs.harvard.edu/abs/2023MNRAS.522..130R} {522, 130}

\bibitem[\protect\citeauthoryear{{Salpeter}}{{Salpeter}}{1955}]{Salpeter1955}
{Salpeter} E.~E.,  1955, \mn@doi [ApJ] {10.1086/145971}, \href
  {http://adsabs.harvard.edu/abs/1955ApJ...121..161S} {121, 161}

\bibitem[\protect\citeauthoryear{{S{\'a}nchez-Bl{\'a}zquez}
  et~al.,}{{S{\'a}nchez-Bl{\'a}zquez} et~al.}{2006}]{Sanchez2006}
{S{\'a}nchez-Bl{\'a}zquez} P.,  et~al., 2006, \mn@doi [MNRAS]
  {10.1111/j.1365-2966.2006.10699.x}, \href
  {http://saaoads.chpc.ac.za/abs/2006MNRAS.371..703S} {371, 703}

\bibitem[\protect\citeauthoryear{{Sarzi} et~al.,}{{Sarzi}
  et~al.}{2006}]{Sarzi2006}
{Sarzi} M.,  et~al., 2006, \mn@doi [MNRAS] {10.1111/j.1365-2966.2005.09839.x},
  \href {http://saaoads.chpc.ac.za/abs/2006MNRAS.366.1151S} {366, 1151}

\bibitem[\protect\citeauthoryear{{Sarzi} et~al.,}{{Sarzi}
  et~al.}{2018}]{Sarzi2018}
{Sarzi} M.,  et~al., 2018, \mn@doi [\aap] {10.1051/0004-6361/201833137}, \href
  {https://ui.adsabs.harvard.edu/abs/2018A&A...616A.121S} {616, A121}

\bibitem[\protect\citeauthoryear{{Schawinski} et~al.,}{{Schawinski}
  et~al.}{2014}]{Schawinski2014}
{Schawinski} K.,  et~al., 2014, \mn@doi [\mnras] {10.1093/mnras/stu327}, \href
  {https://ui.adsabs.harvard.edu/abs/2014MNRAS.440..889S} {440, 889}

\bibitem[\protect\citeauthoryear{{Schlafly} \& {Finkbeiner}}{{Schlafly} \&
  {Finkbeiner}}{2011}]{Schlafly2011}
{Schlafly} E.~F.,  {Finkbeiner} D.~P.,  2011, \mn@doi [\apj]
  {10.1088/0004-637X/737/2/103}, \href
  {http://adsabs.harvard.edu/abs/2011ApJ...737..103S} {737, 103}

\bibitem[\protect\citeauthoryear{{Schlegel}, {Finkbeiner}  \&
  {Davis}}{{Schlegel} et~al.}{1998}]{Schlegel1998}
{Schlegel} D.~J.,  {Finkbeiner} D.~P.,   {Davis} M.,  1998, \mn@doi [ApJ]
  {10.1086/305772}, \href {http://saaoads.chpc.ac.za/abs/1998ApJ...500..525S}
  {500, 525}

\bibitem[\protect\citeauthoryear{{Schweizer}}{{Schweizer}}{1980}]{Schweizer1980}
{Schweizer} F.,  1980, \mn@doi [\apj] {10.1086/157870}, \href
  {https://ui.adsabs.harvard.edu/abs/1980ApJ...237..303S} {237, 303}

\bibitem[\protect\citeauthoryear{{Serra} et~al.,}{{Serra}
  et~al.}{2016}]{Serra2016}
{Serra} P.,  et~al., 2016, in MeerKAT Science: On the Pathway to the SKA. p.~8
  (\mn@eprint {arXiv} {1709.01289}), \mn@doi{10.22323/1.277.0008}

\bibitem[\protect\citeauthoryear{{Serra} et~al.,}{{Serra}
  et~al.}{2019}]{Serra2019}
{Serra} P.,  et~al., 2019, \mn@doi [\aap] {10.1051/0004-6361/201936114}, \href
  {https://ui.adsabs.harvard.edu/abs/2019A&A...628A.122S} {628, A122}

\bibitem[\protect\citeauthoryear{{Serra} et~al.,}{{Serra}
  et~al.}{2023}]{Serra2023}
{Serra} P.,  et~al., 2023, \mn@doi [\aap] {10.1051/0004-6361/202346071}, \href
  {https://ui.adsabs.harvard.edu/abs/2023A&A...673A.146S} {673, A146}

\bibitem[\protect\citeauthoryear{{Skrutskie} et~al.,}{{Skrutskie}
  et~al.}{2006}]{Skrutskie2006}
{Skrutskie} M.~F.,  et~al., 2006, \mn@doi [\aj] {10.1086/498708}, \href
  {https://ui.adsabs.harvard.edu/abs/2006AJ....131.1163S} {131, 1163}

\bibitem[\protect\citeauthoryear{{Su}, {Salo}, {Janz}, {Venhola}  \&
  {Peletier}}{{Su} et~al.}{2022}]{Su2022}
{Su} A.~H.,  {Salo} H.,  {Janz} J.,  {Venhola} A.,   {Peletier} R.~F.,  2022,
  \mn@doi [\aap] {10.1051/0004-6361/202142593}, \href
  {https://ui.adsabs.harvard.edu/abs/2022A&A...664A.167S} {664, A167}

\bibitem[\protect\citeauthoryear{{Taylor} et~al.,}{{Taylor}
  et~al.}{2011}]{Taylor2011}
{Taylor} E.~N.,  et~al., 2011, \mn@doi [\mnras]
  {10.1111/j.1365-2966.2011.19536.x}, \href
  {https://ui.adsabs.harvard.edu/abs/2011MNRAS.418.1587T} {418, 1587}

\bibitem[\protect\citeauthoryear{{Vazdekis}, {S{\'a}nchez-Bl{\'a}zquez},
  {Falc{\'o}n-Barroso}, {Cenarro}, {Beasley}, {Cardiel}, {Gorgas}  \&
  {Peletier}}{{Vazdekis} et~al.}{2010}]{Vazdekis2010}
{Vazdekis} A.,  {S{\'a}nchez-Bl{\'a}zquez} P.,  {Falc{\'o}n-Barroso} J.,
  {Cenarro} A.~J.,  {Beasley} M.~A.,  {Cardiel} N.,  {Gorgas} J.,   {Peletier}
  R.~F.,  2010, \mn@doi [MNRAS] {10.1111/j.1365-2966.2010.16407.x}, \href
  {http://saaoads.chpc.ac.za/abs/2010MNRAS.404.1639V} {404, 1639}

\bibitem[\protect\citeauthoryear{{Vazdekis} et~al.,}{{Vazdekis}
  et~al.}{2015}]{Vazdekis2015}
{Vazdekis} A.,  et~al., 2015, \mn@doi [MNRAS] {10.1093/mnras/stv151}, \href
  {http://adsabs.harvard.edu/abs/2015MNRAS.449.1177V} {449, 1177}

\bibitem[\protect\citeauthoryear{{Venhola} et~al.,}{{Venhola}
  et~al.}{2018}]{Venhola2018}
{Venhola} A.,  et~al., 2018, \mn@doi [\aap] {10.1051/0004-6361/201833933},
  \href {https://ui.adsabs.harvard.edu/abs/2018A&A...620A.165V} {620, A165}

\bibitem[\protect\citeauthoryear{{Venhola} et~al.,}{{Venhola}
  et~al.}{2019}]{Venhola2019}
{Venhola} A.,  et~al., 2019, \mn@doi [\aap] {10.1051/0004-6361/201935231},
  \href {https://ui.adsabs.harvard.edu/abs/2019A&A...625A.143V} {625, A143}

\bibitem[\protect\citeauthoryear{{Wang} \& {Lilly}}{{Wang} \&
  {Lilly}}{2020}]{Wang2020}
{Wang} E.,  {Lilly} S.~J.,  2020, \mn@doi [\apj] {10.3847/1538-4357/ab7b7d},
  \href {https://ui.adsabs.harvard.edu/abs/2020ApJ...892...87W} {892, 87}

\bibitem[\protect\citeauthoryear{{Wang} et~al.,}{{Wang}
  et~al.}{2022}]{Wang2022}
{Wang} D.,  et~al., 2022, \mn@doi [\mnras] {10.1093/mnras/stac2428}, \href
  {https://ui.adsabs.harvard.edu/abs/2022MNRAS.516.3411W} {516, 3411}

\bibitem[\protect\citeauthoryear{{Weibel}, {Wang}  \& {Lilly}}{{Weibel}
  et~al.}{2023}]{Weibel2023}
{Weibel} A.,  {Wang} E.,   {Lilly} S.~J.,  2023, \mn@doi [\apj]
  {10.3847/1538-4357/accffc}, \href
  {https://ui.adsabs.harvard.edu/abs/2023ApJ...950..102W} {950, 102}

\bibitem[\protect\citeauthoryear{{Weinmann}, {van den Bosch}, {Yang}  \&
  {Mo}}{{Weinmann} et~al.}{2006}]{Weinmann2006}
{Weinmann} S.~M.,  {van den Bosch} F.~C.,  {Yang} X.,   {Mo} H.~J.,  2006,
  \mn@doi [\mnras] {10.1111/j.1365-2966.2005.09865.x}, \href
  {https://ui.adsabs.harvard.edu/abs/2006MNRAS.366....2W} {366, 2}

\bibitem[\protect\citeauthoryear{{Wetzel}, {Tinker}, {Conroy}  \& {van den
  Bosch}}{{Wetzel} et~al.}{2013}]{Wetzel2013}
{Wetzel} A.~R.,  {Tinker} J.~L.,  {Conroy} C.,   {van den Bosch} F.~C.,  2013,
  \mn@doi [\mnras] {10.1093/mnras/stt469}, \href
  {https://ui.adsabs.harvard.edu/abs/2013MNRAS.432..336W} {432, 336}

\bibitem[\protect\citeauthoryear{{Wilkinson}, {Maraston}, {Goddard}, {Thomas}
  \& {Parikh}}{{Wilkinson} et~al.}{2017}]{Wilkinson2017}
{Wilkinson} D.~M.,  {Maraston} C.,  {Goddard} D.,  {Thomas} D.,   {Parikh} T.,
  2017, \mn@doi [\mnras] {10.1093/mnras/stx2215}, \href
  {https://ui.adsabs.harvard.edu/abs/2017MNRAS.472.4297W} {472, 4297}

\bibitem[\protect\citeauthoryear{{Woo} et~al.,}{{Woo} et~al.}{2013}]{Woo2013}
{Woo} J.,  et~al., 2013, \mn@doi [\mnras] {10.1093/mnras/sts274}, \href
  {https://ui.adsabs.harvard.edu/abs/2013MNRAS.428.3306W} {428, 3306}

\bibitem[\protect\citeauthoryear{{Worthey}}{{Worthey}}{1994}]{Worthey1994}
{Worthey} G.,  1994, \mn@doi [\apjs] {10.1086/192096}, \href
  {https://ui.adsabs.harvard.edu/abs/1994ApJS...95..107W} {95, 107}

\bibitem[\protect\citeauthoryear{{Yoon}, {Chung}, {Smith}  \&
  {Jaff{\'e}}}{{Yoon} et~al.}{2017}]{Yoon2017}
{Yoon} H.,  {Chung} A.,  {Smith} R.,   {Jaff{\'e}} Y.~L.,  2017, \mn@doi [\apj]
  {10.3847/1538-4357/aa6579}, \href
  {https://ui.adsabs.harvard.edu/abs/2017ApJ...838...81Y} {838, 81}

\bibitem[\protect\citeauthoryear{{Zabel} et~al.,}{{Zabel}
  et~al.}{2019}]{Zabel2019}
{Zabel} N.,  et~al., 2019, \mn@doi [\mnras] {10.1093/mnras/sty3234}, \href
  {https://ui.adsabs.harvard.edu/abs/2019MNRAS.483.2251Z} {483, 2251}

\bibitem[\protect\citeauthoryear{{Zabludoff} \& {Mulchaey}}{{Zabludoff} \&
  {Mulchaey}}{1998}]{Zabludoff1998}
{Zabludoff} A.~I.,  {Mulchaey} J.~S.,  1998, \mn@doi [\apj] {10.1086/305355},
  \href {https://ui.adsabs.harvard.edu/abs/1998ApJ...496...39Z} {496, 39}

\bibitem[\protect\citeauthoryear{{Zabludoff}, {Zaritsky}, {Lin}, {Tucker},
  {Hashimoto}, {Shectman}, {Oemler}  \& {Kirshner}}{{Zabludoff}
  et~al.}{1996}]{Zabludoff1996}
{Zabludoff} A.~I.,  {Zaritsky} D.,  {Lin} H.,  {Tucker} D.,  {Hashimoto} Y.,
  {Shectman} S.~A.,  {Oemler} A.,   {Kirshner} R.~P.,  1996, \mn@doi [\apj]
  {10.1086/177495}, \href
  {https://ui.adsabs.harvard.edu/abs/1996ApJ...466..104Z} {466, 104}

\bibitem[\protect\citeauthoryear{{Zinn}, {Middelberg}, {Norris}  \&
  {Dettmar}}{{Zinn} et~al.}{2013}]{Zinn2013}
{Zinn} P.~C.,  {Middelberg} E.,  {Norris} R.~P.,   {Dettmar} R.~J.,  2013,
  \mn@doi [\apj] {10.1088/0004-637X/774/1/66}, \href
  {https://ui.adsabs.harvard.edu/abs/2013ApJ...774...66Z} {774, 66}

\bibitem[\protect\citeauthoryear{{de Blok} et~al.,}{{de Blok}
  et~al.}{2018}]{deBlok2018}
{de Blok} W.~J.~G.,  et~al., 2018, \mn@doi [\apj] {10.3847/1538-4357/aad557},
  \href {https://ui.adsabs.harvard.edu/abs/2018ApJ...865...26D} {865, 26}

\bibitem[\protect\citeauthoryear{{de Jong}}{{de Jong}}{1996}]{deJong1996}
{de Jong} R.~S.,  1996, \mn@doi [\aap] {10.48550/arXiv.astro-ph/9601005}, \href
  {https://ui.adsabs.harvard.edu/abs/1996A&A...313...45D} {313, 45}

\makeatother
\end{thebibliography}



\clearpage

\appendix

\section{Best-fitting FIREFLY combinations of single-burst stellar population models}
\label{SFHs}

We show the best-fitting FIREFLY combinations of single-burst stellar population models to our spectra here. As discussed in Section \ref{stelpopmodels}, we focus on the differences between the inner (left) and outer apertures (right). This is complementary to the mass-weighted SSP-equivalent age gradients represented in Fig. \ref{fig:Agesplots}. In all cases stellar templates below 1 Gyr were excluded, except for NGC 1316 (as discussed in Section \ref{stelpopmodels}). 

\begin{figure}
\centering
\subfloat{\includegraphics[scale=0.28, trim=90 240 80 290, clip]{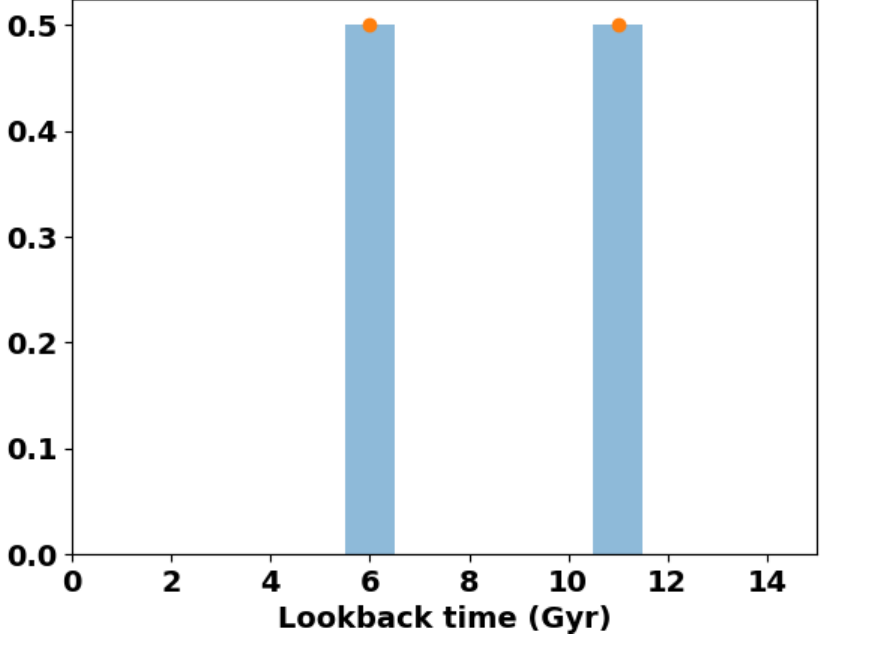}}
\subfloat{\includegraphics[scale=0.28, trim=90 240 80 290, clip]{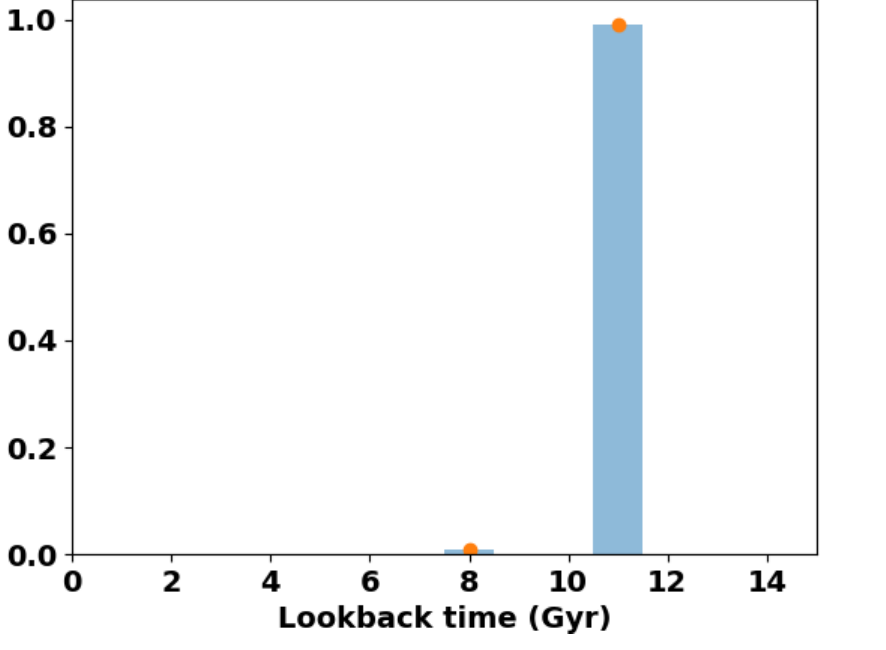}}
   \caption{NGC 1326B.}
\label{fig:SFH_NGC1326B}
\end{figure}

\begin{figure}
\centering
\subfloat{\includegraphics[scale=0.28, trim=90 240 80 290, clip]{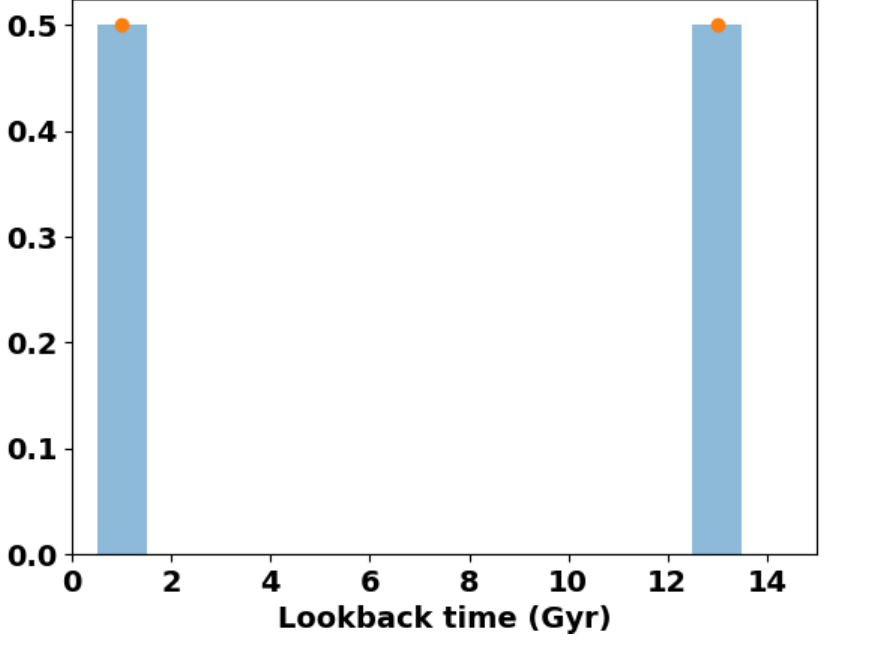}}
\subfloat{\includegraphics[scale=0.28, trim=90 240 80 290, clip]{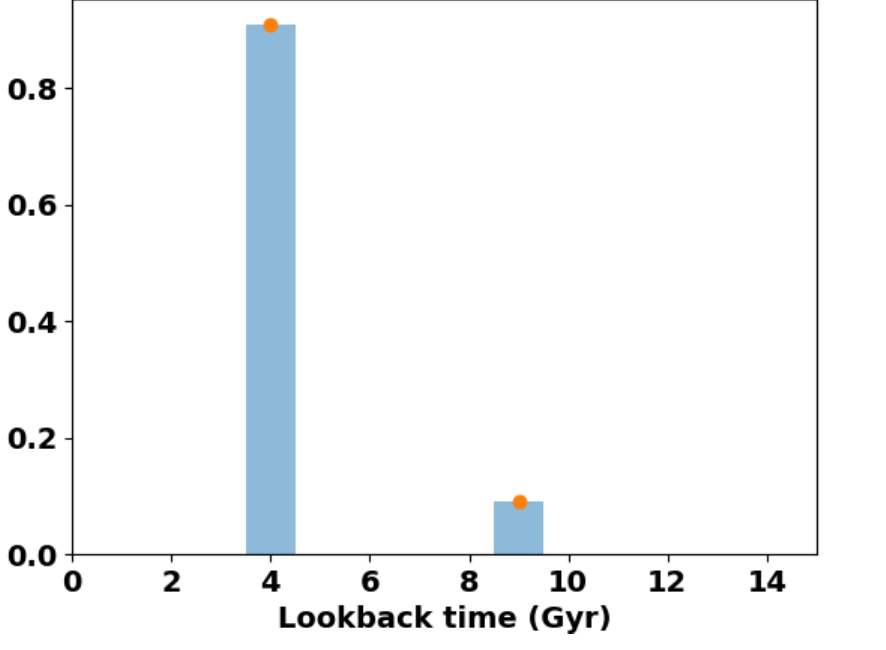}}
   \caption{NGC 1310.}
\label{fig:SFH_NGC1310}
\end{figure}

\begin{figure}
\centering
\subfloat{\includegraphics[scale=0.28, trim=90 240 80 290, clip]{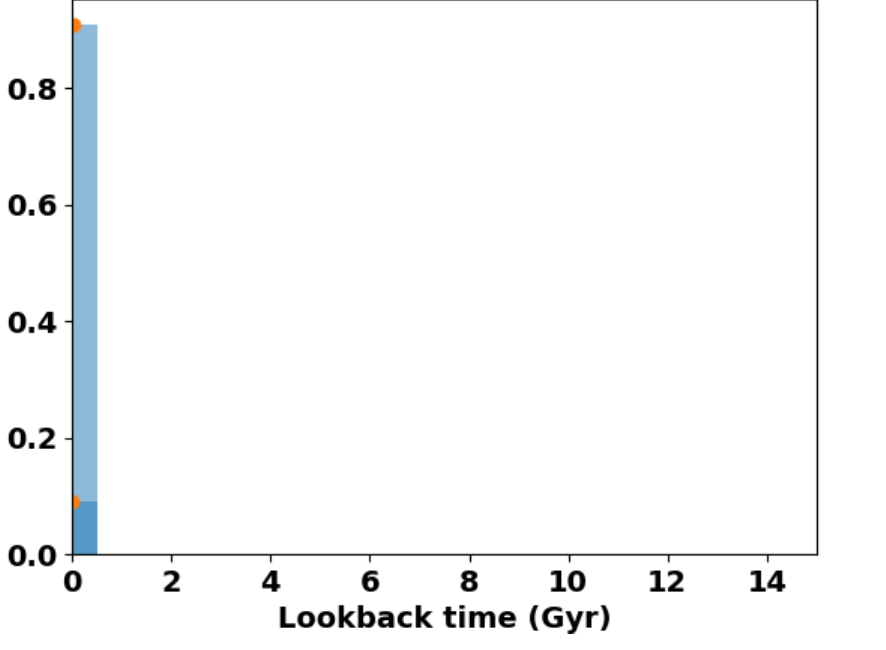}}
\subfloat{\includegraphics[scale=0.28, trim=90 240 80 290, clip]{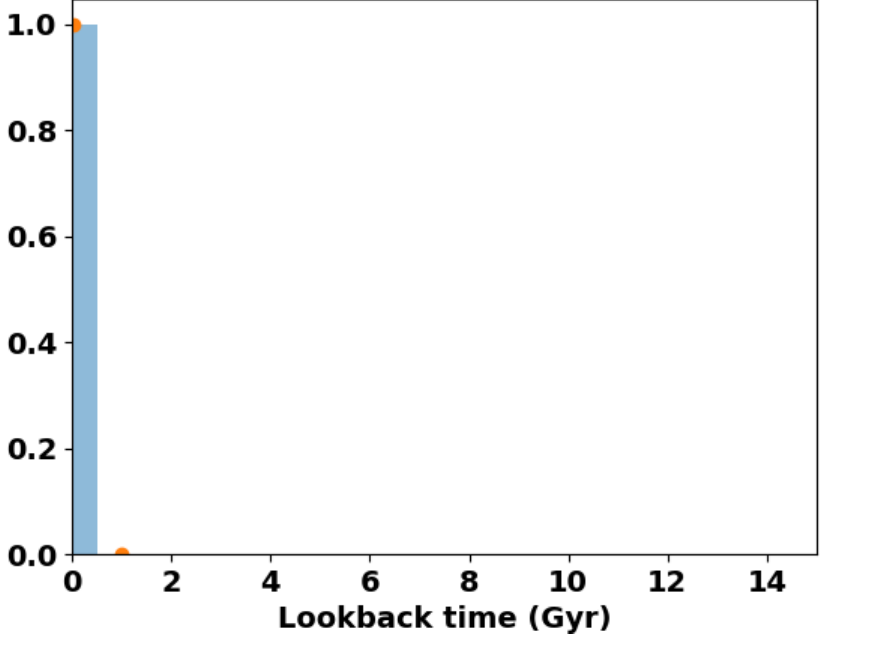}}
   \caption{NGC 1316.}
\label{fig:SFH_NGC1316}
\end{figure}

\begin{figure}
\centering
\subfloat{\includegraphics[scale=0.28, trim=90 240 80 290, clip]{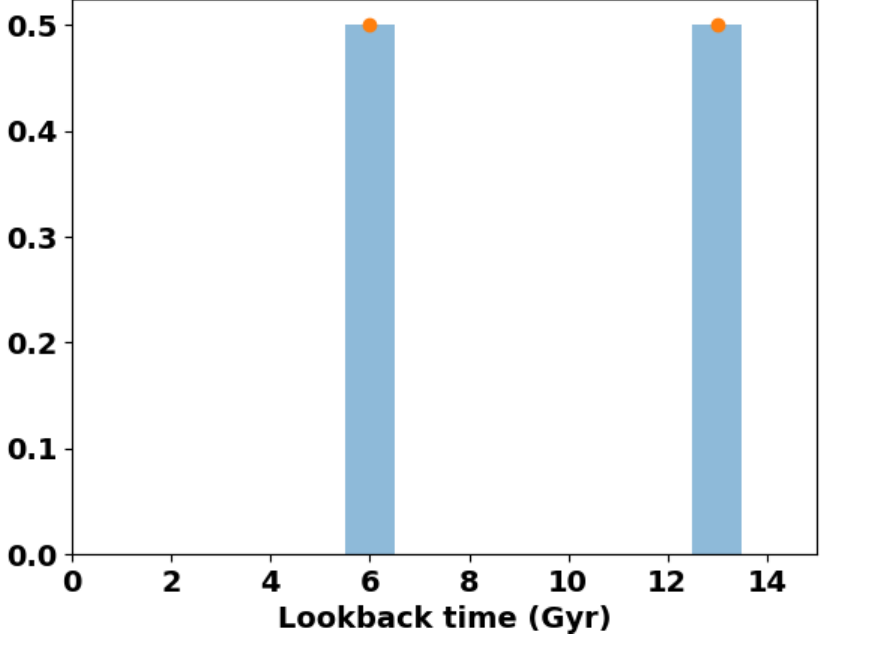}}
\subfloat{\includegraphics[scale=0.28, trim=90 240 80 290, clip]{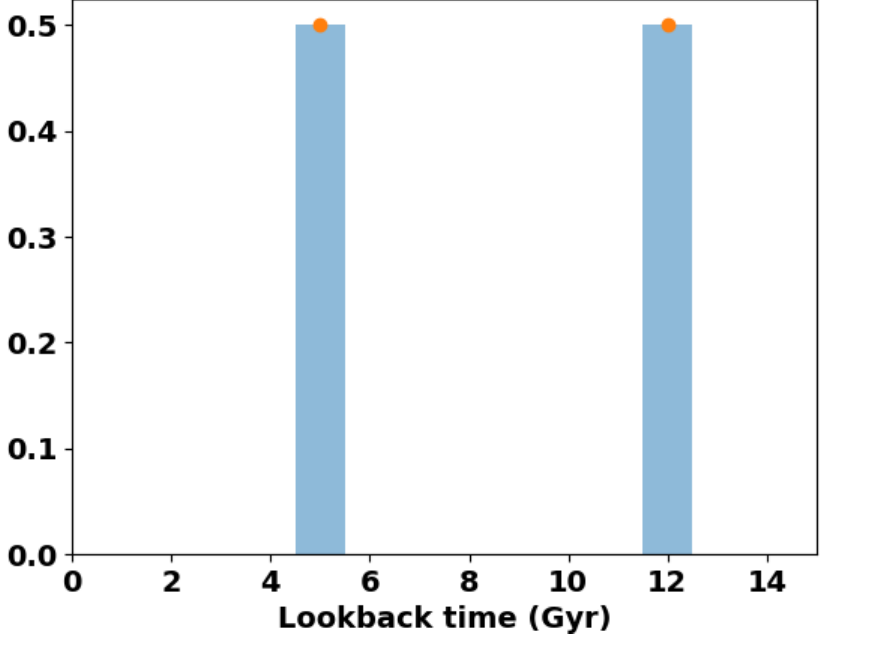}}
   \caption{ESO 301-IG11.}
\label{fig:SFH_ESO 301-IG11}
\end{figure}

\begin{figure}
\centering
\subfloat{\includegraphics[scale=0.28, trim=90 240 80 290, clip]{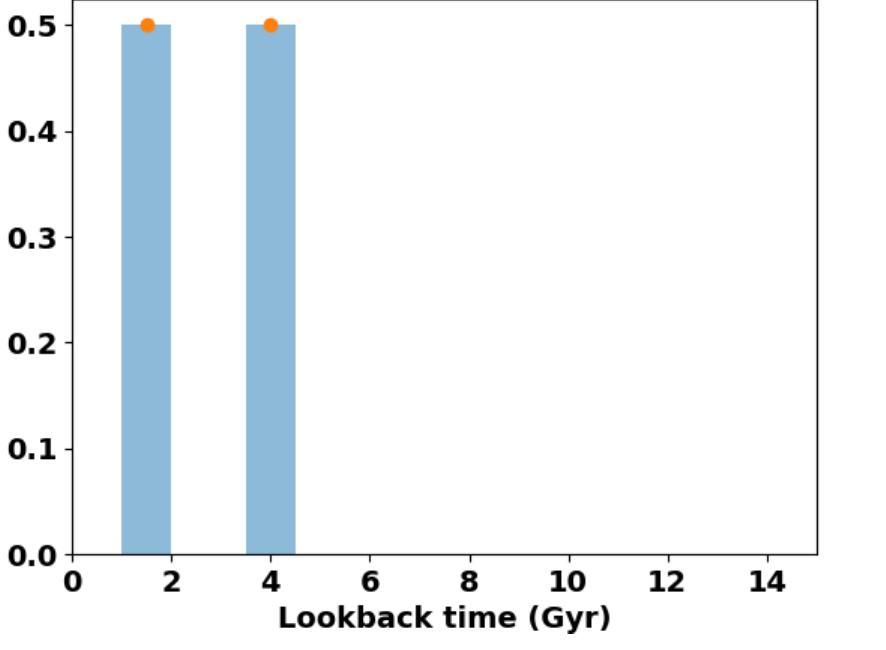}}
\subfloat{\includegraphics[scale=0.28, trim=90 240 80 290, clip]{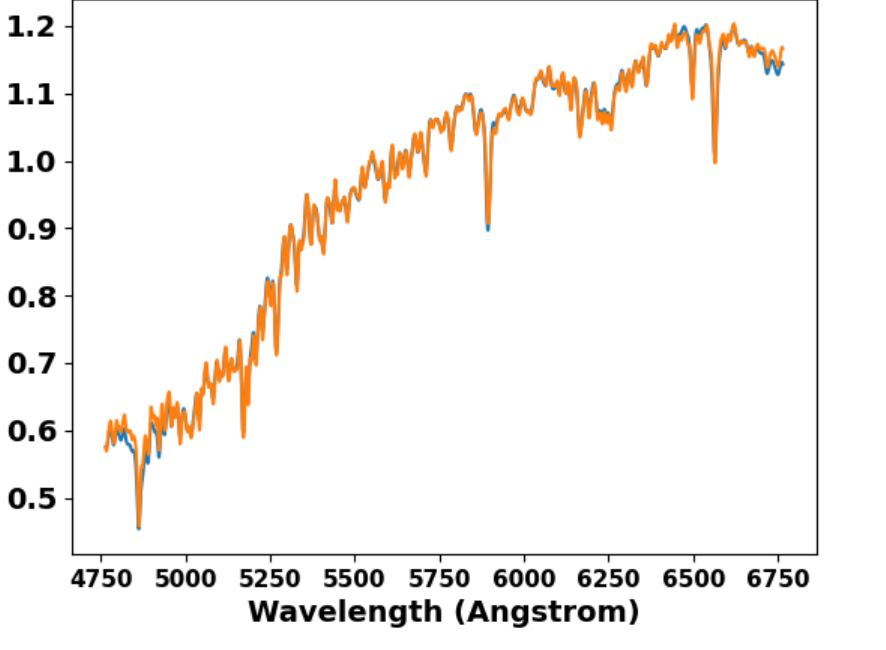}}
   \caption{NGC 1326. As we only extracted a central aperture for NGC 1326, we show the spectral fit and star formation history of the central spectrum instead of the star formation history of the central versus outer apertures.}
\label{fig:SFH_NGC1326}
\end{figure}

\begin{figure}
\centering
\subfloat{\includegraphics[scale=0.28, trim=90 240 80 290, clip]{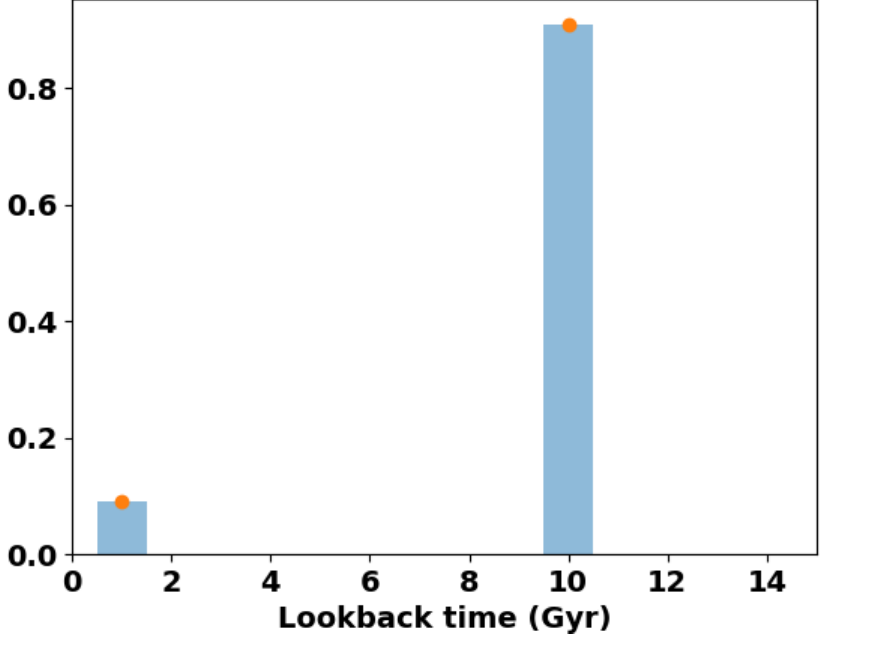}}
\subfloat{\includegraphics[scale=0.28, trim=90 240 80 290, clip]{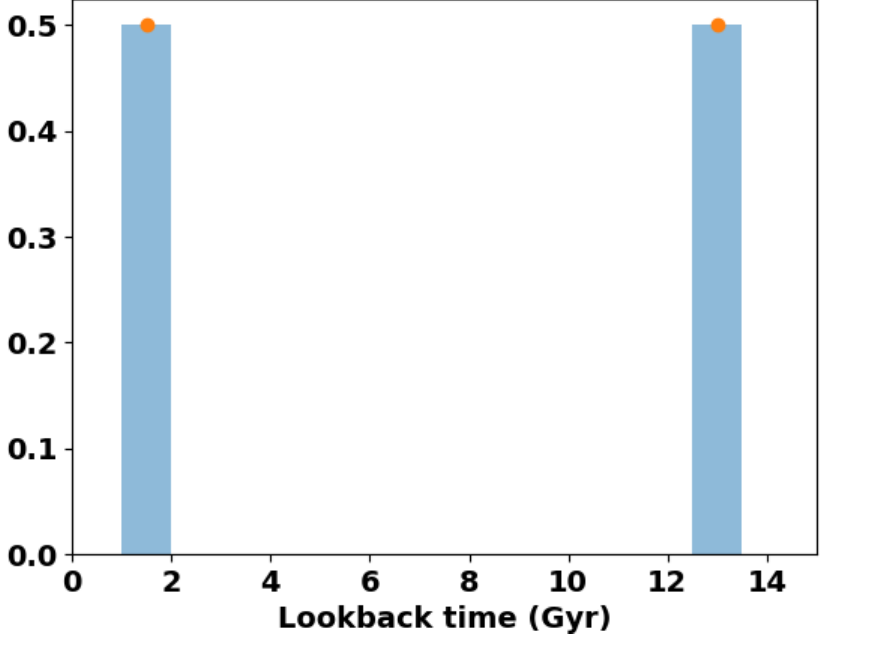}}
   \caption{FCC 35.}
\label{fig:SFH_FCC35}
\end{figure}

\begin{figure}
\centering
\subfloat{\includegraphics[scale=0.28, trim=90 240 80 290, clip]{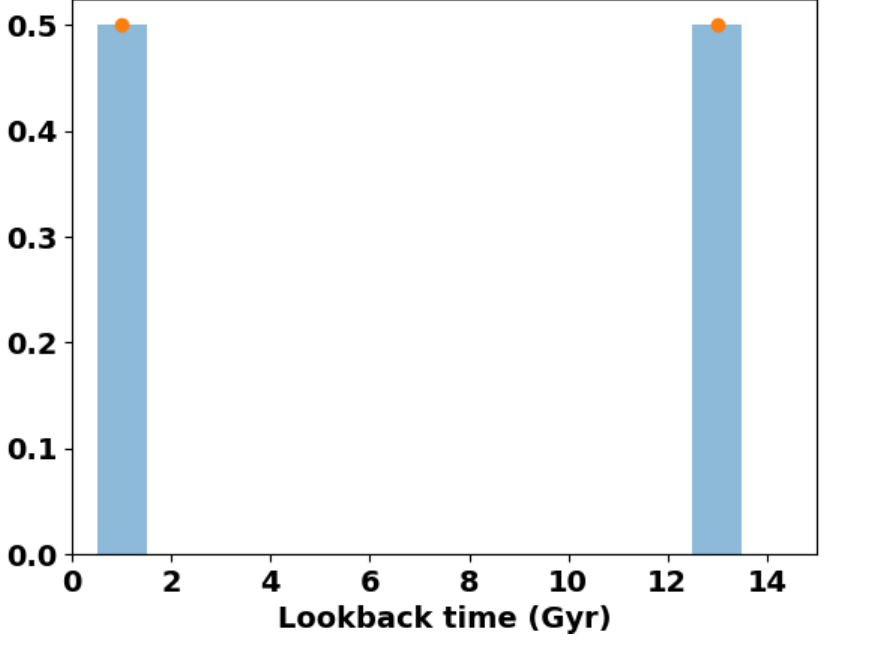}}
\subfloat{\includegraphics[scale=0.28, trim=90 240 80 290, clip]{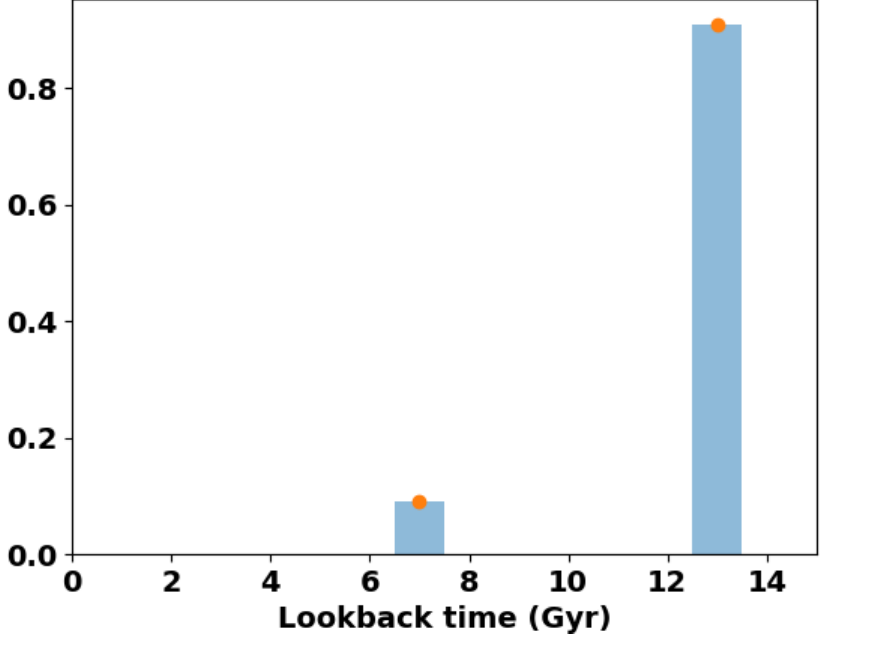}}
   \caption{NGC 1317.}
\label{fig:SFH_NGC1317}
\end{figure}

\begin{figure}
\centering
\subfloat{\includegraphics[scale=0.28, trim=90 240 80 290, clip]{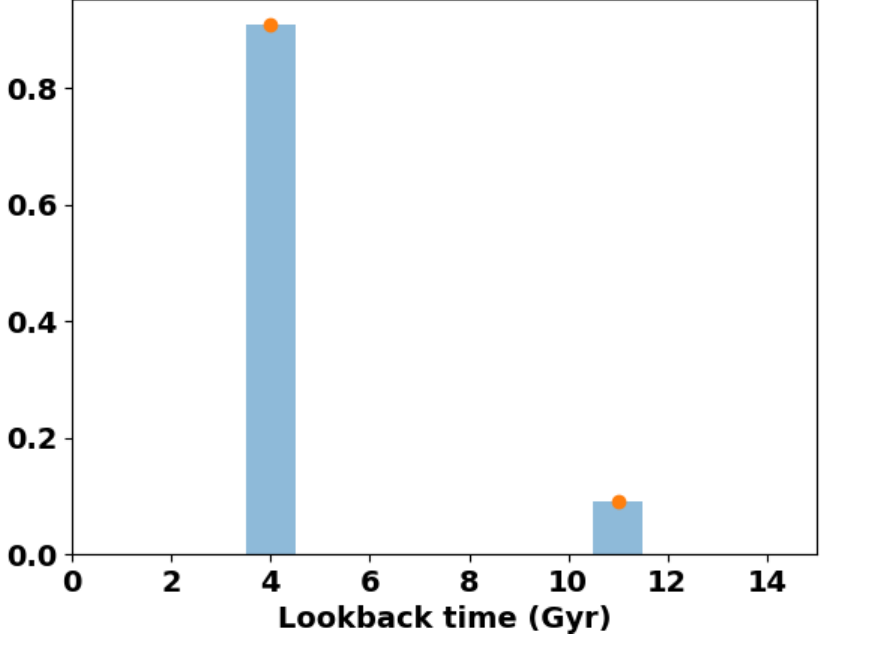}}
\subfloat{\includegraphics[scale=0.28, trim=90 240 80 290, clip]{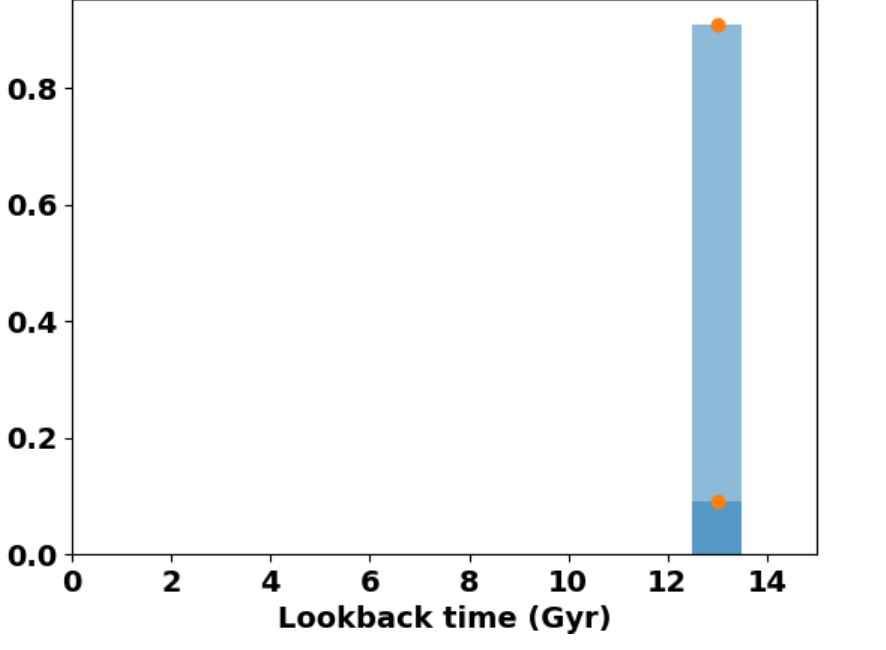}}
   \caption{NGC 1316C.}
\label{fig:SFH_NGC1316C}
\end{figure}

\begin{figure}
\centering
\subfloat{\includegraphics[scale=0.28, trim=90 240 80 290, clip]{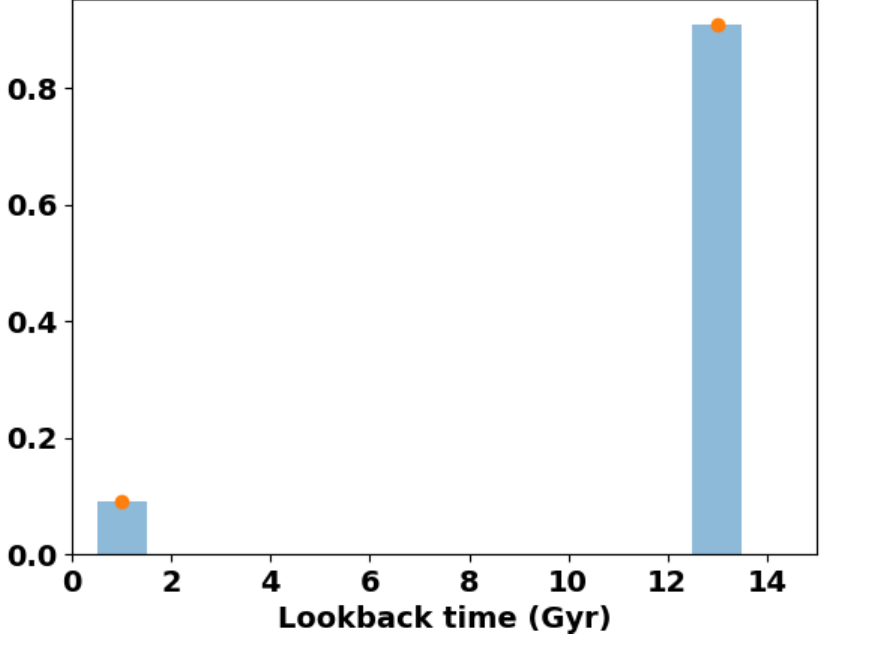}}
\subfloat{\includegraphics[scale=0.28, trim=90 240 80 290, clip]{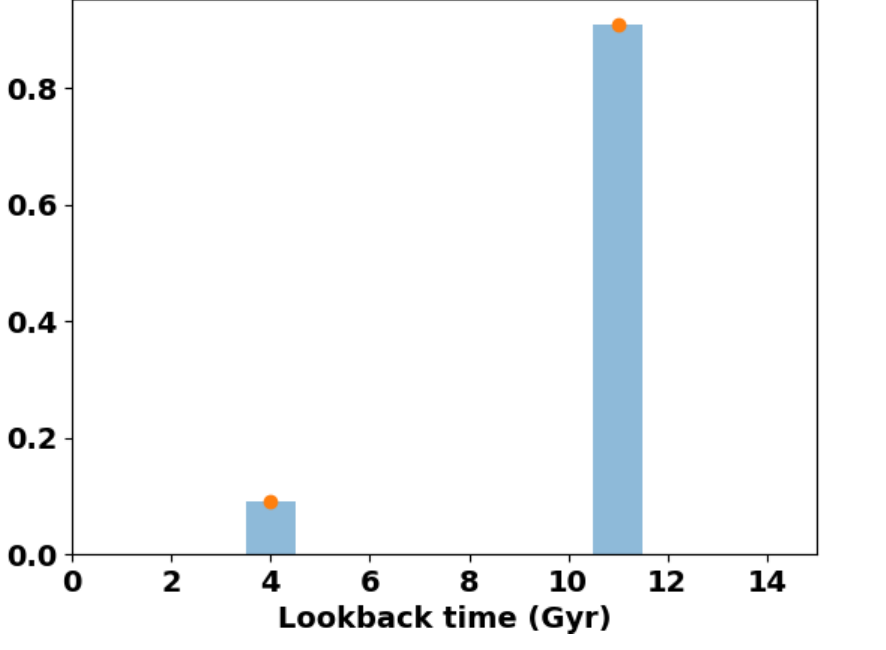}}
   \caption{FCC 46.}
\label{fig:SFH_FCC46}
\end{figure}

\begin{figure}
\centering
\subfloat{\includegraphics[scale=0.28, trim=90 240 80 290, clip]{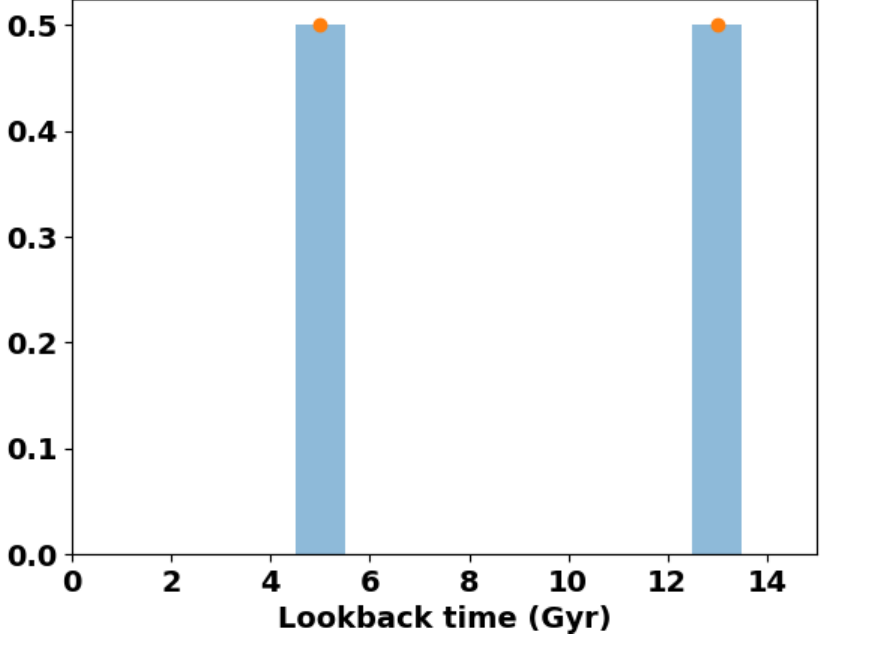}}
\subfloat{\includegraphics[scale=0.28, trim=90 240 80 290, clip]{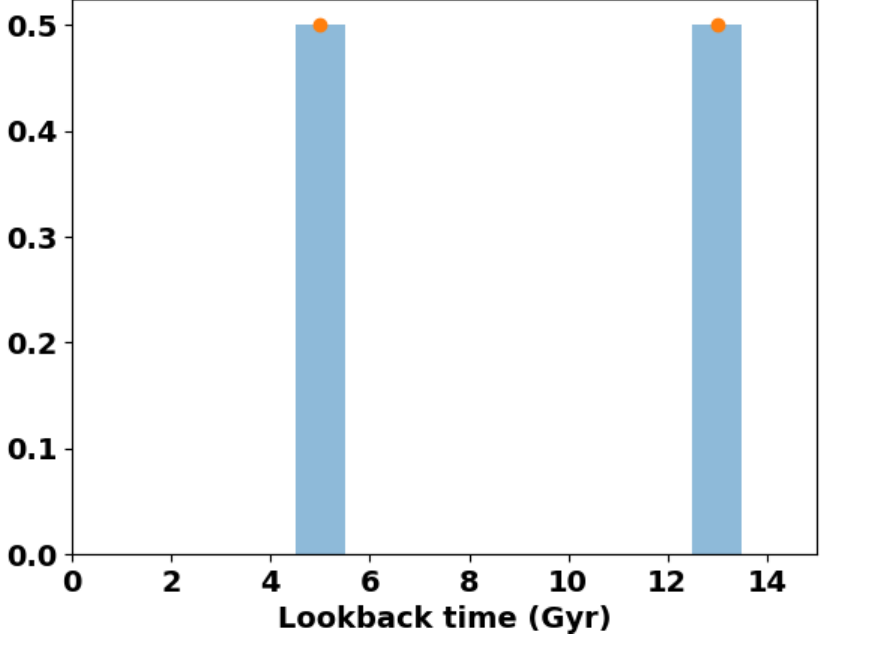}}
   \caption{NGC 1341.}
\label{fig:SFH_NGC1341}
\end{figure}


\bsp 
\label{lastpage}
\end{document}